\pdfoutput=1

\documentclass[preprint,fleqn,5p,numbers,sort&compress]{elsarticle}

\usepackage{graphicx}
\usepackage{amssymb,amsmath}
\usepackage{natbib}
\usepackage{hyperref}
\usepackage{mathtools,cuted}
\usepackage{commath}
\usepackage{siunitx}
\usepackage{gensymb}
\hypersetup{pdftitle = The title of my PDF, pdfauthor = My name, pdfsubject= The subject, pdfkeywords = keyword1 keyword2 keyword3}
\hypersetup{colorlinks = true, linkcolor = blue, anchorcolor = red, citecolor = blue, filecolor = red, pagecolor = red, urlcolor = blue}
\journal{International Journal of Non-Linear Mechanics}
\begin{document}
\begin{frontmatter}
\title{On the fractal basins of convergence of the libration points in the axisymmetric five-body problem: the convex configuration}
\author[mss]{Md Sanam Suraj\corref{cor1}}
\ead{mdsanamsuraj@gmail.com, mdsanamsuraj@aurobindo.du.ac.in}
\cortext[cor1]{Corresponding author}
\author[ps]{Prachi Sachan}
\author[eez]{Euaggelos E. Zotos}
\author[am]{Amit Mittal}
\author[rag]{Rajiv Aggarwal}
\address[mss]{Department of Mathematics,
Sri Aurobindo College, University of Delhi, Delhi-110017, India}
\address[ps]{Department of Mathematics,
University of Delhi, Delhi-110007, India}
\address[eez]{Department of Physics, School of Science,
Aristotle University of Thessaloniki, GR-541 24, Thessaloniki, Greece}
\address[am]{Department of Mathematics, ARSD College,
University of Delhi, Delhi-110021, India}
\address[rag]{Department of Mathematics, Deshbandhu College,
University of Delhi, Delhi-110019, India}
\begin{abstract}
In the present work, the Newton-Raphson basins of convergence, corresponding to the coplanar libration points (which act as numerical attractors), are unveiled in the axisymmetric five-body problem, where convex configuration is considered. In particular, the four primaries are set in axisymmetric central configuration, where the motion is governed only by mutual gravitational attractions. It is observed that the total number libration points are either eleven, thirteen or fifteen for different combination of the angle parameters. Moreover, the stability analysis revealed that the all the libration points are linearly stable for all the studied combination of angle parameters. The multivariate version of the Newton-Raphson iterative scheme is used to reveal the structures of the basins of convergence, associated with the libration points, on various types of two-dimensional configuration planes. In addition, we present how the basins of convergence are related with the corresponding number of required iterations. It is unveiled that in almost every cases, the basins of  convergence corresponding to the collinear libration point $L_2$ have infinite extent. Moreover, for some combination of the angle parameters, the collinear libration points $L_{1,2}$ have also infinite extent.  In addition, it can be observed that the domains of convergence, associated with the collinear libration point $L_1$, look like exotic bugs with many legs and antennas
whereas the domains of convergence, associated with $L_{4,5}$ look like butterfly wings for some combinations of angle parameters. Particularly, our numerical investigation suggests that the evolution of the attracting domains in this dynamical system is very complicated, yet a worthy studying problem.
\end{abstract}
\end{frontmatter}
\section{Introduction}
\label{intro}

Over the years, the problem of $N$-bodies has fascinated many researchers and scientists. In these days, one of the most important and well-studied dynamical models in celestial mechanics is the restricted five-body problem, where the fifth body, which is always referred as a test particle, is considered massless and it does not influence the motion of the four primaries which move in circular orbits around their common center of mass. A plethora of articles are available on the planar central configurations of $N$-bodies with $N = 4, 5$ and 7 (e.g., \cite{erd16}, \cite{ham05}, \cite{ham07}, \cite{mel13}). A series of papers are available on the restricted three and four body problem (e.g., \cite{AM14a,AM14b,AM15a,AM15b,AM15c, AG16,AMSB18}, \cite{EAK16}).

The gravitational five-body problem has been introduced by \cite{oll88}, where the motion of the fifth body of negligible mass was discussed, with respect to other bodies, known as primaries. It was assumed, that the three primaries, with equal masses, move in circular orbits on the same plane, around their common center of mass, while a mass of $\beta > 0$ times the mass of the equal primaries is situated at the center of mass (origin). Obviously, this model is reduced to the restricted four-body problem when $\beta = 0$ (that is when the central primary is missing). Furthermore, it was revealed that there exist nine libration points in which three of them are stable for $\beta > 43.18$, while all the other libration points are linearly unstable, for smaller values of the parameter $\beta$.

On the five-body problem there is a plethora of previous studies, such as \cite{kul11} and \cite{mar13} on the rhomboidal five-body problem, \cite{pap07} on the five-body problem, where some or even all the primary bodies are sources of radiation and \cite{gao17} on the axisymmetric five-body problem, when the four primaries are maintaining the axisymmetric central configuration.

In the present paper, we will discuss the basins of convergence, by applying the multivariate version of the Newton-Raphson method in the mathematical model of axisymmetric five-body problem. The analysis of the basins of convergence, associated with the libration points, provides various information regarding the intrinsic properties of the dynamical model. This should be true taking into account that the corresponding iterative scheme contains both the first as well as the second order partial derivative of the effective potential. Therefore it combines the dynamics of the test particle's orbit with the corresponding stability properties. Additionally, the basins of convergence provide the optimal initial conditions which can be used for numerically obtaining the position of the libration points (note that there are no analytic formulae for the position of the libration points). All the above-mentioned reasons explain and justify we need to know the convergence properties of a dynamical system.

The layout of the present paper is as follows: the most crucial properties of the dynamical system are described in Section \ref{Properties of the dynamical system}. The section \ref{the libration points and stability} explains the parametric evolution of the positions as well as of the stability of the libration points. The following section contains the main numerical results, corresponding to the features of the Newton-Raphson basins of convergence, where as in Section \ref{concluding remarks}, the main results are analyzed and most important conclusions are emphasized. The paper ends with the Section \ref{Future work} in which an outline of future related works is given.

\begin{figure}[!t]
\centering
\resizebox{\hsize}{!}{\includegraphics{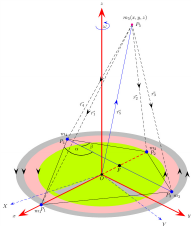}}
\caption{The convex configuration of the axisymmetric five-body problem, with the synodic coordinate system $Oxyz$ and the inertial frame $OXYZ$. More details are given in the text. (Color figure online).}
\label{Fig:1}
\end{figure}

\section{Properties of the dynamical system}
\label{Properties of the dynamical system}
In the present paper, we consider the scenario according to which a fifth body with infinitesimal mass $m_5$, always referred as the test particle, moves under the mutual gravitational attraction of four primaries, which are set in basic axisymmetric central configuration, as presented in \cite{erd16}.

We have taken $P_i, i=1,2,...,4$, with coordinates $(x_i, y_i, z_i)$, four point-like bodies with masses $m_i$, in the planar reference plane in which the $Ox$-axis is taken in such manner that the two primaries $m_{1,2}$ lie on it, while the $Oy$-axis is perpendicular to $Ox$-axis and passes through the center of mass of the four-body system with coordinates $(0, g, 0)$. The other two bodies, with masses $m_{3,4}$, are situated in the different half-planes determined by the $Ox$-axis. When we consider the case, i.e., $g=0, y_{3,4} \neq 0$, the center of mass coincides with the origin $O$ of the coordinate system and the $Ox-$axis becomes the axis of symmetry. Thus, the considered case leads to a symmetric configuration where two equal masses $m_{3,4}$ are situated opposite to each other and outside the axis of symmetry, while two bodies $m_{1,2}$ are situated on it. Moreover, the formed four-sided polygon is called convex configuration when the line joining the masses $m_{3,4}$ intersects the $Ox$-axis at the point $F$ which lies between the masses $m_{1,2}$ (see Fig. \ref{Fig:1}). In the present scenario, we assume that $(m_2 \geq m_1)$, where as $(m_1 \geq m_2)$ will be the mirror configuration.

According to \cite{erd16}, there are three possible configurations of four bodies with an axis of symmetry passing through the center of two bodies ($P_1$ and $P_2$), where as the two other bodies of equal masses ($P_3$ and $P_4$) are situated symmetrically with respect to this axis. Furthermore, the point $F$ located on the axis of symmetry, is the middle point of line joining the center of primaries $P_3$ and $P_4$. In addition, let us take $O'$, the center of mass of $P_1, P_3$ and $P_4$ and $O$ is the center of mass of the entire system. On the basis of the position of the points $F$, $O$, and $O'$, the following three configurations are possible: (i) the convex configuration: when the point $F$ lies between the primaries $P_1$  and $P_2$. (ii) first concave configuration: the primary $P_2$ lies between $F$ and $O' $, and (iii) second concave configuration: the primary $P_2$ lies between $O'$ and $P_1$.

In the present paper, we have considered only the first case, i.e., the convex configuration. According to \cite{gao17}, the expression of the time-independent effective potential, in a synodic coordinates system $Oxyz$ is
\begin{equation}\label{Eq:1}
\Omega(x, y, z)=\frac{1}{\Delta}\sum_{i=1}^{4}\frac{m_i}{r_i}+\frac{1}{2}\big(x^{2}+y^{2}\big),
\end{equation}
where $(x,y,z)$ are the coordinates of the test particle, and $(x_1, y_1, z_1)=(a, 0, 0)$,  $(x_2, y_2, z_2)=(b, 0, 0),$
$(x_3, y_3, z_3)=(c, 1, 0)$, and $(x_4, y_4, z_4)=(c, -1,0)$, while $r_i$ are the distances of the test particle from the primaries $P_i$, respectively and
\begin{align*}
r_{i}&=\sqrt{(x-x_i)^{2}+(y-y_i)^{2}+(z-z_i)^{2}},\\
m_3 &=m_4=m,\\
a&=(1-m_{1})\tan \alpha+m_{2}\tan \beta,\\
b&=-m_{1}\tan \alpha -(1-m_{2})\tan \beta,\\
m_{1}&=\frac{(b_{1}+a_{0}-b_{0})b_{0}}{a_{0}b_{1}+a_{1}b_{0}-a_{1}b_{1}},\\
m_{2}&=\frac{(a_{1}+b_{0}-a_{0})a_{0}}{a_{0}b_{1}+a_{1}b_{0}-a_{1}b_{1}},\\
a_0&=\big(\cos^3\alpha-\frac{1}{8}\big)\tan\alpha,\\
b_0&=\big(\cos^3\beta-\frac{1}{8}\big)\tan\beta,\\
a_1&=\big(\frac{1}{8}-\cos^3\alpha-\cos^3\beta \big)\tan\beta+\frac{1}{(\tan\alpha+\tan \beta)^2}\\
   &-\frac{1}{8}\tan\alpha,\\
b_1&=\big(\frac{1}{8}-\cos^3\alpha-\cos^3\beta \big)\tan\alpha+\frac{1}{(\tan\alpha+\tan\beta)^2}\\
   &-\frac{1}{8}\tan\beta,
\end{align*}
As discussed in \cite{erd16}, the three possible central configurations, in which two are concave configurations while the third one is convex configuration, are subjected to the angle coordinates $\alpha$ and $\beta$ and therefore must satisfy some conditions. In the present paper, where we consider the case of convex configuration only, the conditions to be simultaneously satisfied are as follows:
\begin{align}\label{Eq:2}
\alpha+2\beta& >90^\circ,\nonumber\\
\alpha&<60^\circ,\nonumber\\
\beta&<\alpha.
\end{align}
Moreover, the quantity $\Delta$ is the angular velocity of the synodic frame and given by the formula:
\begin{equation}\label{Eq:3}
\Delta=\frac{m_2/|P_1P_2|^2+2m\sin \alpha /|P_1P_3|^2}{P_1O}.
\end{equation}
Using the transformation from the inertial to the synodic coordinate system, while scaling the physical quantities, the equations of motion of the test particle in the rotating frame of reference read as:
\begin{subequations}
\begin{eqnarray}
\label{Eq:4a}
\ddot{x}-2\dot{y}&=&\Omega_x,\\
\label{Eq:4b}
\ddot{y}+2\dot{x}&=&\Omega_y,\\
\label{Eq:4c}
\ddot{z}&=&\Omega_z,
\end{eqnarray}
\end{subequations}
where
\begin{subequations}
\begin{eqnarray}
\label{Eq:5a}
\Omega_x&=&x-\frac{1}{\Delta}\sum _{i=1}^{4}\frac{m_i(x-x_i)}{r_i^{3}},\\
\label{Eq:5a}
\Omega_y&=&y-\frac{1}{\Delta}\sum _{i=1}^{4}\frac{m_i(y-y_i)}{r_i^{3}},\\
\label{Eq:5a}
\Omega_z&=&-\frac{z}{\Delta}\sum _{i=1}^{4}\frac{m_i}{r_i^{3}}.
\end{eqnarray}
\end{subequations}

In the same vein, the second-order partial derivatives of the effective potential function (which will be required to examine the stability of libration points as well as for the multivariate Newton-Raphson iterative scheme) are
\begin{subequations}
\begin{eqnarray}
\label{Eq:6a}
\Omega_{xx}&=&1-\frac{1}{\Delta}\sum_{i=1}^{4}\bigg(\frac{m_i}{r_i^{3}}-\frac{3m_i(x-x_i)^{2}}{r_i^{5}}\bigg),\\
\label{Eq:6b}
\Omega_{yy}&=&1-\frac{1}{\Delta}\sum_{i=1}^{4}\bigg(\frac{m_i}{r_i^{3}}-\frac{3m_i(y-y_i)^{2}}{r_i^{5}}\bigg),\\
\label{Eq:6c}
\Omega_{zz}&=&-\frac{1}{\Delta}\sum_{i=1}^{4}\bigg(\frac{m_i}{r_i^{3}}-\frac{3m_i(z-z_i)^{2}}{r_i^{5}}\bigg),
\end{eqnarray}
\begin{eqnarray}
\label{Eq:6d}
\Omega_{xy}&=&\frac{1}{\Delta}\sum_{i=1}^{4}\frac{3m_i(x-x_i)(y-y_i)}{r_i^{5}}=\Omega_{yx},\\
\label{Eq:6e}
\Omega_{yz}&=&\frac{1}{\Delta}\sum_{i=1}^{4}\frac{3m_i(y-y_i)(z-z_i)}{r_i^{5}}=\Omega_{zy},\\
\label{Eq:6e}
\Omega_{zx}&=&\frac{1}{\Delta}\sum_{i=1}^{4}\frac{3m_i(z-z_i)(x-x_i)}{r_i^{5}}=\Omega_{xz}.
\end{eqnarray}
\end{subequations}

The system of equations (\ref{Eq:4a}-\ref{Eq:4c}) admits the so-called Jacobi integral of motion (which corresponds to the total orbital energy) and written as:
\begin{equation}\label{Eq:7}
 J(x,y,z,\dot{x}, \dot{y}, \dot{z})=2\Omega(x,y,z)-(\dot{x}^2+\dot{y}^2+\dot{z}^2)=C,
\end{equation}
where $\dot{x}, \dot{y}$, and $\dot{z}$ are the velocities, whereas the Jacobian constant is represented by $C$  and it is conserved.
\begin{figure*}[!t]
\label{Fig:2n}
\centering
\resizebox{\hsize}{!}{\includegraphics{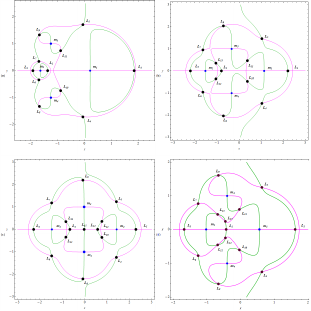}}
\caption{Positions (black dots) and numbering of the libration points $(L_i; i = 1,..., 11 \text{ or }  13 \text{ or }   15) $ through the intersections of $\Omega_x = 0$ (magenta) and $\Omega_y = 0$(green), when (a):  $\alpha=55.5\degree$ and $\beta=21\degree$ (eleven libration points), (b): $\alpha=55.5\degree$ and $\beta=50\degree$ (thirteen libration points), (c): $\alpha=55.5\degree$ and $\beta=55\degree$ (fifteen libration points), and (d): $\alpha=43.5\degree$ and $\beta=31\degree$ (fifteen libration points). The black dots pinpoint the libration points, while the blue dots represent the centers $(P_i; i = 1, 2, 3, 4) $ of the primaries. (Color figure online).}
\end{figure*}
\section{The parametric evolution of the libration points and their linear stability}
\label{the libration points and stability}

For the existence of the libration points, the necessary and sufficient conditions are
\begin{equation}\label{Eq:8}
 \dot{x}=\dot{y}=\dot{z}=\ddot{x}=\ddot{y}=\ddot{z}=0.
\end{equation}
The positions of the libration points can be depicted easily by solving the following system of equations
\begin{equation}\label{Eq:9}
  \Omega_x=0, \Omega_y=0, \Omega_z=0.
\end{equation}
\begin{figure*}[!t]
\label{Fig:3}
\begin{center}
(a)\includegraphics[scale=6.5]{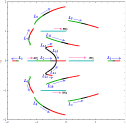}
(b)\includegraphics[scale=7]{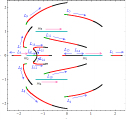}
(c)\includegraphics[scale=6.5]{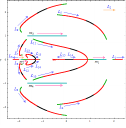}
\caption{The parametric evolution of the positions of the libration points, $L_i, i = 1,...,15$, in the axisymmetric five-body
problem with convex configuration, when (a) $\alpha=43.5\degree$, and $\beta \in(24\degree,  43.5\degree)$; (b) $\alpha=55.5\degree$, and $\beta \in(22\degree,  55.5\degree)$; and (c) $\alpha=59\degree$, and $\beta \in(16\degree,  59\degree)$. The
arrows indicate the movement direction of the libration points as
the value of the angle parameter $\beta$ increases. (Color figure online).}
\end{center}
\end{figure*}
In this manuscript, we restrict our study only to the coplanar libration points on the configuration $(x, y)$ plane. The angle parameters $\alpha$ and $\beta$ determine the total number of libration points in the coplanar axisymmetric restricted problem of five bodies. We have considered three sets of values of the parameter $\alpha$ and consequently the acceptable range of the parameter $\beta$ is as follows
\begin{enumerate}
\item when $\alpha=43.5\degree$,
\begin{enumerate}
\item[-]$\beta\in(23.25\degree, 29.685\degree]$, there exist 11 libration points,
\item[-]$\beta\in[29.686\degree, 34.474\degree]$, there exist 15 libration points, and
\item[-]$\beta\in[34.475\degree, 43.5\degree)$, there exist 13 libration points,
\end{enumerate}
\item when $\alpha=55.5\degree$,
\begin{enumerate}
\item[-]$\beta\in(17.25\degree, 23.656\degree]$, there exist 11 libration points,
\item[-]$\beta\in[23.657\degree, 24.280\degree]$, and $\beta\in[50.744\degree, 55.5\degree)$, there exist 15 libration points, and
\item[-]$\beta\in[24.281\degree, 50.743\degree]$, there exist 13 libration points,
\end{enumerate}
\item when $\alpha=59\degree$,
\begin{enumerate}
\item[-]$\beta\in(15.5\degree, 21.680\degree]$, and $\beta\in[57.948\degree, 59\degree)$, there exist 11 libration points,
\item[-]$\beta\in[49.627\degree, 54.146\degree]$, there exist 15 libration points and
\item[-]$\beta \in[21.681\degree, 49.626\degree]$, and $\beta \in[54.147\degree, 57.947\degree]$, there exist 13 libration points.
\end{enumerate}
\end{enumerate}

We observe that there exist either eleven, thirteen or fifteen libration points, depending on the various combinations of the angle parameters $\alpha$ and $\beta$. In Fig. \ref{Fig:2n}, we depict how the intersections of the equation $\Omega_x=0,$ (the magenta line) and $\Omega_y=0$ (the green line) define the positions of the libration points on the configuration $(x, y)$ plane. The numerically evaluated libration points are depicted by the black dots while the blue dots represent the centre of four primaries. It is observed that the total number of the libration points as well as their positions are changed for the different combination of the parameters $\alpha$ and $\beta$.

The evolution of the positions of the libration points, as a function of angle parameter $\beta$ for fixed value of angle parameter $\alpha$ is unveiled in Fig. \ref{Fig:3}.  Here, it observed that the centers of the four primaries move along the straight lines from left to right, as the value of the parameter $\beta$ increases. In panel (a), the evolution of the positions of the libration points are depicted for fixed $\alpha=43.5\degree$, while $\beta\in(23.25\degree, 43.5\degree)$. We observe that the green, black, and red colors show the intervals of $\beta$ for which eleven, fifteen and thirteen libration points exist, respectively. It can be noticed that the non-collinear libration points are symmetrical, with respect to the $x-$axis. Furthermore, as soon as the value of $\beta\approx 29.686\degree$, two pairs of libration points emerge from two points near the $x-$axis and between the primaries $m_{1,2}$. Two of them, $L_{13, 15}$ lie on the upper half of the $x-$axis, while the other two, $L_{12, 14}$ lie on the lower half of the $x-$axis. We see that as the value of $\beta$ increases, the libration points $L_{14, 15}$ move towards the $x-$axis and for $\beta=34.475\degree$ the libration points $L_{14, 15}$ move on collision course with libration point $L_2$. These libration points mutually annihilate when the collision occurs. On the other hand, the libration points $L_{12, 13}$ move away from the $x-$axis, as $\beta$ increases. Moreover, the libration points $L_{6, 7, 8, 9}$ move away, while libration points $L_{4, 5, 10, 11}$ move towards the $x-$axis, as $\beta$ increases. In Fig. \ref{Fig:3}b, the evolution of the libration points are illustrated for $\alpha=55.5\degree$ and $\beta\in(17.25\degree, 55.5\degree)$. We may observe that the movement of the libration points are the same as in the previous panel. However, for $\alpha=59\degree$ and $\beta\in(15.5\degree, 59\degree)$, (see panel (c)) there exist 15 libration points for two types of combinations. In first case, three collinear and twelve non-collinear libration points exist, where in the second case five collinear libration points and ten non-collinear libration points exist.

It is very informative to reveal the linear stability of the libration points by shifting the origin of the reference frame to the exact position of the libration point. For this, we apply the transformation:
\begin{subequations}
\begin{eqnarray}
\label{Eq:10a}
  x&=&x_0+\xi,\\
  \label{Eq:10b}
  y&=&y_0+\eta, \quad \xi, \eta <<1.
\end{eqnarray}
\end{subequations}
Now expanding the equations of motion (\ref{Eq:4a}-\ref{Eq:4c}) and neglecting higher order terms, the linearized system with respect to $\xi$ and $\eta$ can be written as:
\begin{equation}\label{Eq:11}
 \dot{\Xi}=\Lambda\Xi, \quad\Xi=(\xi, \eta, \dot{\xi}, \dot{\eta})^T,
\end{equation}
where $\Xi$ denotes the state vector of the test particle, with respect to the libration point, where as the $\Lambda$ represents the time-independent coefficient matrix of the variational equation which is read as
\begin{equation}\label{Eq:12}
 \Lambda= \left[
             \begin{array}{cccc}
               0 & 0  & 1 & 0 \\
               0 & 0 & 0 & 1 \\
               \Omega_{xx}^0 & \Omega_{xy}^0  & 0 & 2 \\
               \Omega_{yx}^0 & \Omega_{yy}^0 & -2 & 0\\
             \end{array}
           \right]
\end{equation}
where the superscript "0" at the partial derivatives of the second order denotes evaluation at the position of the libration point $(x_0, y_0)$.

The characteristic equation of the linear system of equations (\ref{Eq:11}) is obtained by setting $\varphi=\lambda^2$, i.e.,
\begin{equation}\label{Eq:13}
 k_2 \varphi^2+k_1 \varphi+k_0=0.
\end{equation}
where
\begin{align*}
  k_2 & =1, \\
  k_1 & =4-\Omega_{xx}^0- \Omega_{yy}^0,\\
  k_0 & =\Omega_{xx}^0\Omega_{yy}^0-(\Omega_{xy}^0)^2.
\end{align*}
The necessary and sufficient conditions for a libration point to be linearly stable is that all roots of the characteristic equation for $\lambda$ are pure imaginary. This happens only when the following conditions are satisfied simultaneously,
\begin{align}\label{Eq:14}
  k_1 & >0,  \quad k_0 >0, \quad
  D=k_1^2-4k_0k_2>0.
\end{align}

We have numerically calculated the eigenvalues of the characteristic equation (\ref{Eq:13}) for the libration points evaluated for $\alpha=43.5\degree, 55.5\degree,$ and $59\degree$ and for the corresponding permissible ranges of $\beta$. We observe that for none of the combinations of the angle variables $\alpha$ and $\beta$, the roots of the characteristic equation are pure imaginary. Therefore we conclude that none of the libration points are stable, for any combination of the angle coordinates, in the axisymmetric five-body problem.

\begin{figure*}[!t]
\label{Fig:4n}
\begin{center}
(a)\includegraphics[scale=.4]{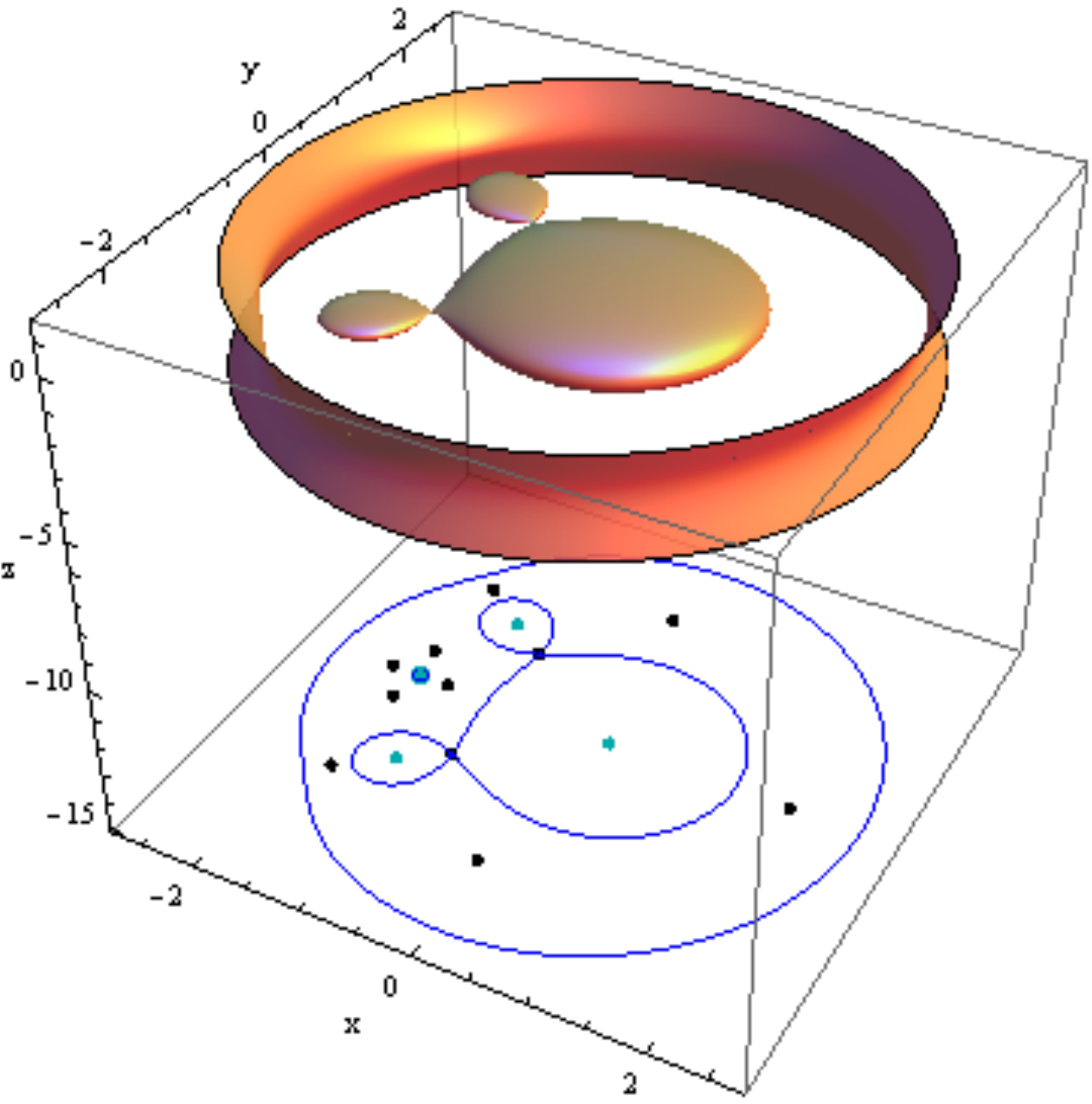}
(b)\includegraphics[scale=.4]{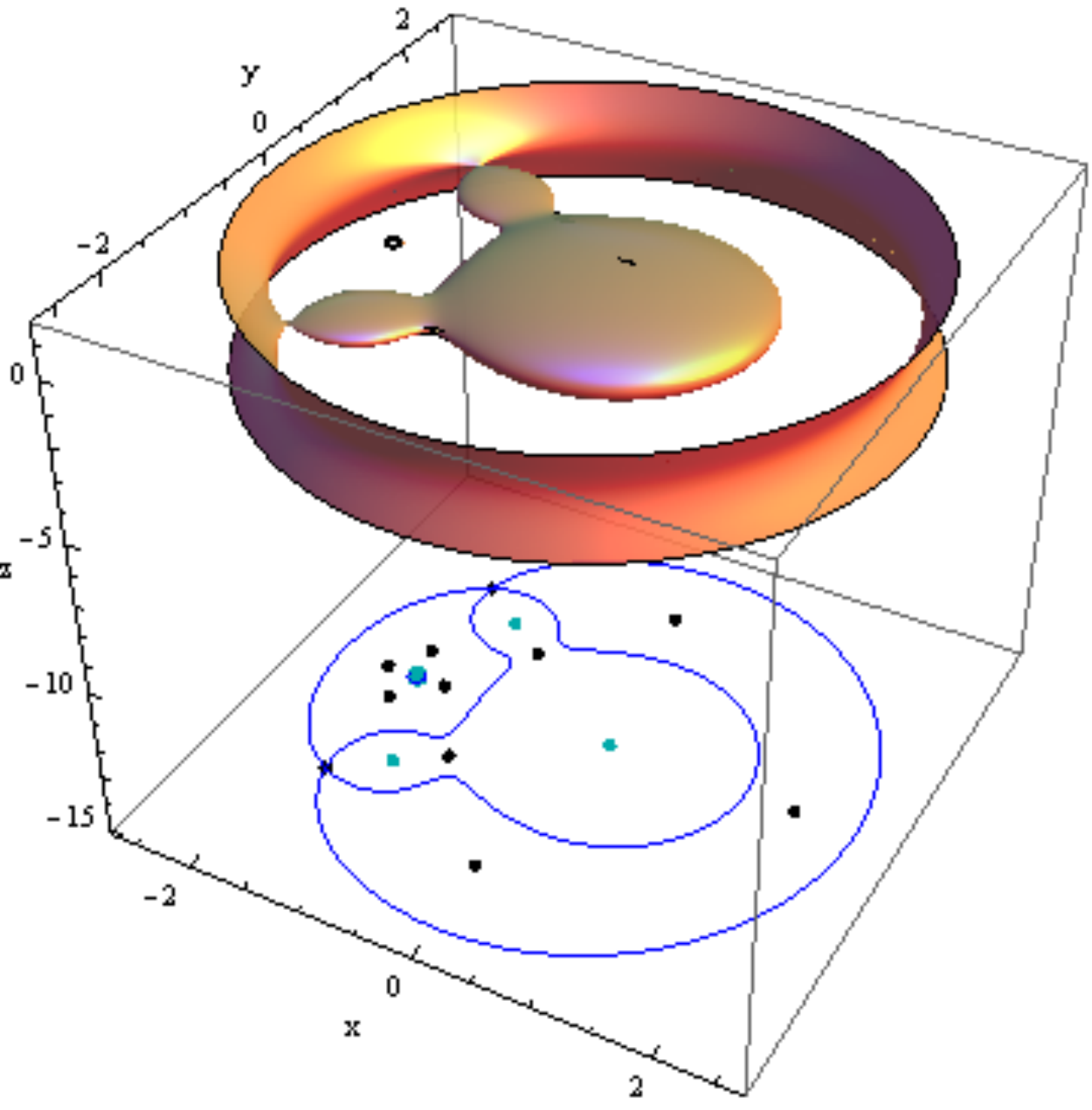}
(c)\includegraphics[scale=.4]{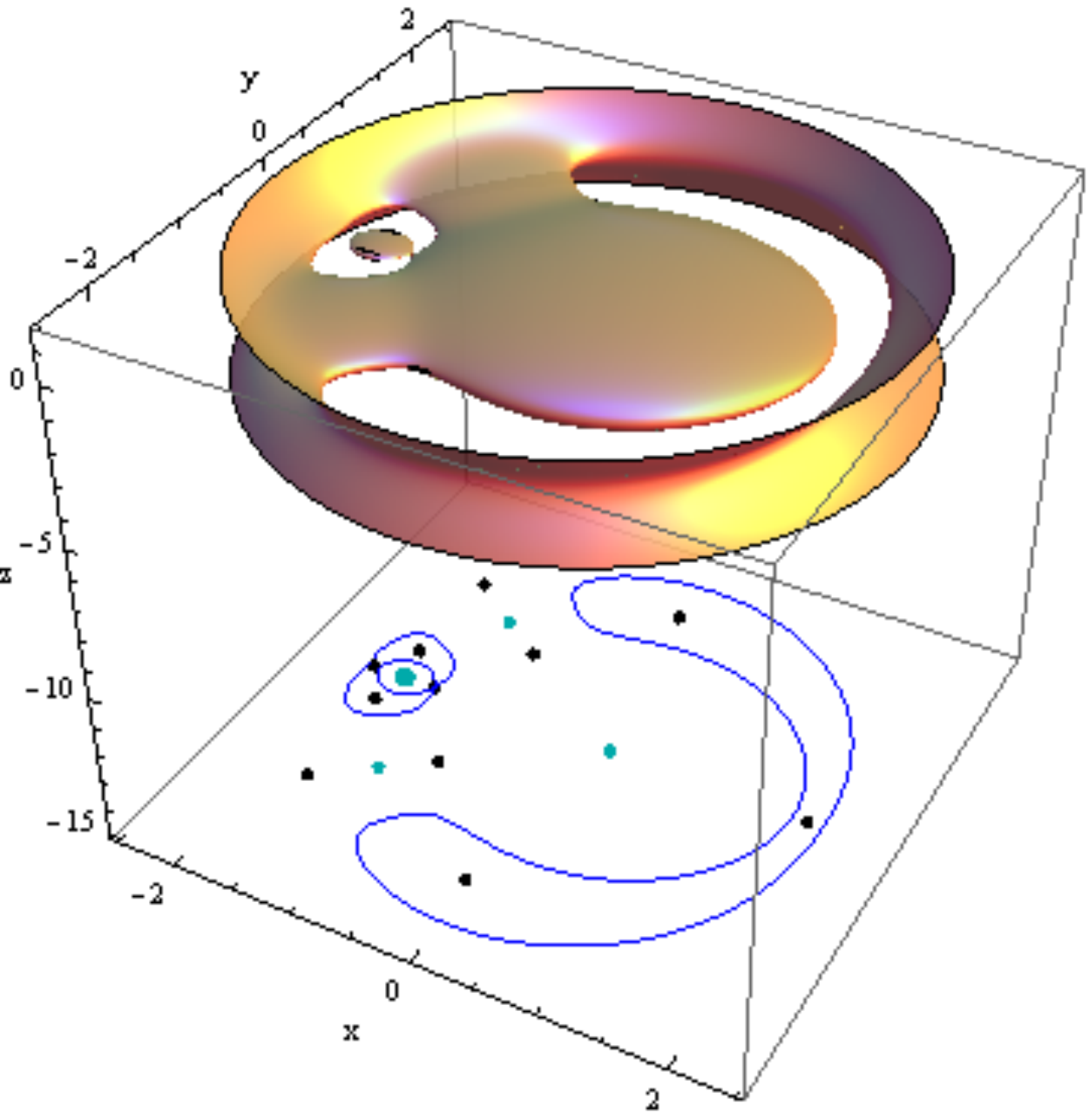}
(d)\includegraphics[scale=.4]{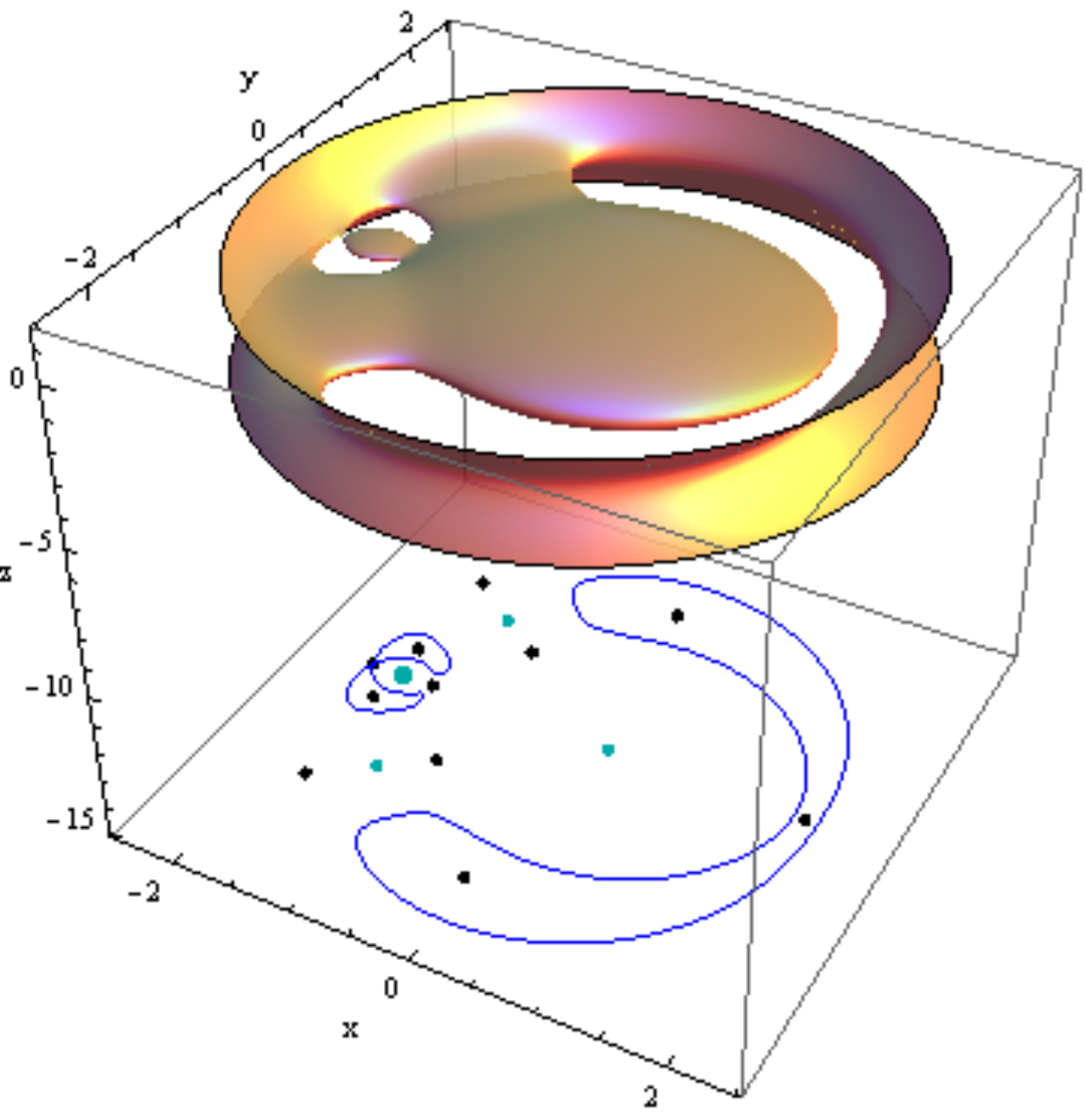}
(e)\includegraphics[scale=.4]{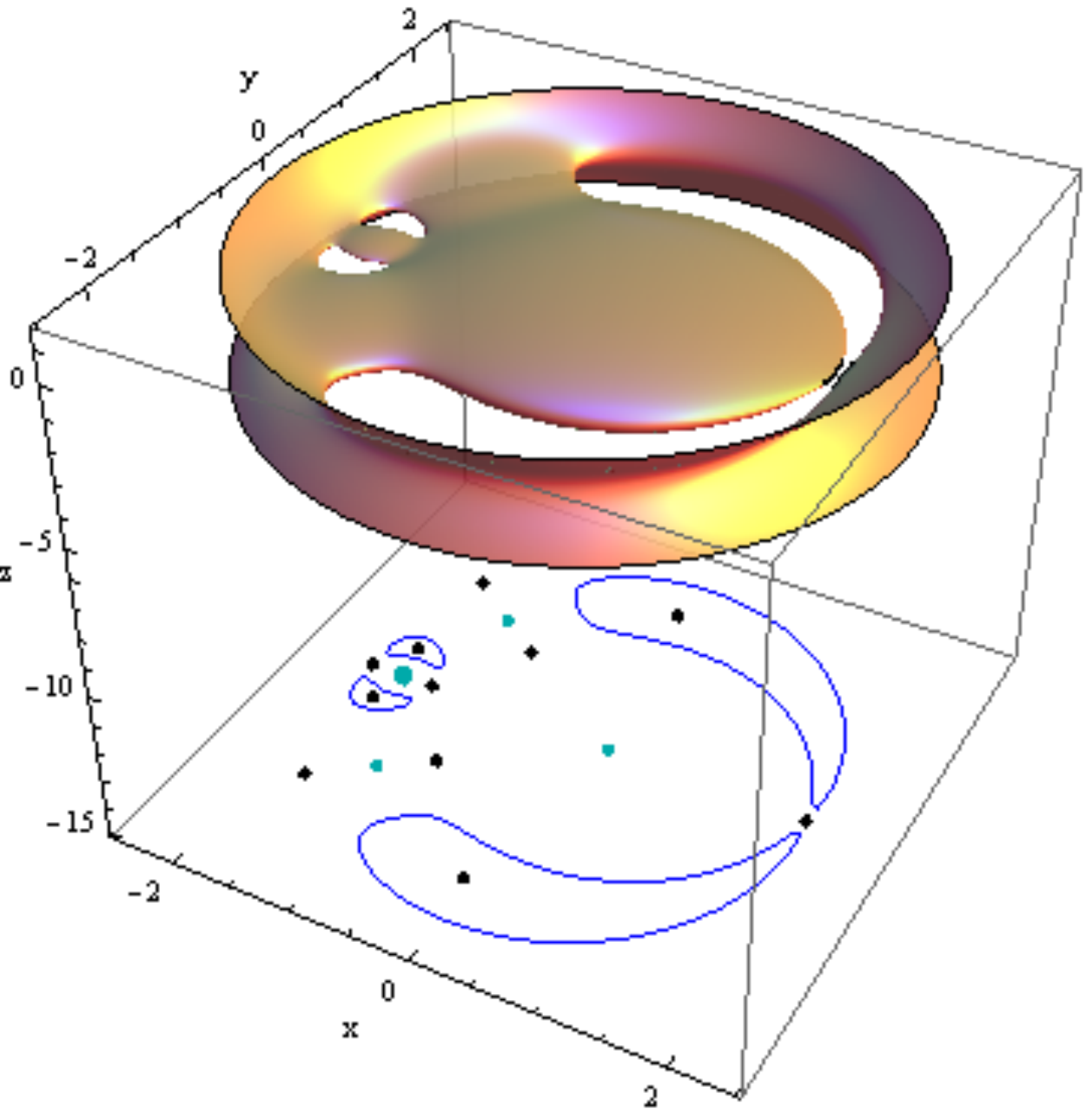}
(f)\includegraphics[scale=.4]{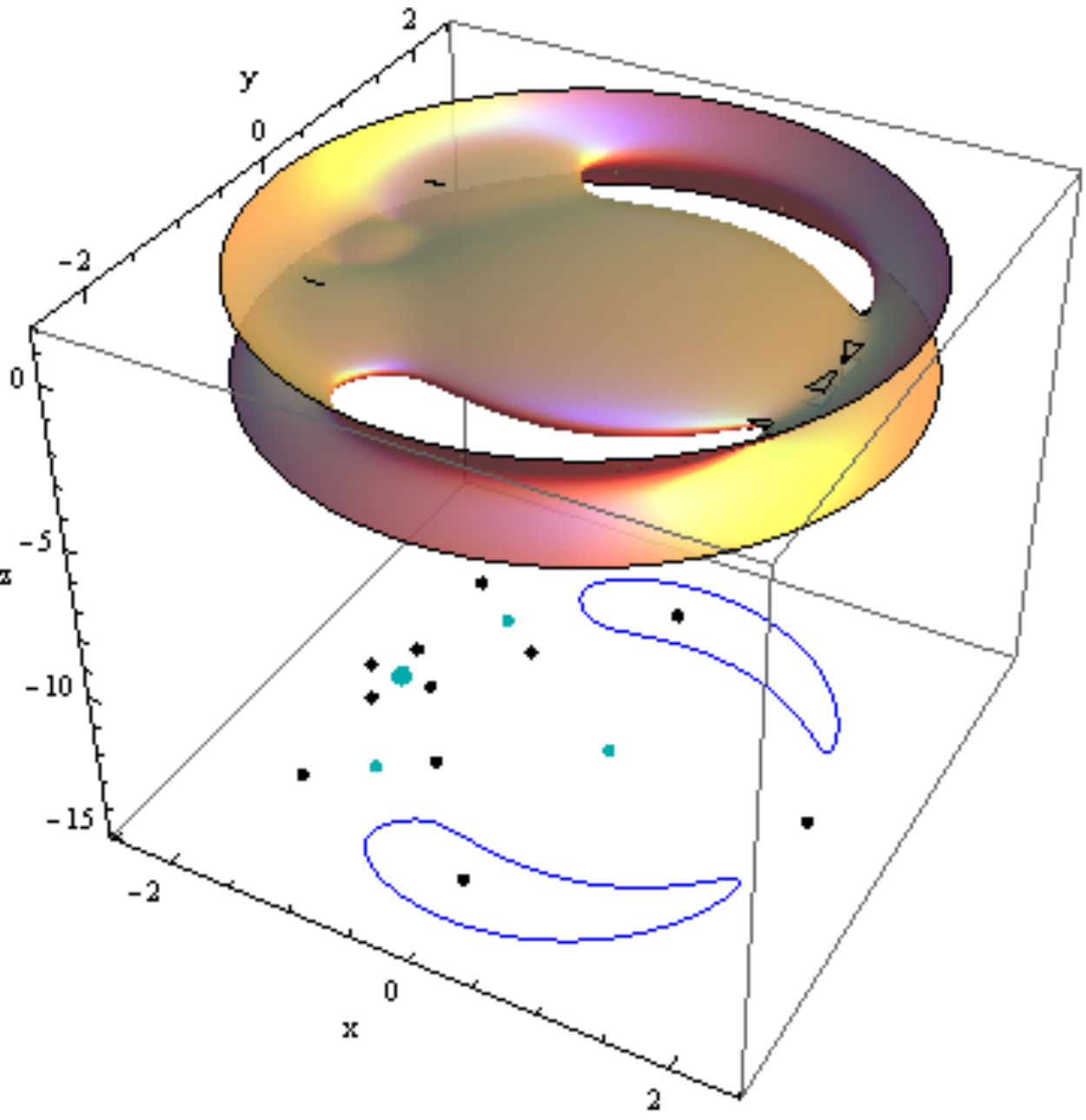}
(g)\includegraphics[scale=.4]{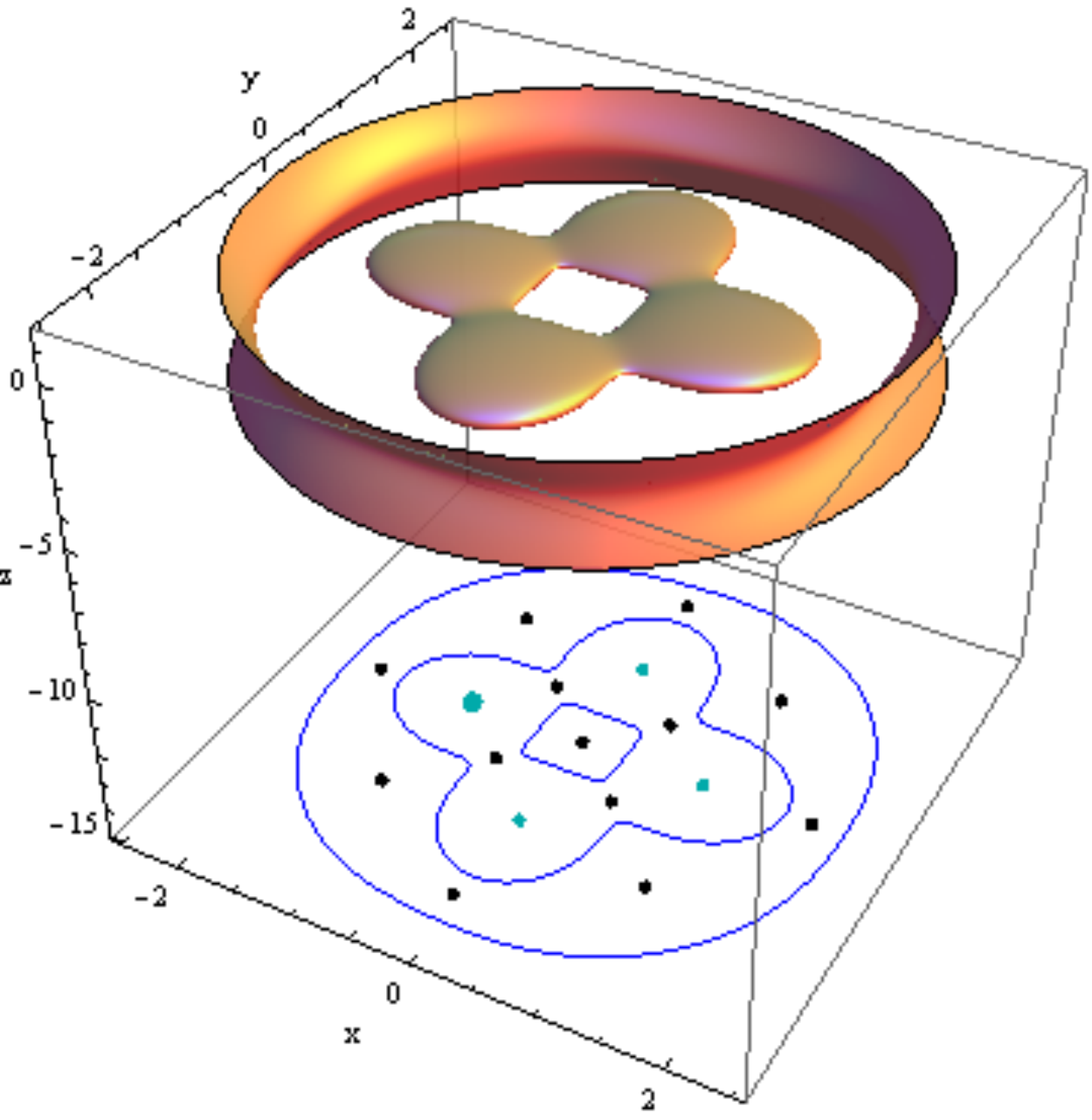}
(h)\includegraphics[scale=.4]{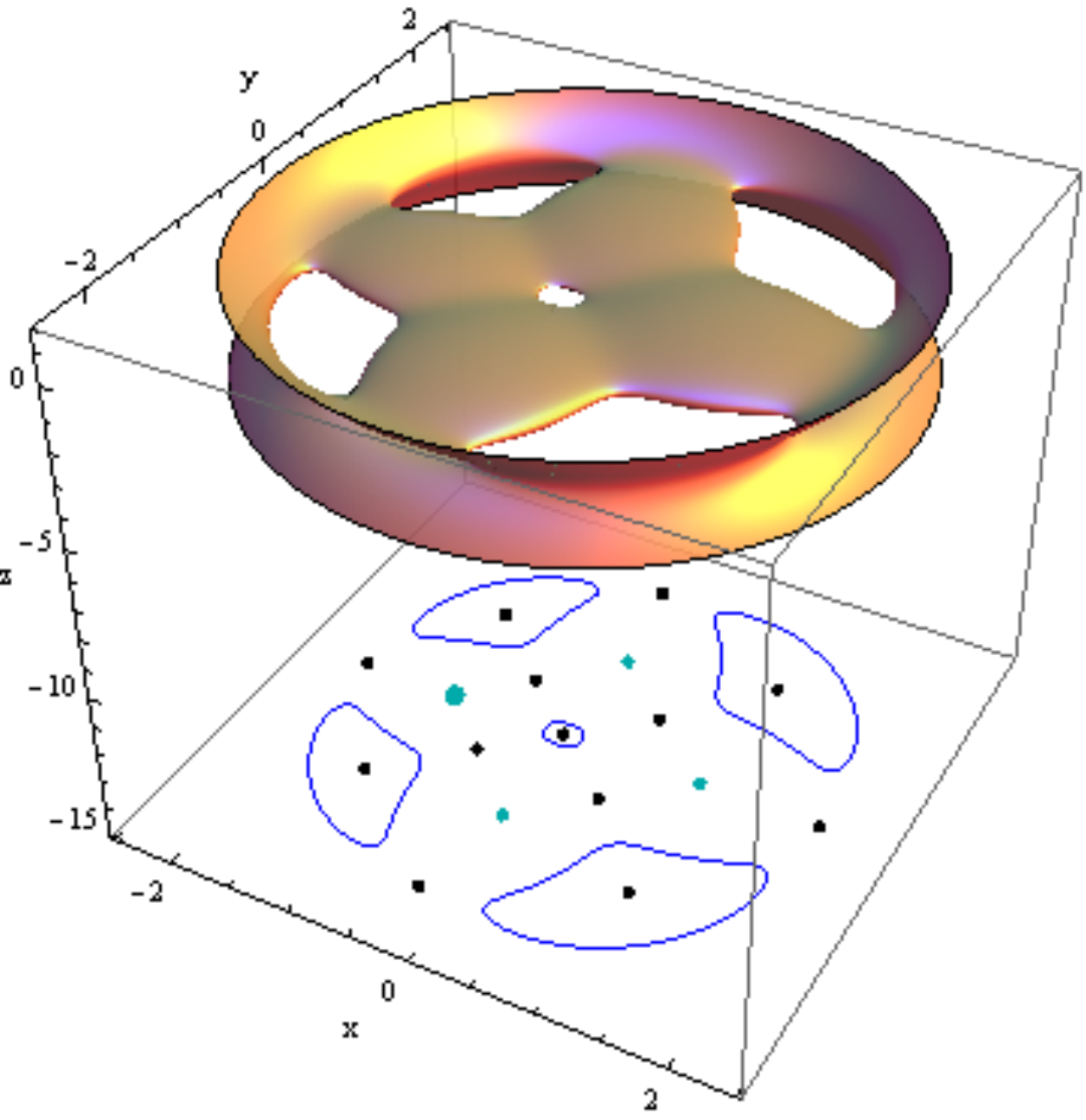}
(i)\includegraphics[scale=.4]{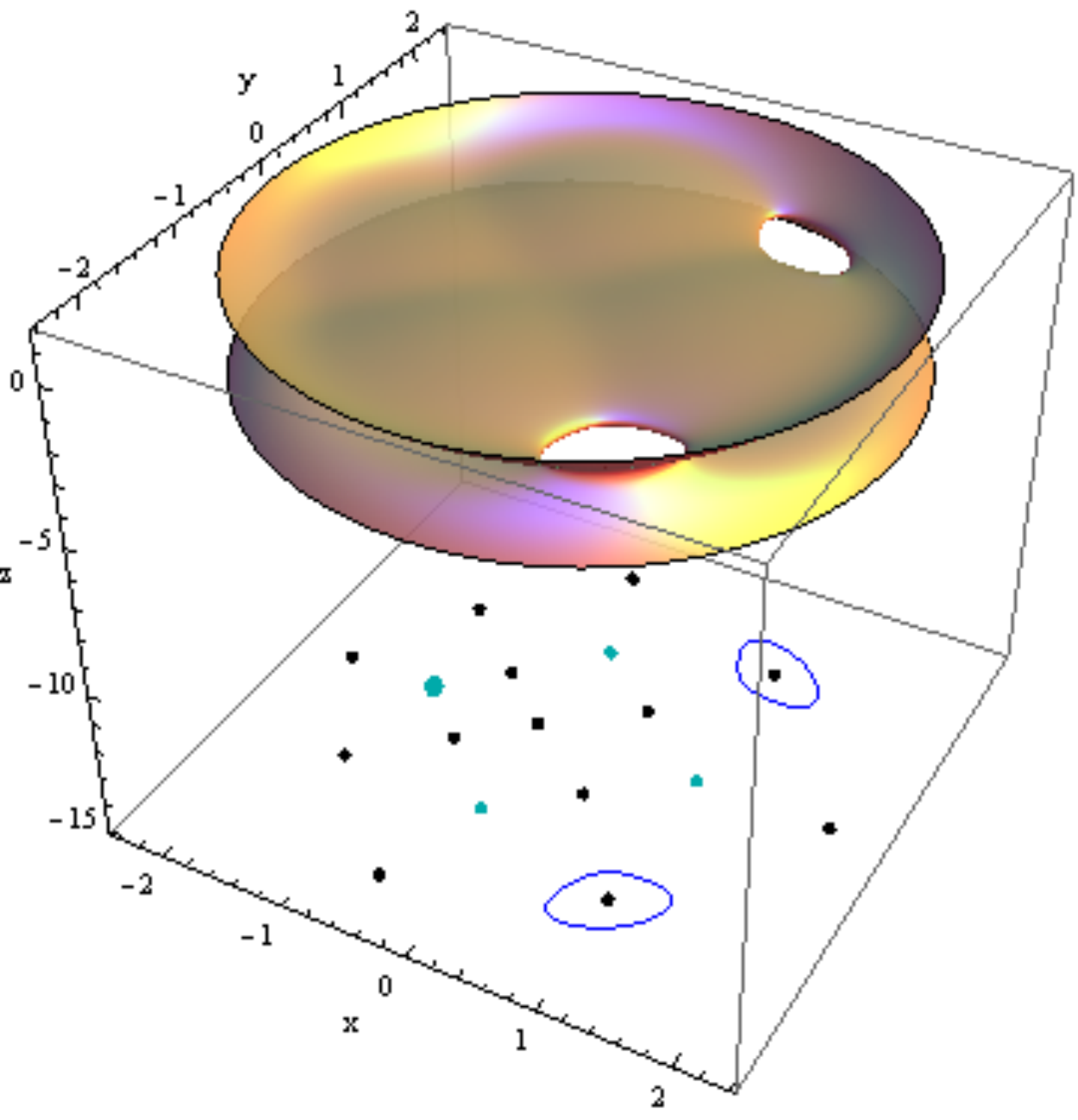}
\caption{Evolution of the structure of 3-dimensional ZVSs for the axisymmetric five-body problem, as a function of the Jacobi constant $C$. The projections of these surfaces onto the configuration $(x, y)$ plane are depicted at the bottom of the corresponding bounding box as blue solid lines, while the black and cyan dots represent the positions of the libration points and primaries bodies, respectively. When $\alpha=45.5\degree, \beta=21\degree$: (a-\emph{upper left}) $C=10.6824$; (b-\emph{upper middle}) $C=10.4356$; (c-\emph{upper right}) $C=9.74949$; (d-\emph{lower left}) $C=9.73105$; (e-\emph{lower middle}) $C=9.70163$; and (f-\emph{lower right}) $C=9.64123$. The evolution of the ZVSs as a function of $\alpha$ for $C=9.01847$ and $\beta=45\degree$, while (g) $\alpha=46\degree$; (h) $\alpha=48.5\degree$; and (i) $\alpha=51\degree$. (Color figure online).}
\end{center}
\end{figure*}

In Fig. \ref{Fig:4n}(a-i), we present the evolution of the structure of the three dimensional ZVSs for several values of the Jacobi constant $C$ and for the angle coordinates $\alpha$ and $\beta$. It is observed that as the value of the Jacobi constant $C$ decreases (see Fig. \ref{Fig:4n}(a-f)) various doorways appear through which the test particle can enter the several energetically allowed regions of motion. In \ref{Fig:4n}(g-i), the ZVSs are presented for various value of $\alpha$, where it is seen that the regions of the forbidden motion decrease, as the value of $\alpha$ increases.

\section{The Newton-Raphson basins of convergence}
\label{The Newton-Raphson basins of convergence}
The Newton-Raphson method is considered as one of the most fast and accurate iterative method for solving systems of non-linear equations. The associated multivariate iterative scheme is
\begin{equation}\label{Eq:401}
 \textbf{x}_{n+1}=\textbf{x}_n-\tilde{J}^{-1}f(\textbf{x}_n),
\end{equation}
where $f(\textbf{x}_n)$ represents the system of equations, while the corresponding inverse Jacobian matrix is denoted by $\tilde{J}^{-1}$. For the present model, the system of the equations reads as
\begin{subequations}
\begin{eqnarray}
\label{Eq:402a}
\Omega_{x}(x,y)&=&0,\\
\label{Eq:402b}
  \Omega_{y}(x,y)&=&0.
\end{eqnarray}
\end{subequations}
Moreover, for each coordinate $(x, y)$, the multivariate version of the iterative scheme (\ref{Eq:401}) may decompose into two parts as
\begin{subequations}
\begin{eqnarray}
\label{Eq:403a}
x_{n+1} &= x_n - \left( \frac{\Omega_x \Omega_{yy} - \Omega_y \Omega_{xy}}{\Omega_{yy} \Omega_{xx} - \Omega^2_{xy}} \right)_{(x,y) = (x_n,y_n)}, \\
\label{Eq:403b}
y_{n+1} &= y_n - \left( \frac{\Omega_y \Omega_{xx}-\Omega_x \Omega_{yx}}{\Omega_{yy} \Omega_{xx} - \Omega^2_{xy}} \right)_{(x,y) = (x_n,y_n)}.
\end{eqnarray}
\end{subequations}
\begin{figure*}[!t]
\centering
(a)\includegraphics[scale=1.5]{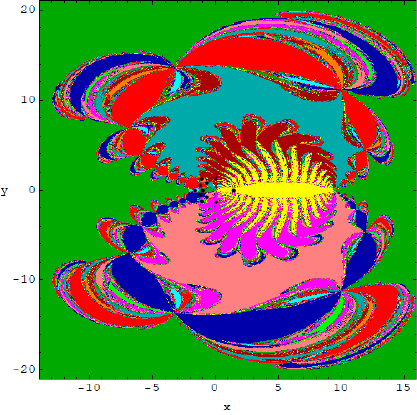}
(b)\includegraphics[scale=5]{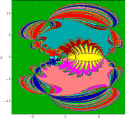}\\
(c)\includegraphics[scale=5]{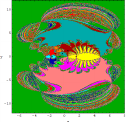}
(d)\includegraphics[scale=5]{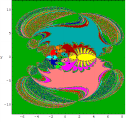}
\caption{The Newton-Raphson basins of attraction on the configuration $(x,y)$ plane, for the case when eleven libration points exist for fixed value of $\alpha=43.5 \degree$ and for:
(a) $\beta=24 \degree$; (b) $\beta=26 \degree$;
(c) $\beta=28 \degree$; (d) $\beta=29.6 \degree$. The color code denoting the attractors is as follows:
$L_1\emph{(yellow)}$; $L_2\emph{(darker green)}$; $L_3\emph{ (cyan)}$; $L_4\emph{ (magenta)}$;
$L_5\emph{(crimson)}$; $L_6\emph{(blue)}$; $L_7\emph{(red)}$; $L_8\emph{(green)}$; $L_9
\emph{(orange)}$; $L_{10} \emph{(pink)}$; $L_{11} \emph{(teal)}$; non-converging points (white).
 (Color figure online).}
\label{NR_Fig_1}
\end{figure*}

Recently, the above mentioned iterative scheme has been used to unveil the basins of convergence in various types of dynamical systems, such as the restricted five-body problem (e.g., \cite{Zot18a}), the restricted four-body problem (\cite{Sur17, Sur18a}, \cite{Z17a, Z17b}), the Sitnikov problem (e.g., \cite{Zot18c, Zot18d}), the collinear four-body problem, in the Copenhagen case with a repulsive quasi-homogeneous Manev-type potential (e.g., \cite{Sur18b}), the out-of-plane libration points in the case of the few-body problem (e.g., \cite{Sur18c}, \cite{Zot18b}) and of course the restricted three-body problem (e.g., \cite{Z16, Z17c}).

\begin{figure*}[!t]
\centering
(a)\includegraphics[scale=.35]{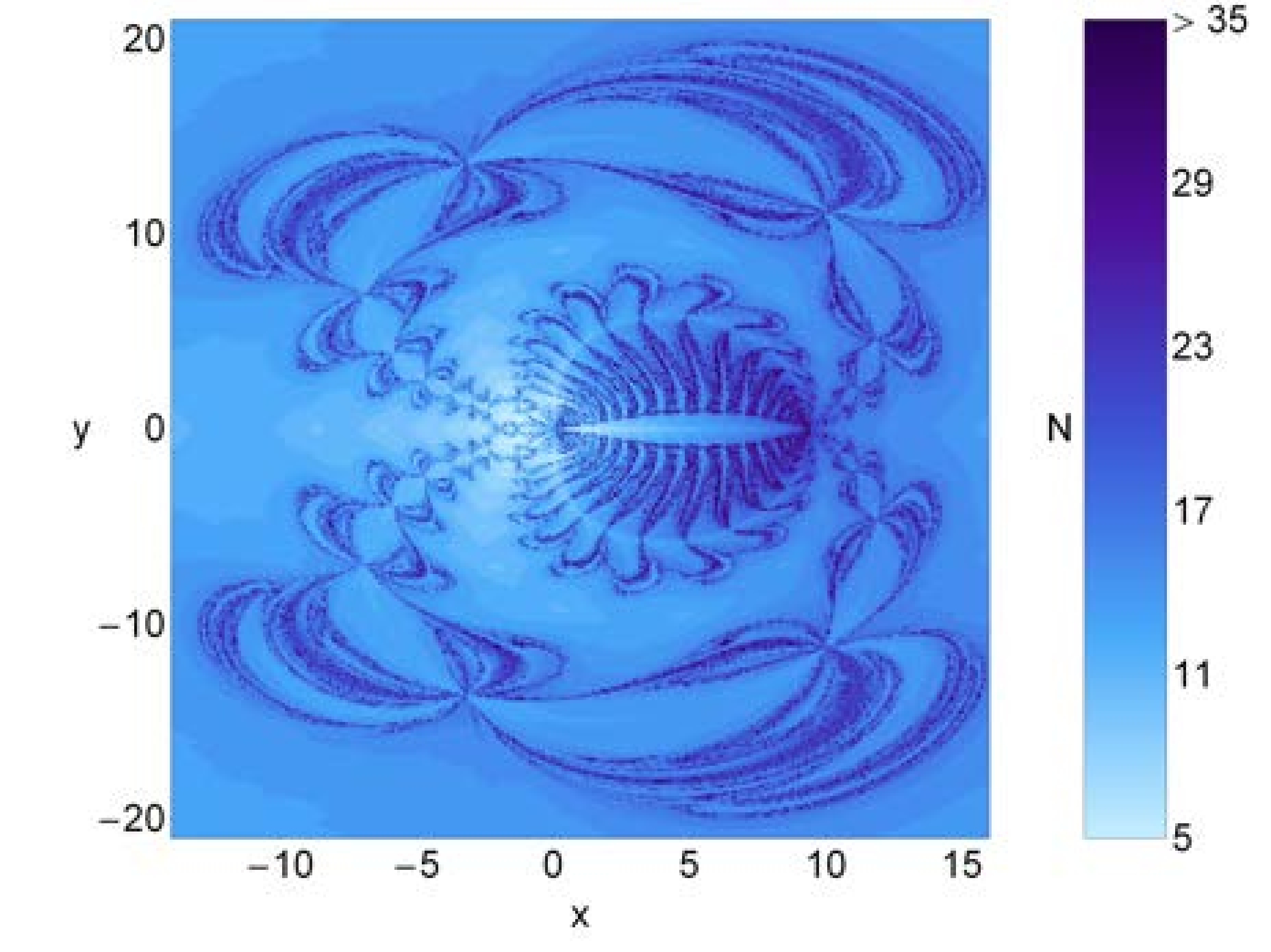}
(b)\includegraphics[scale=.35]{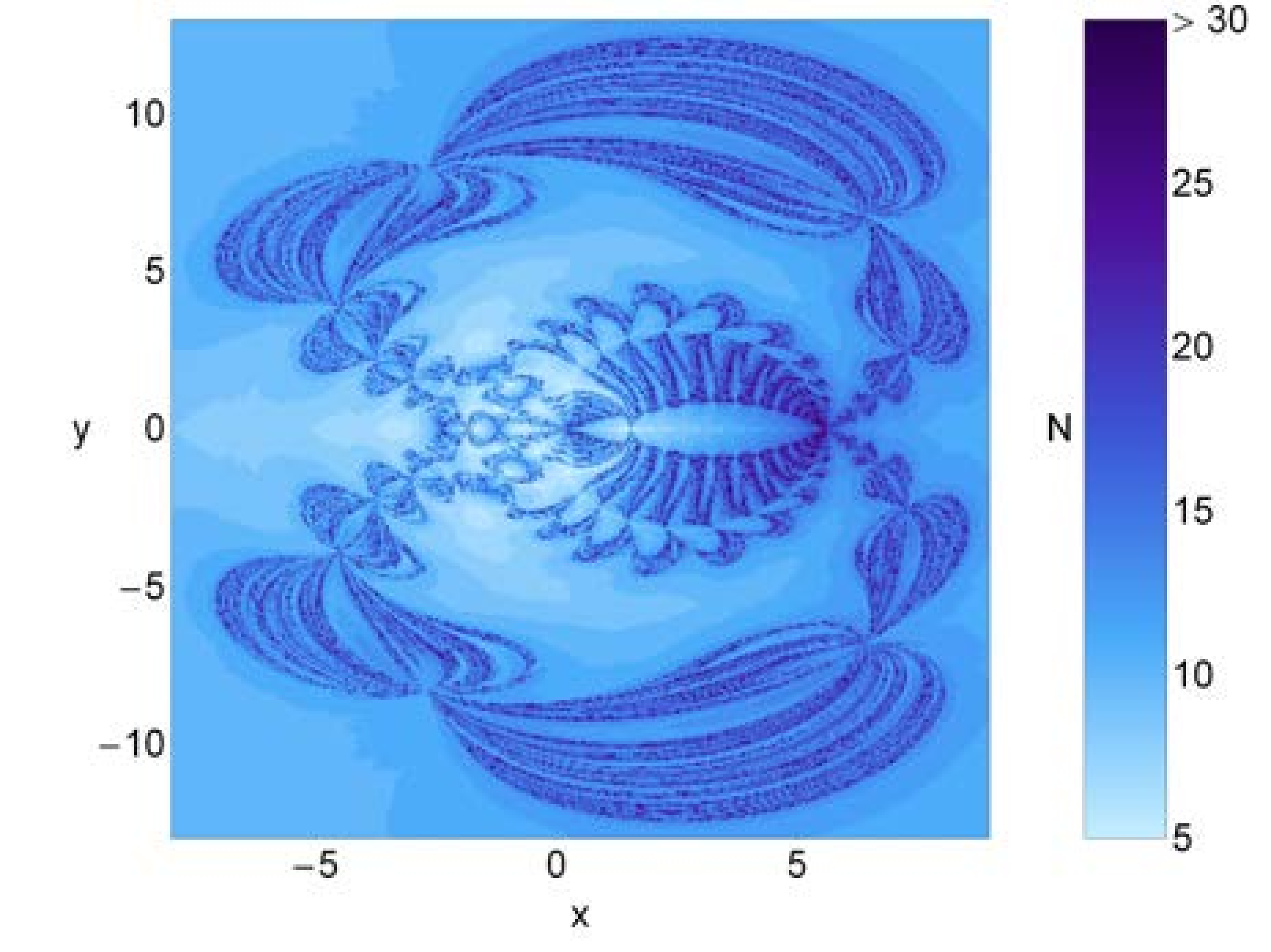}\\
(c)\includegraphics[scale=.35]{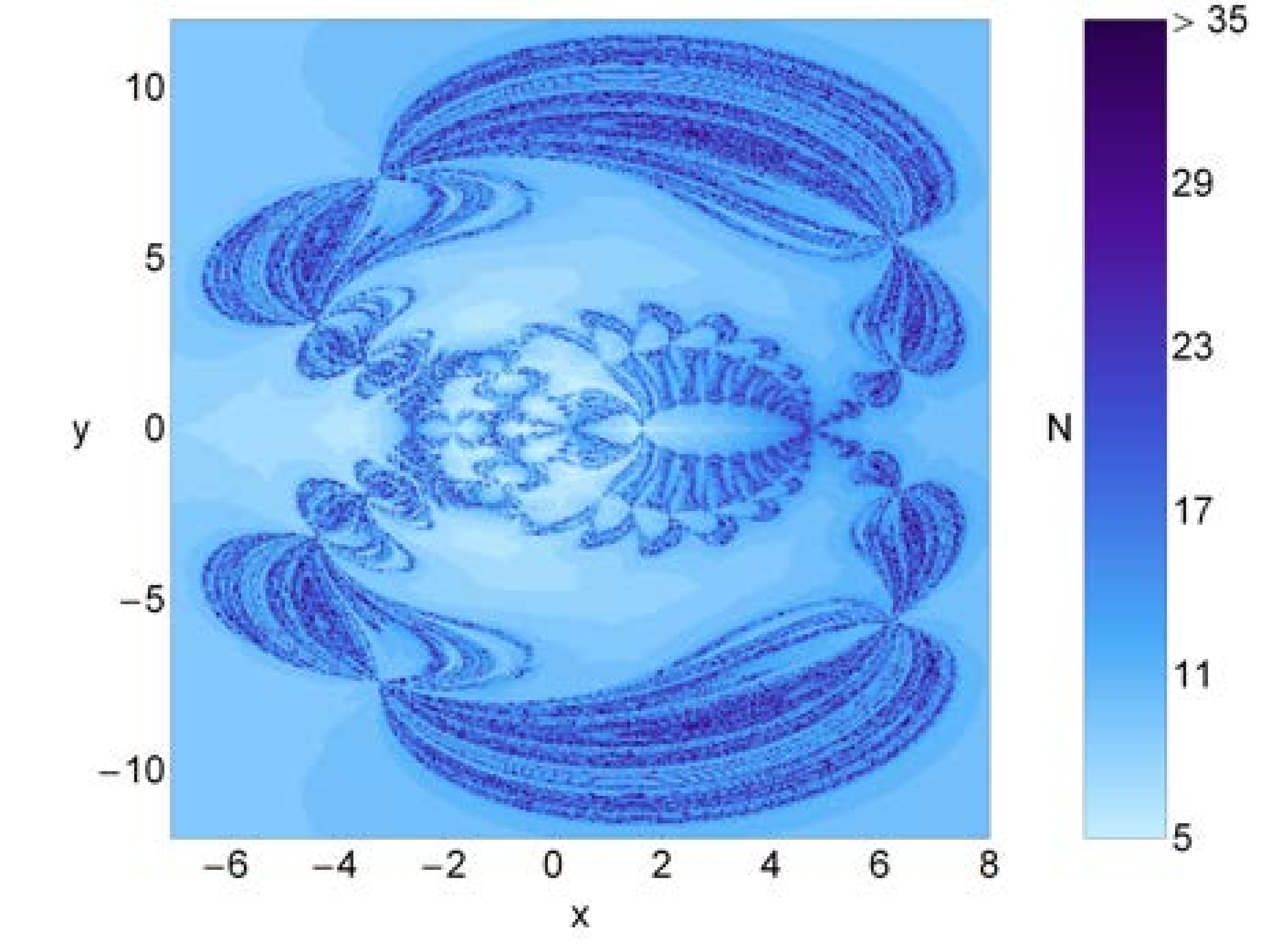}
(d)\includegraphics[scale=.35]{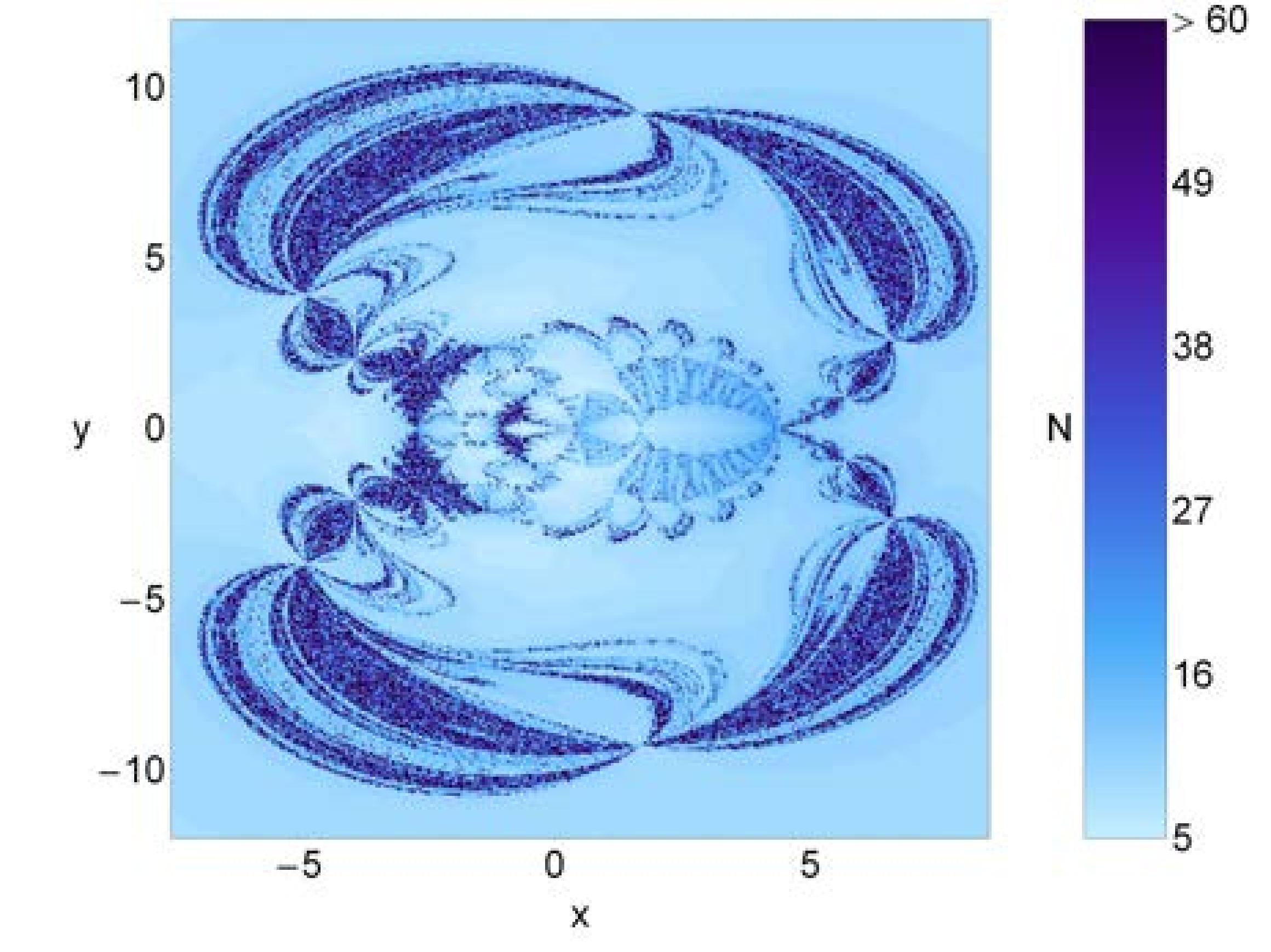}
\caption{The corresponding distributions of number $N$ of the required iterations for obtaining the Newton-Raphson basins of convergence, shown in Fig. \ref{NR_Fig_1}(a-d). (Color figure online).}
\label{NR_Fig_1a}
\end{figure*}
\begin{figure*}[!t]
\centering
(a)\includegraphics[scale=.4]{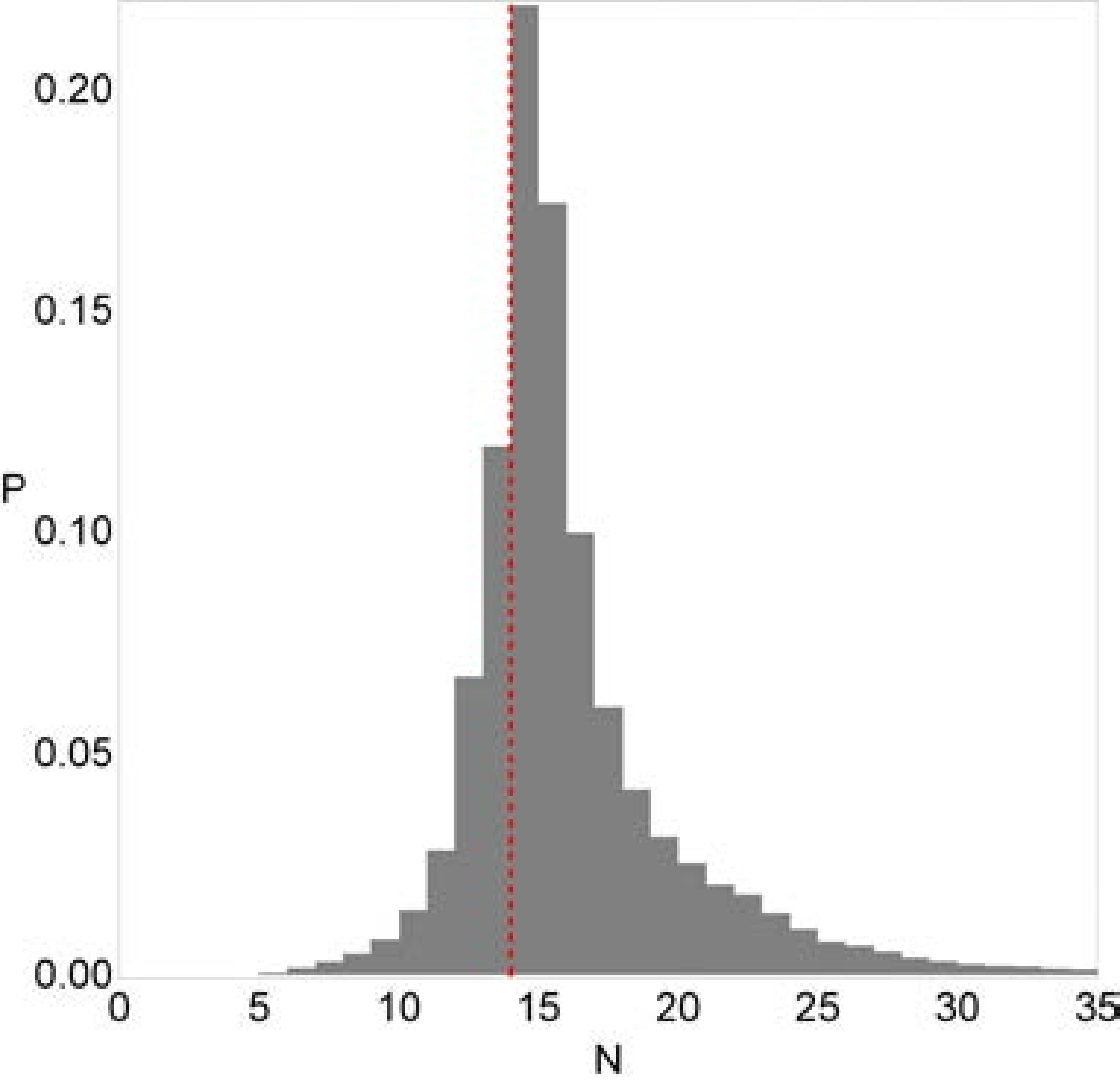}
(b)\includegraphics[scale=.4]{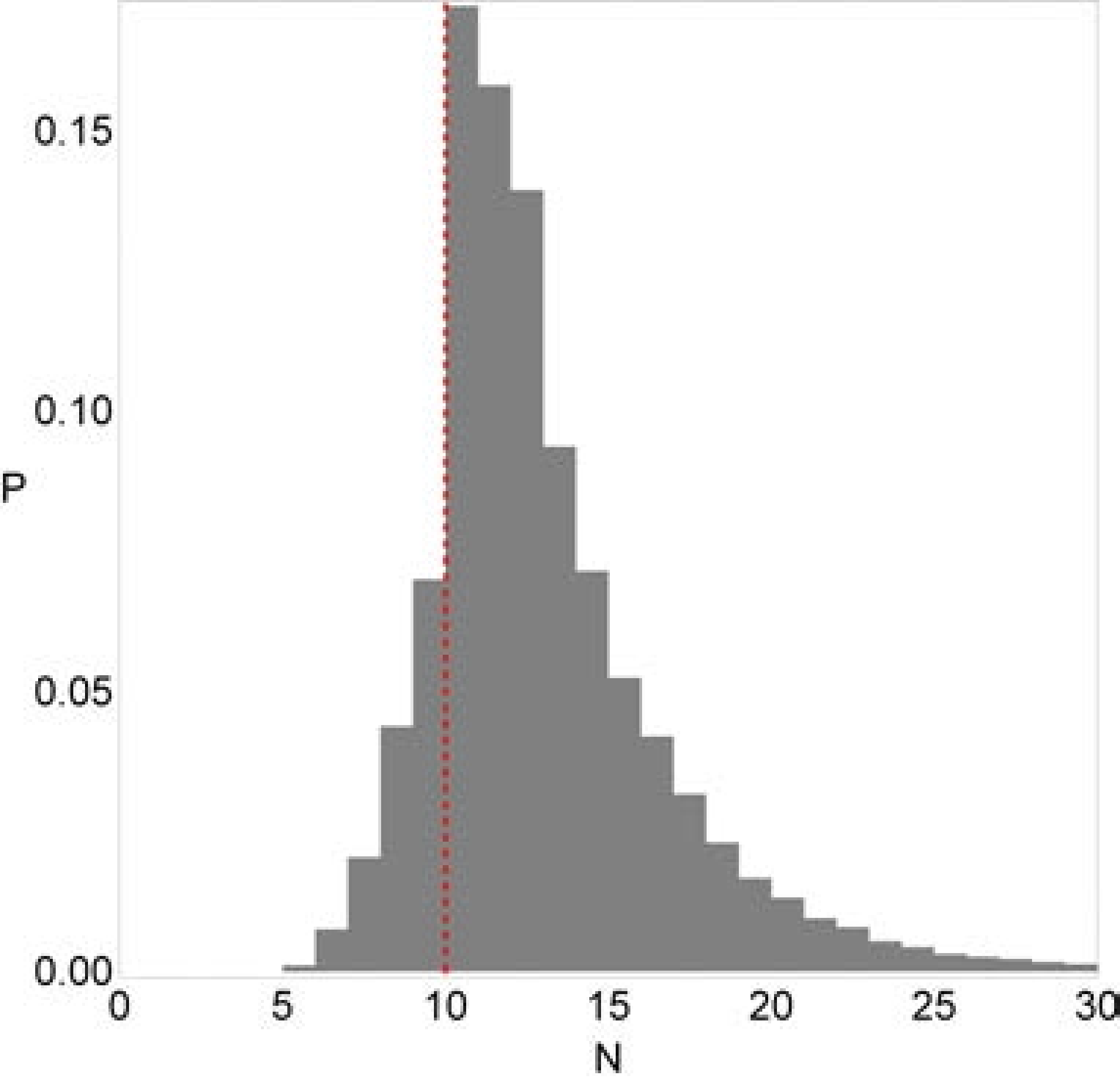}\\
(c)\includegraphics[scale=.4]{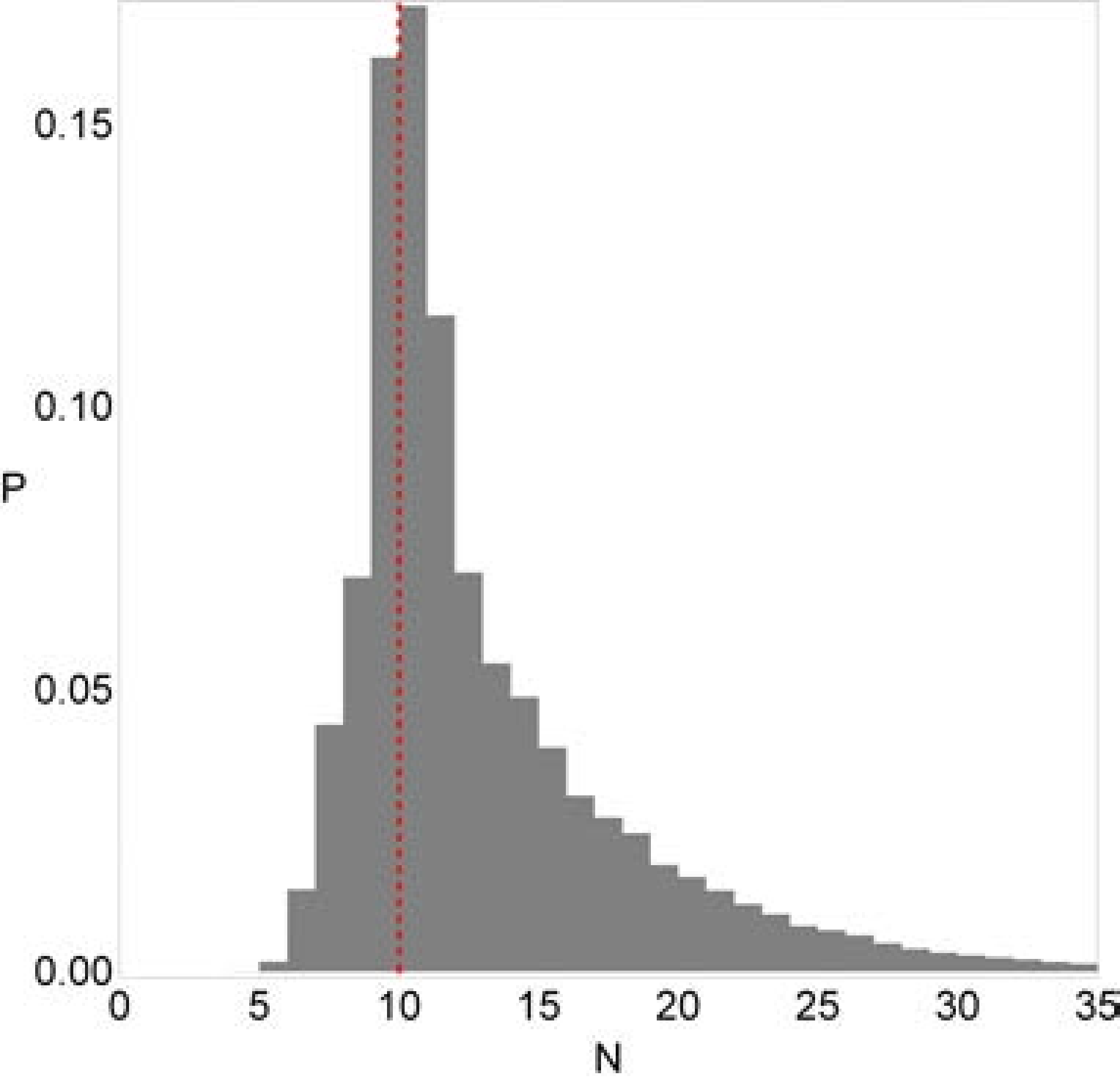}
(d)\includegraphics[scale=.4]{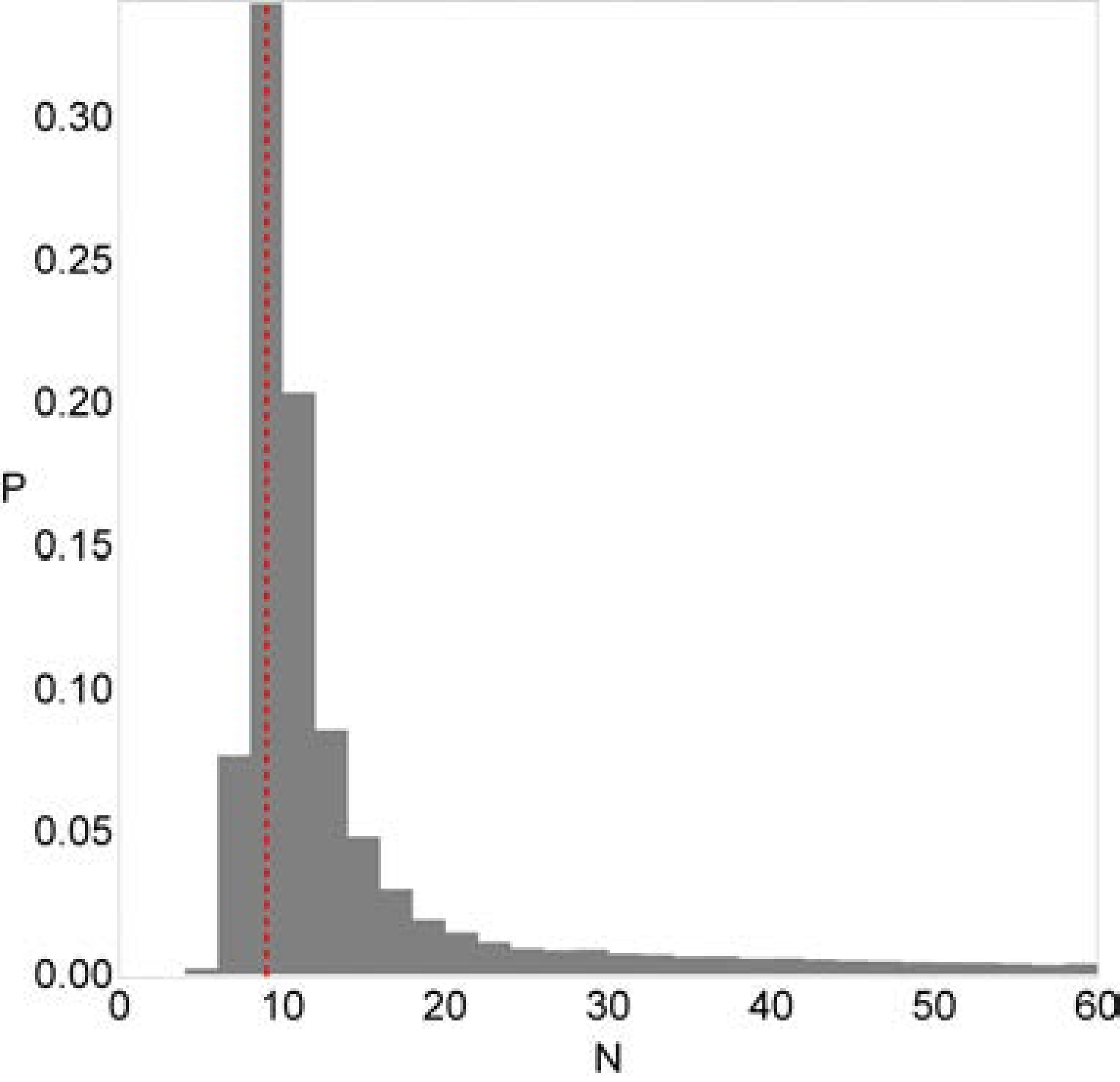}
\caption{The corresponding probability distributions of the required number of iterations for obtaining the Newton-Raphson basins of convergence, shown in Fig. \ref{NR_Fig_1}(a-d). The vertical, dashed, red line indicates, in each case, the most probable number $N^*$ of iterations. (Color figure online)}
\label{NR_Fig_1b}
\end{figure*}
\begin{figure*}[!t]
\centering
(a)\includegraphics[scale=5]{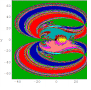}
(b)\includegraphics[scale=5]{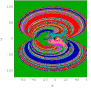}\\
(c)\includegraphics[scale=5]{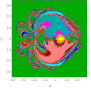}
(d)\includegraphics[scale=5]{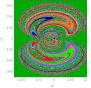}
\caption{The Newton-Raphson basins of attraction on the configuration $(x,y)$ plane for the case when eleven libration points exist for fixed value of $\alpha=55.5 \degree$ (a) $\beta=20 \degree$, (b) $\beta=21 \degree$, (c) $\beta=22 \degree$, (d) $\beta=23 \degree$. The color code is same as in Fig.  \ref{NR_Fig_1} (Color figure online).}
\label{NR_Fig_2}
\end{figure*}
\begin{figure*}[!t]
\centering
(a)\includegraphics[scale=.35]{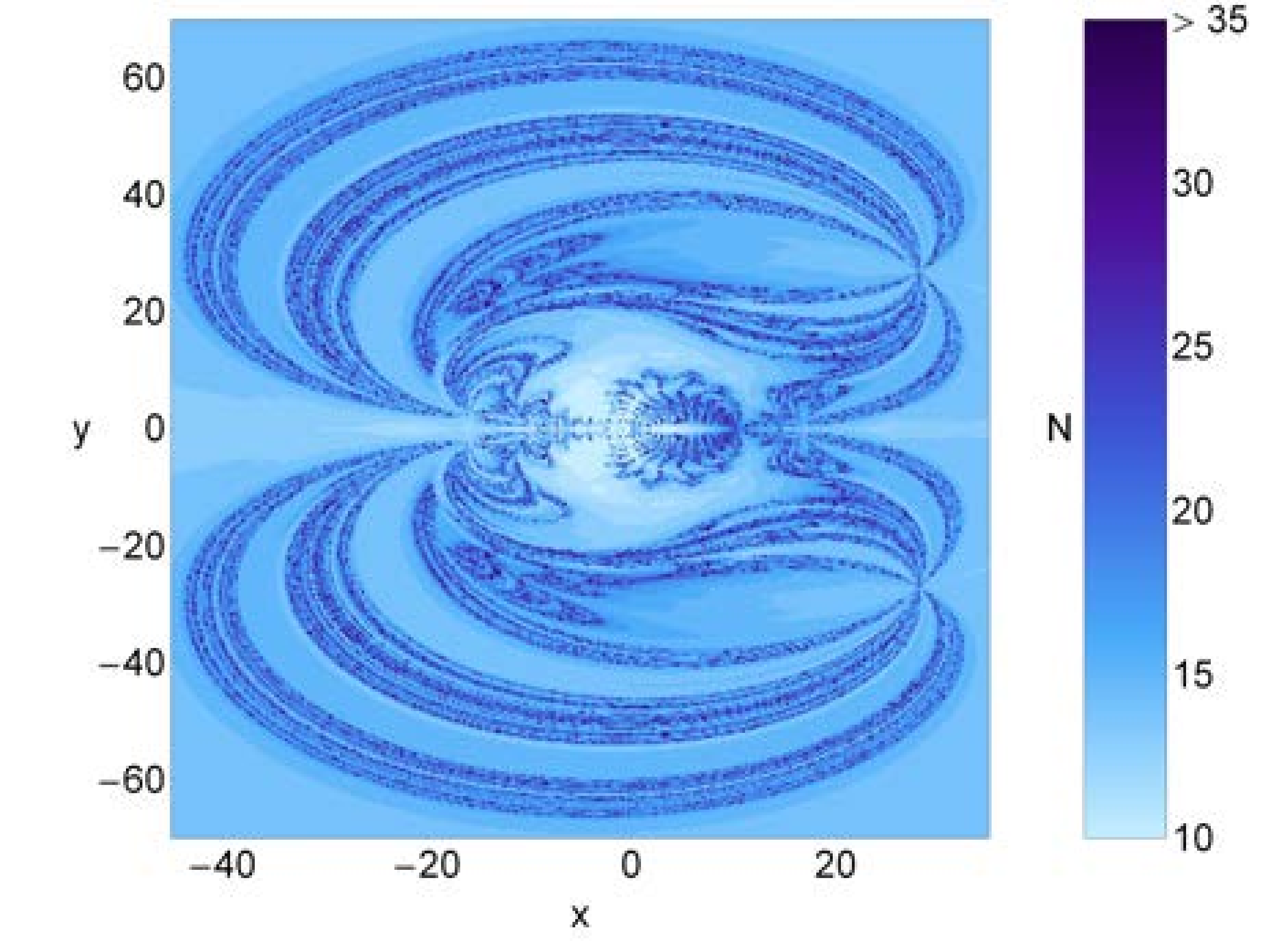}
(b)\includegraphics[scale=.35]{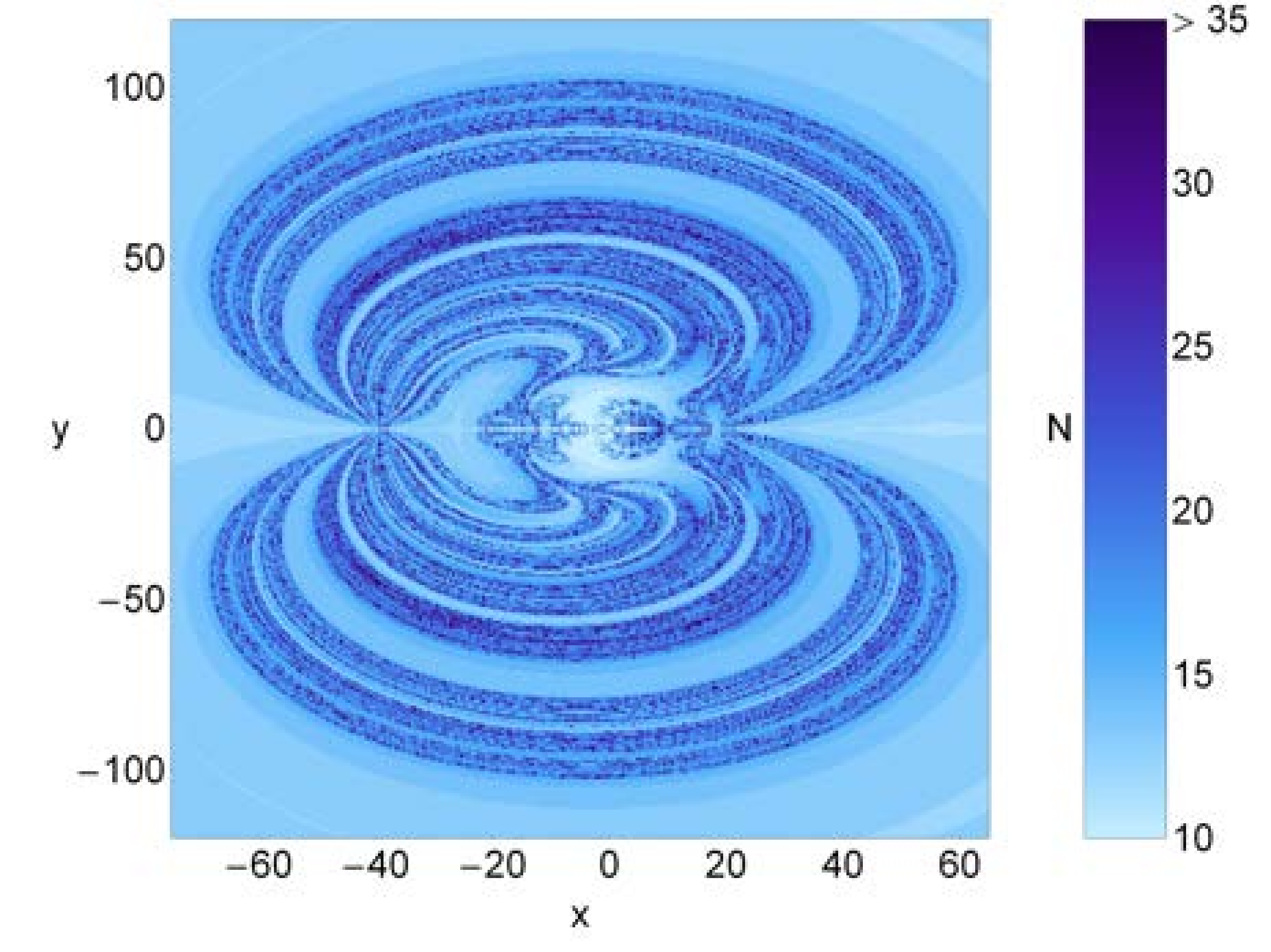}
(c)\includegraphics[scale=.35]{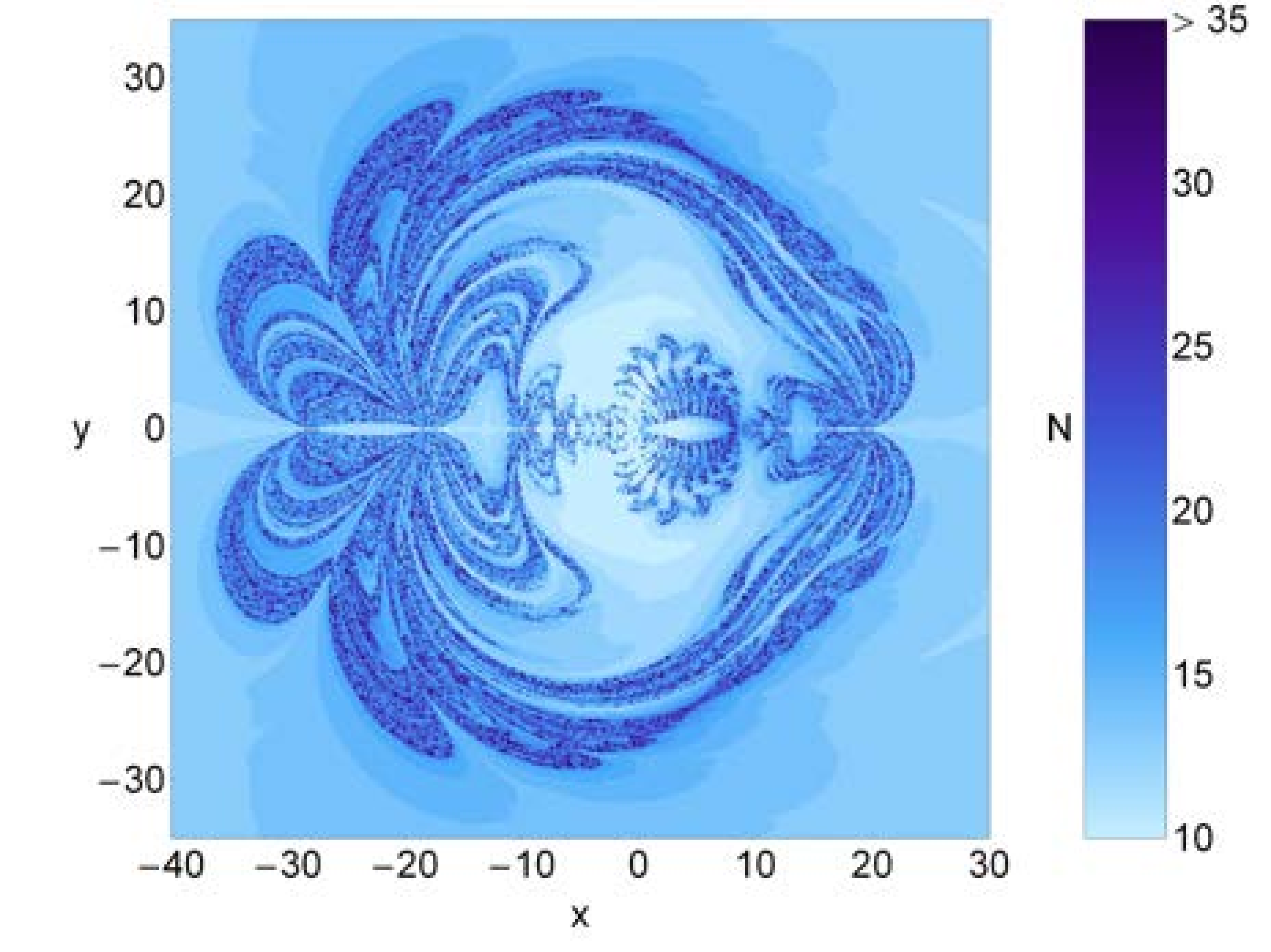}
(d)\includegraphics[scale=.35]{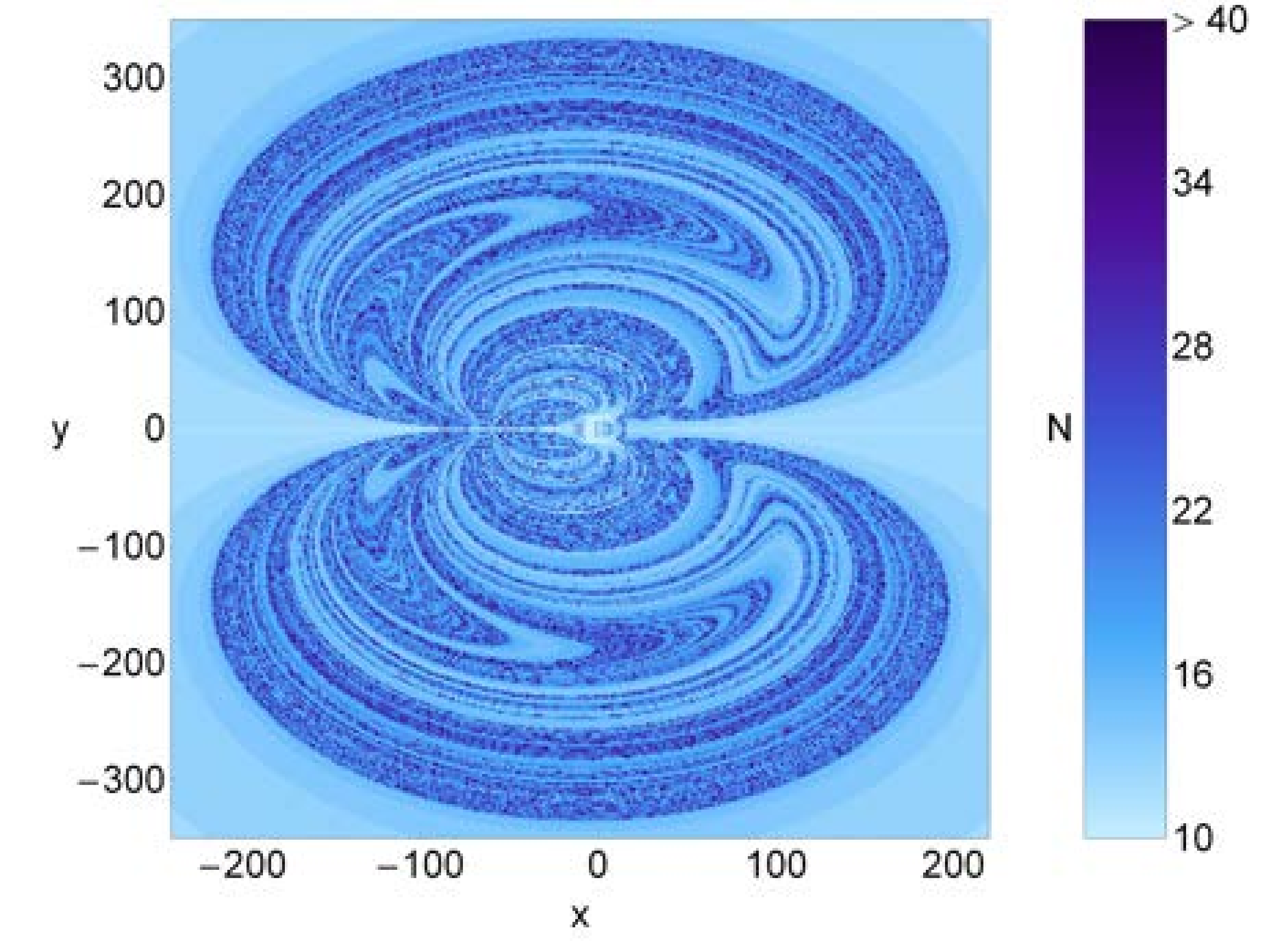}
\caption{The corresponding distributions of the number $N$ of the required iterations for obtaining the Newton-Raphson basins of convergence, shown in Fig. \ref{NR_Fig_2}(a-d). (Color figure online).}
\label{NR_Fig_2a}
\end{figure*}
\begin{figure*}[!t]
\centering
(a)\includegraphics[scale=.4]{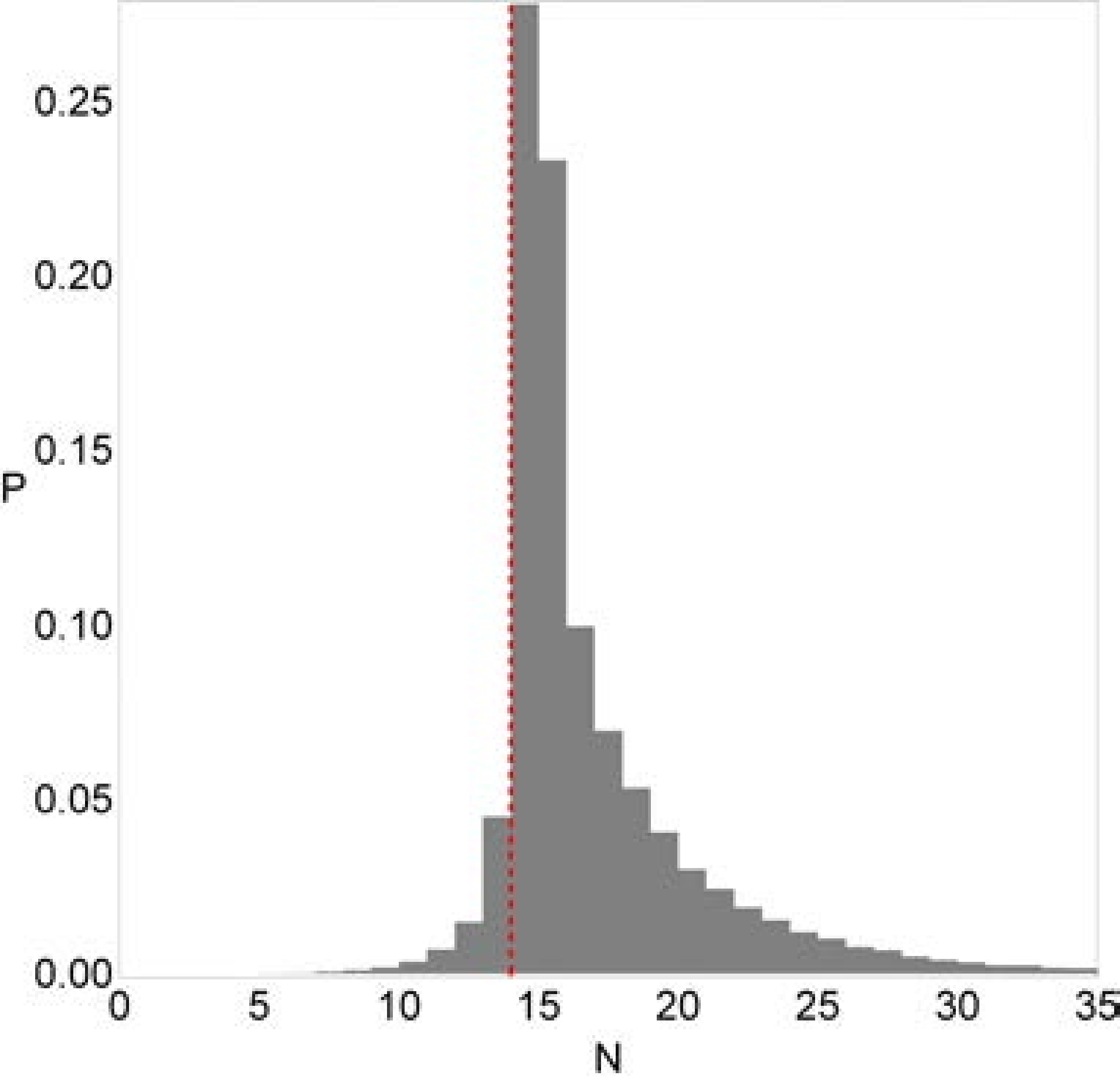}
(b)\includegraphics[scale=.4]{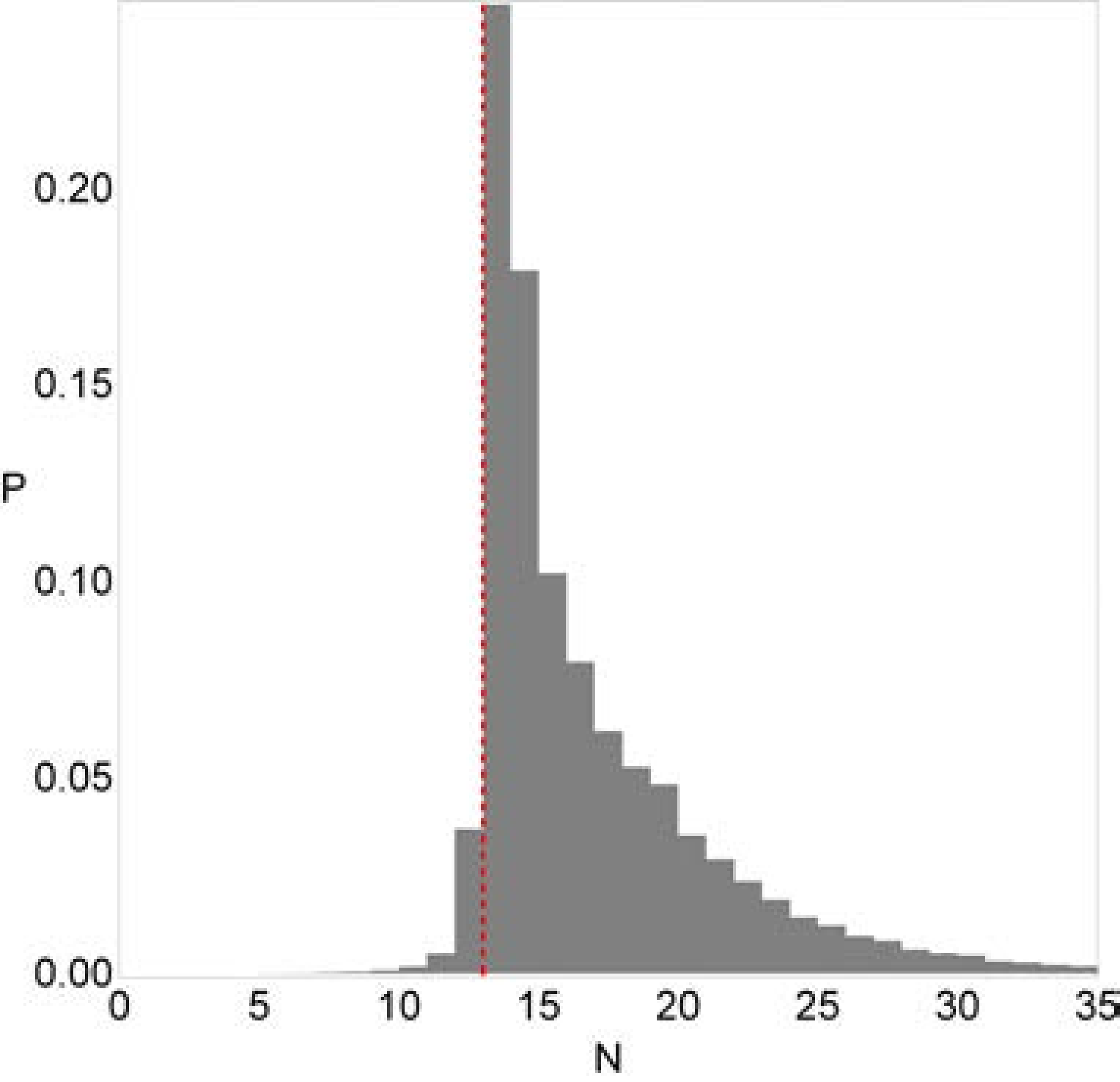}\\
(c)\includegraphics[scale=.4]{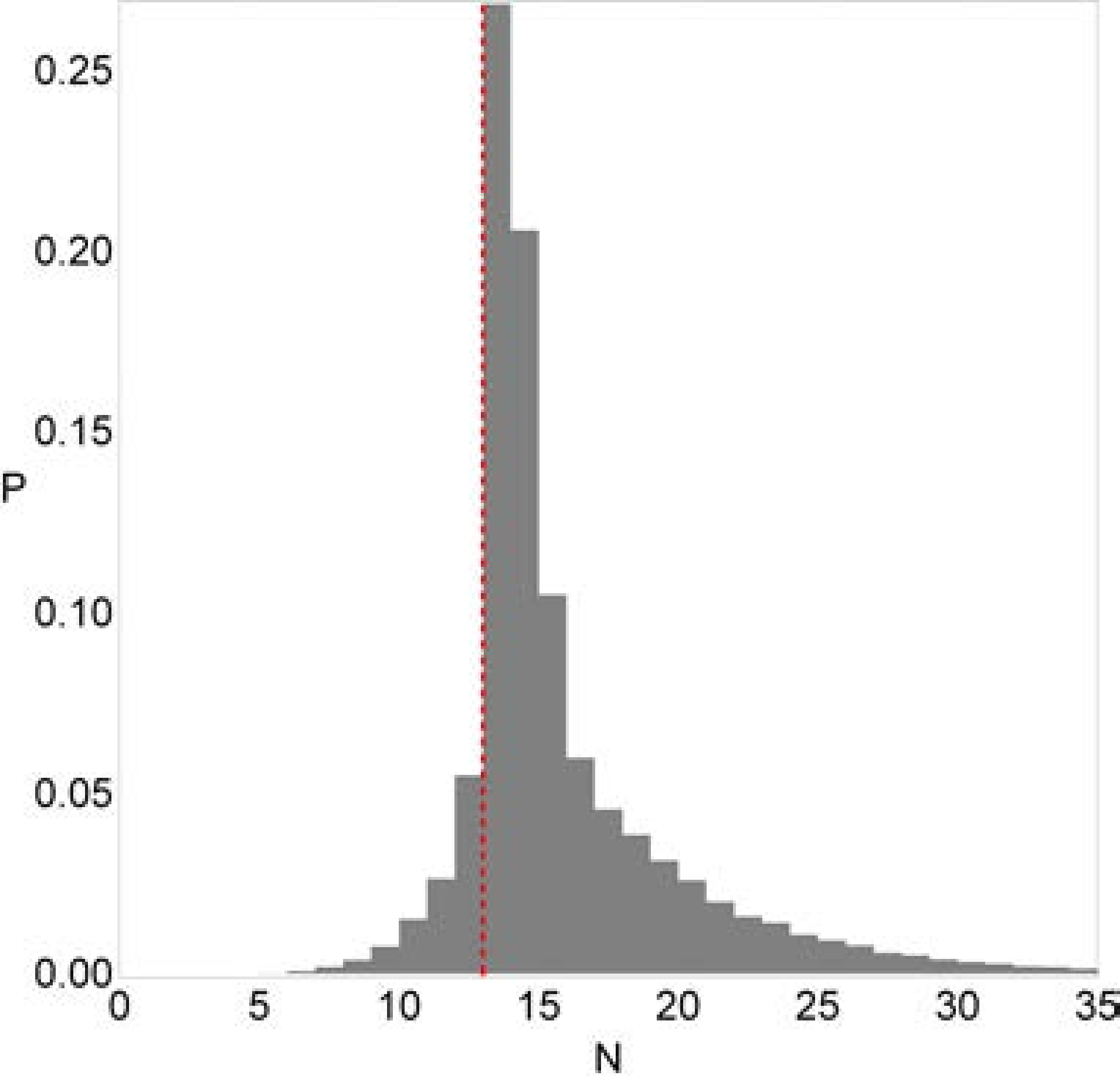}
(d)\includegraphics[scale=.4]{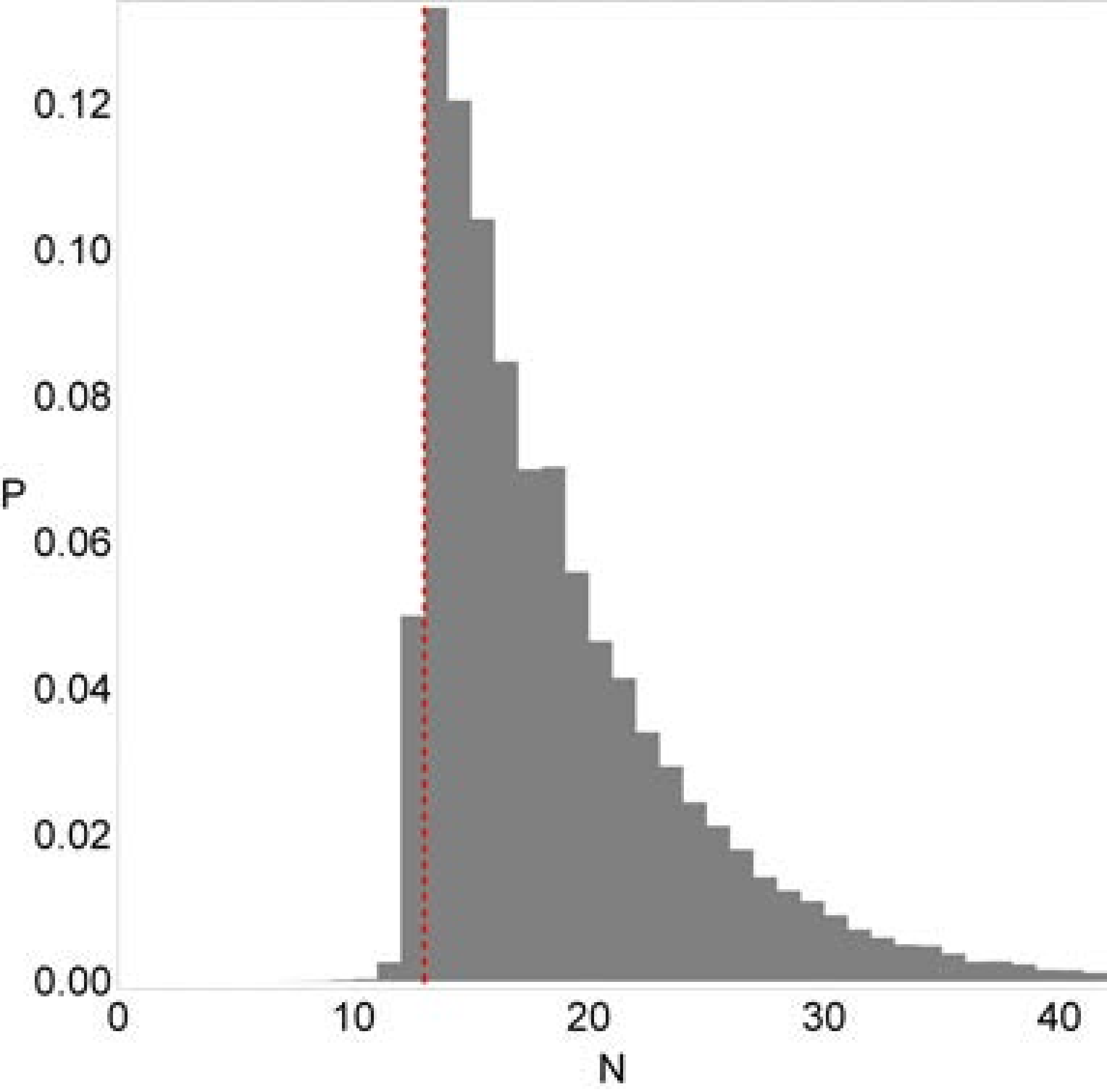}
\caption{The corresponding probability distributions of the required iterations for obtaining the Newton-Raphson basins of attraction shown in Fig. \ref{NR_Fig_2}(a-d). (Color figure online).}
\label{NR_Fig_2b}
\end{figure*}
\begin{figure*}[!t]
\centering
(a)\includegraphics[scale=.65]{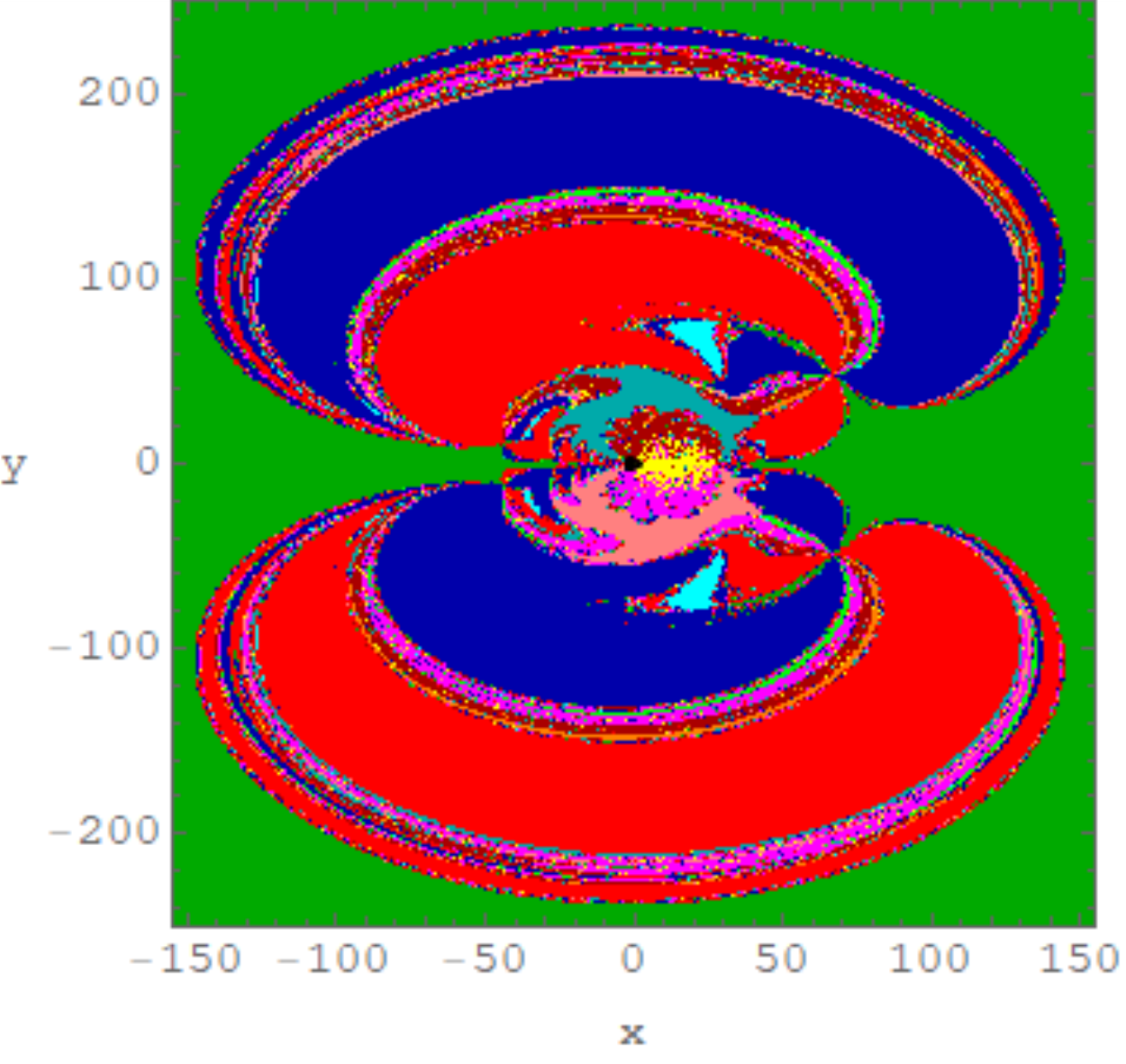}
(b)\includegraphics[scale=6]{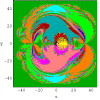}\\
(c)\includegraphics[scale=.35]{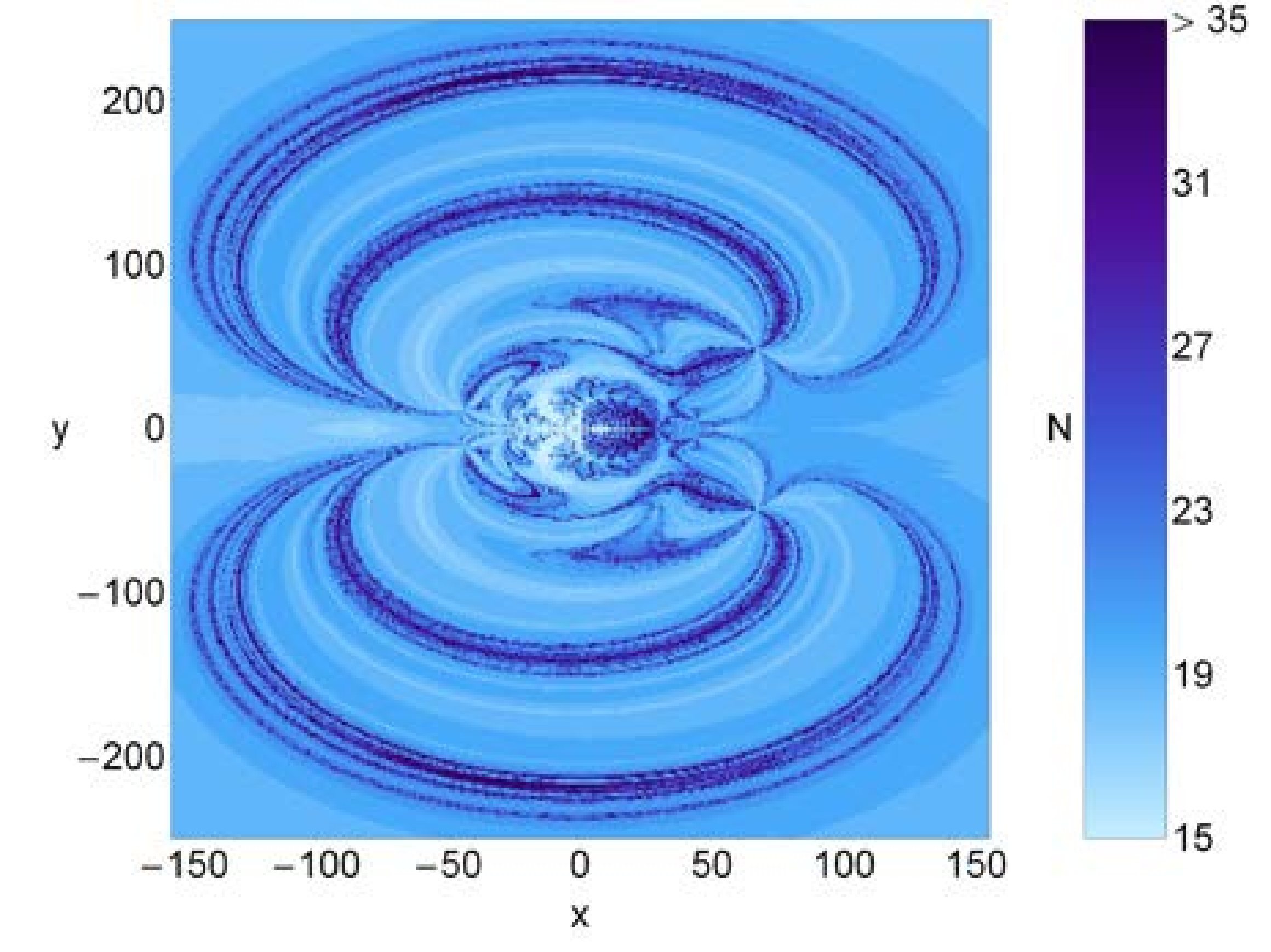}
(d)\includegraphics[scale=.35]{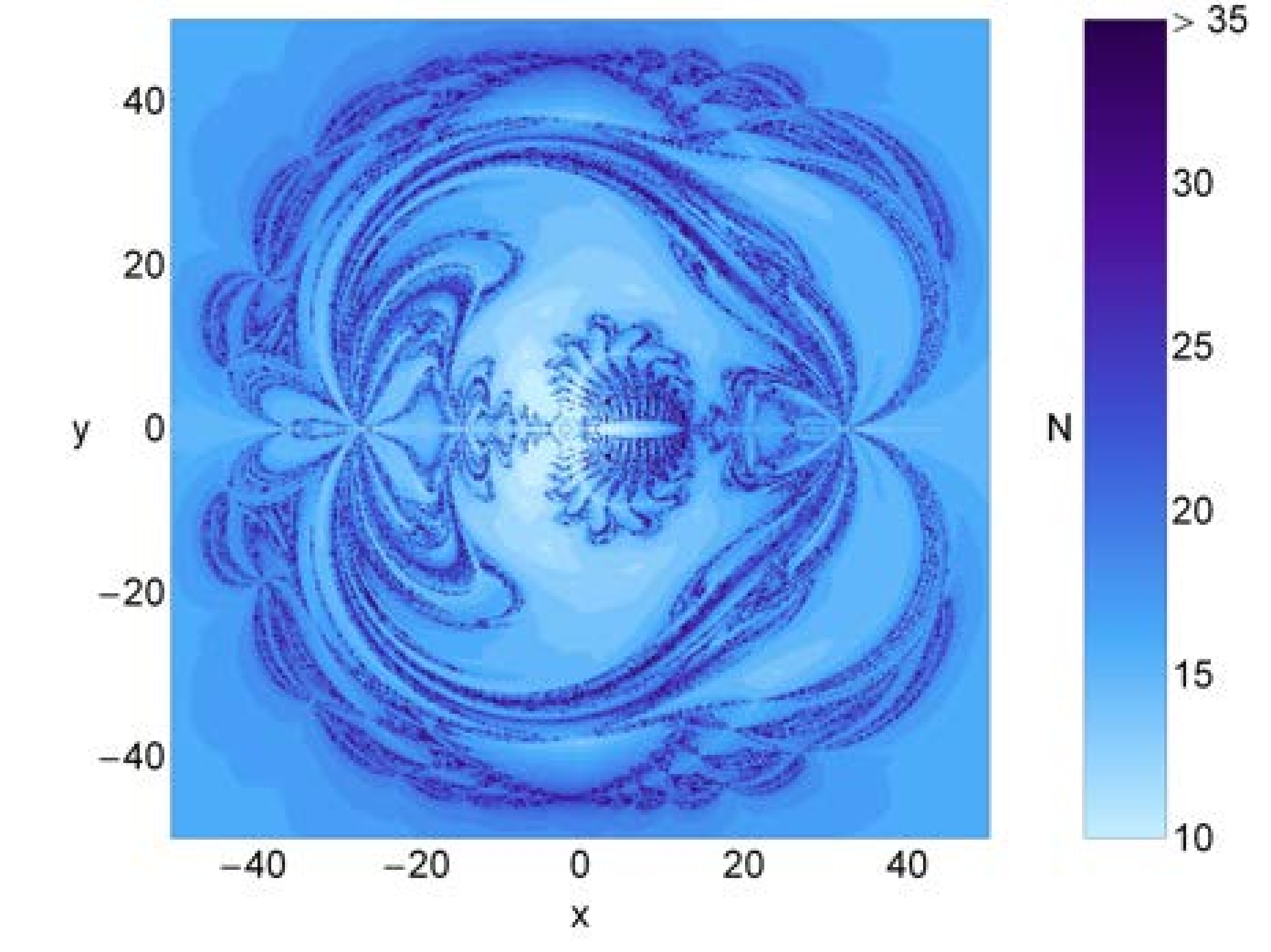}\\
(e)\includegraphics[scale=.4]{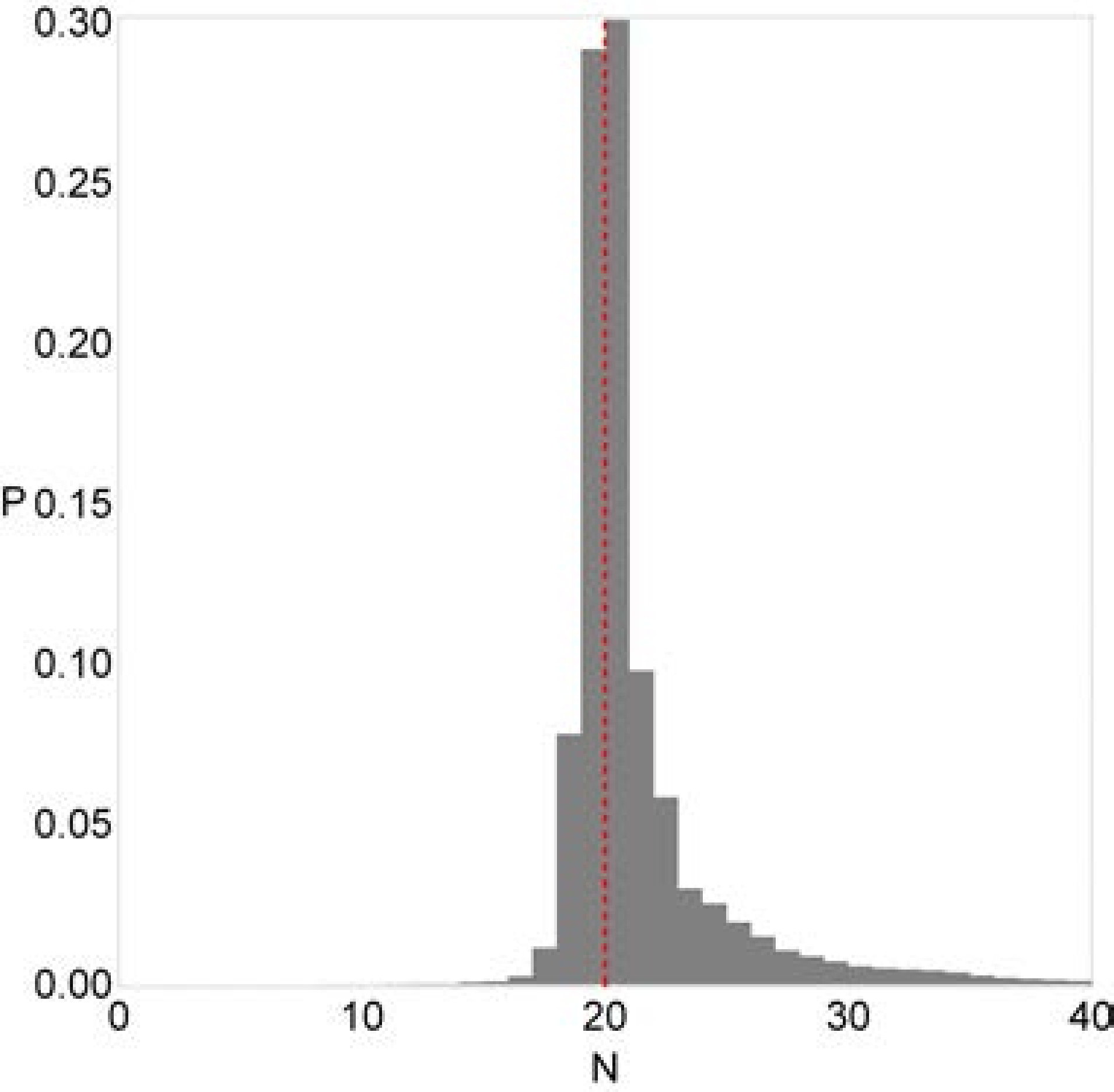}
(f)\includegraphics[scale=.4]{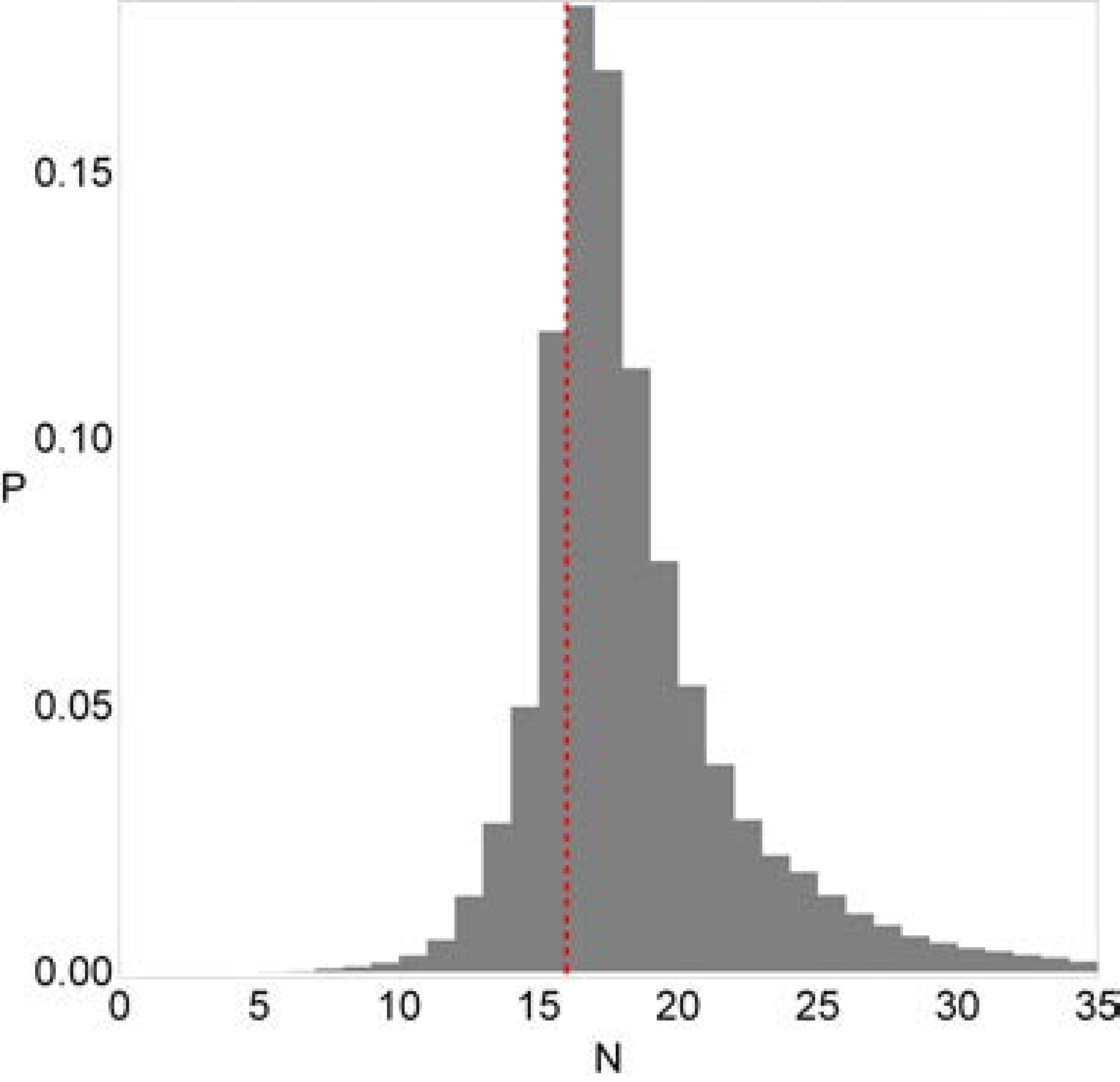}
\caption{The Newton-Raphson basins of convergence on the configuration $(x, y)$ plane for fixed value of $\alpha=59 \degree$ (a) $\beta=16 \degree$, (b) $\beta=18 \degree$, when 11 libration points exist. The color code is same as in Fig. \ref{NR_Fig_1}. (Second row): The distribution of the corresponding number $N$ of the required iterations for obtaining the Newton-Raphson basins of convergence.  (Third row): The corresponding probability distributions of the required iterations for obtaining the Newton-Raphson basins of convergence. (Color figure online).}
\label{NR_Fig_3}
\end{figure*}
\begin{figure*}[!t]
\centering
(a)\includegraphics[scale=.45]{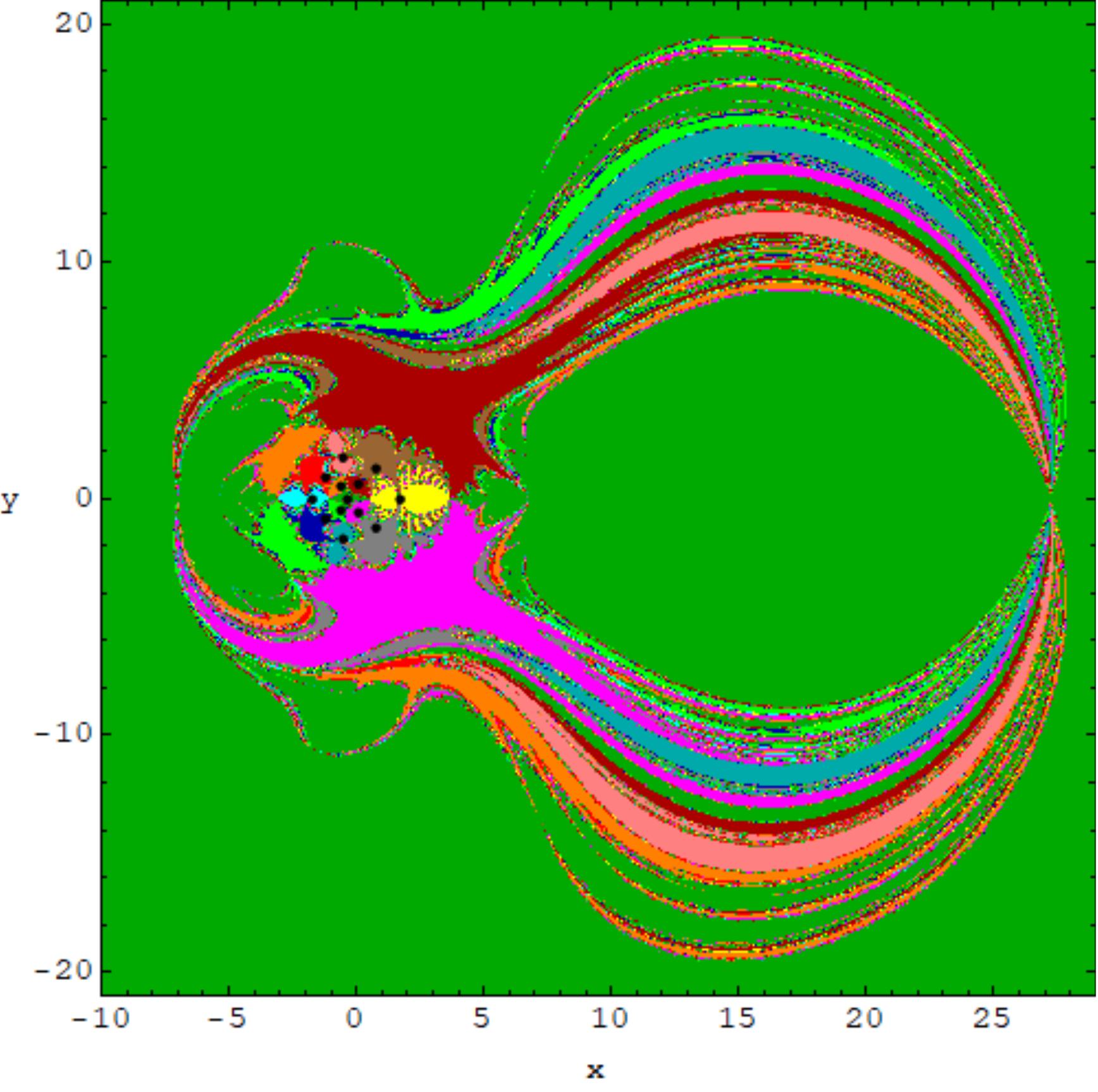}
(b)\includegraphics[scale=6]{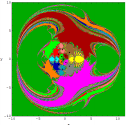}\\
(c)\includegraphics[scale=.45]{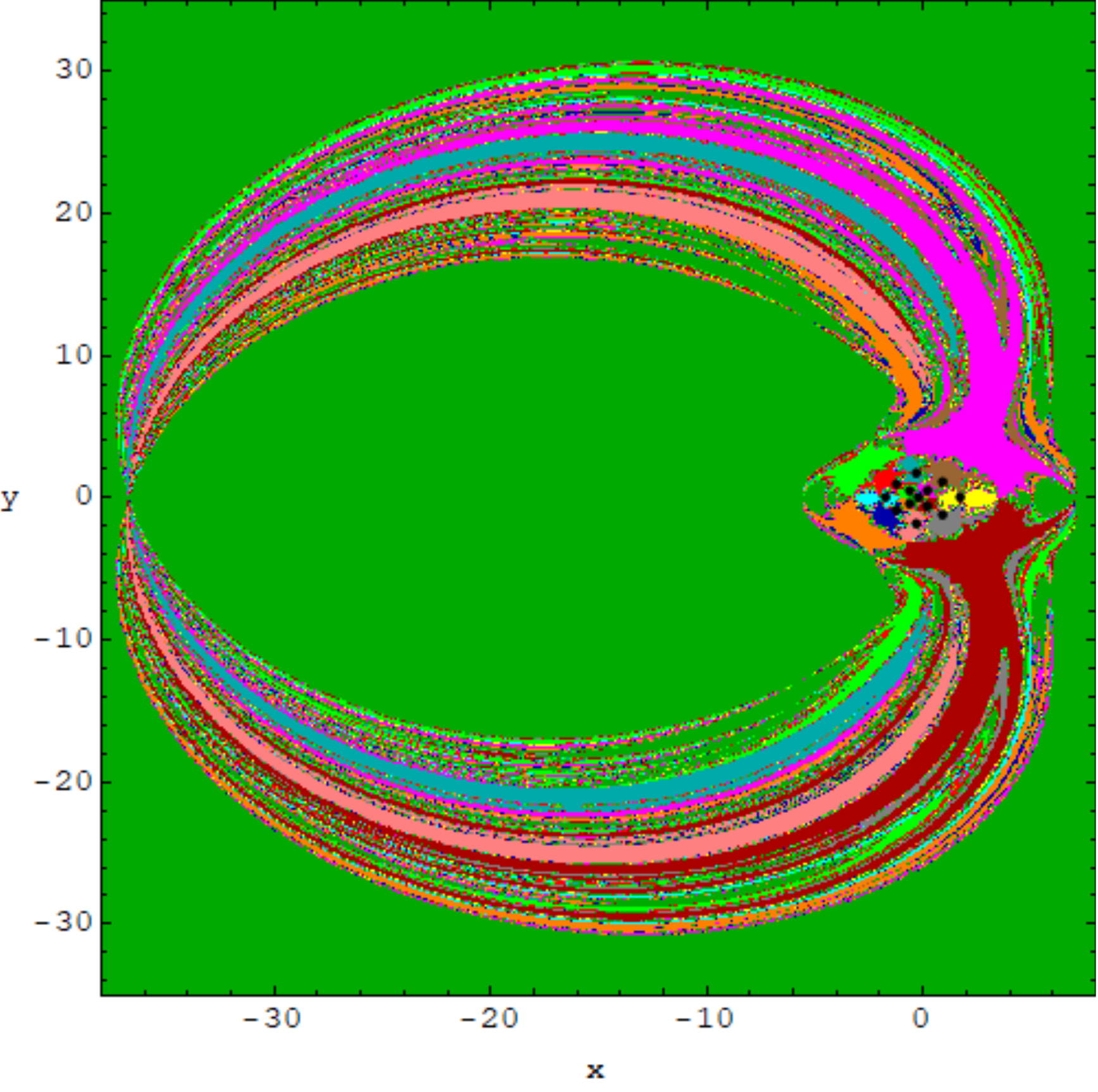}
(d)\includegraphics[scale=6]{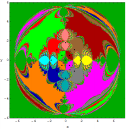}
\caption{The Newton-Raphson basins of attraction on the configuration $(x, y)$ plane for fixed value of $\alpha=43.5 \degree$ (a) $\beta=34.888\degree$, (b) $\beta=36 \degree$, (c) $\beta=38 \degree$, (d) $\beta=43 \degree$, when 13 libration points exist. The color code for libration points $L_1,...,L_{11}$ is same as in Fig.\ref{NR_Fig_1} while $L_{12}$ (\emph{gray}), $L_{13}$ (\emph{brown}) and non-converging points (white). (Color figure online).}
\label{NR_Fig_4}
\end{figure*}
\begin{figure*}[!t]
\centering
(a)\includegraphics[scale=.35]{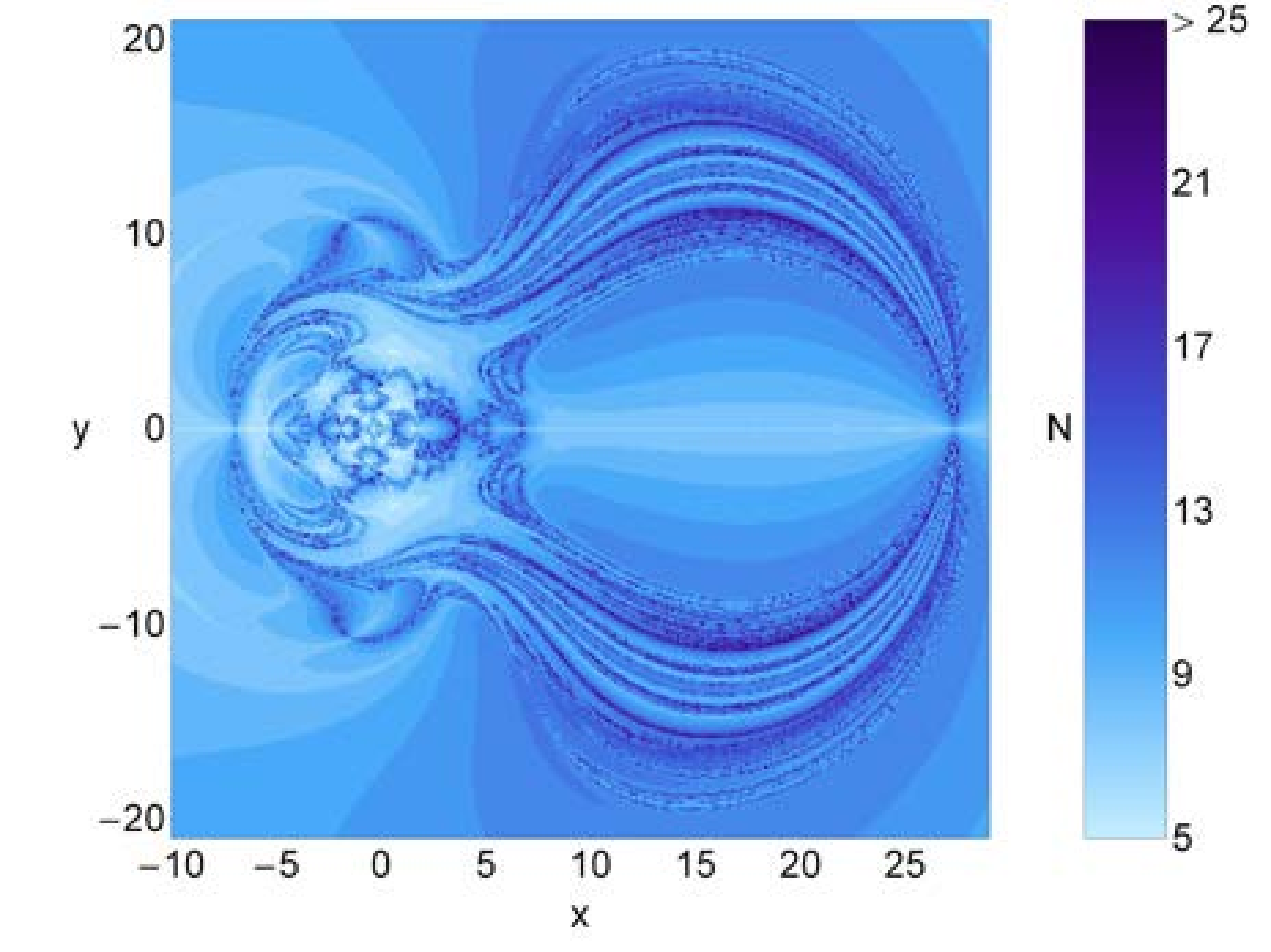}
(b)\includegraphics[scale=.35]{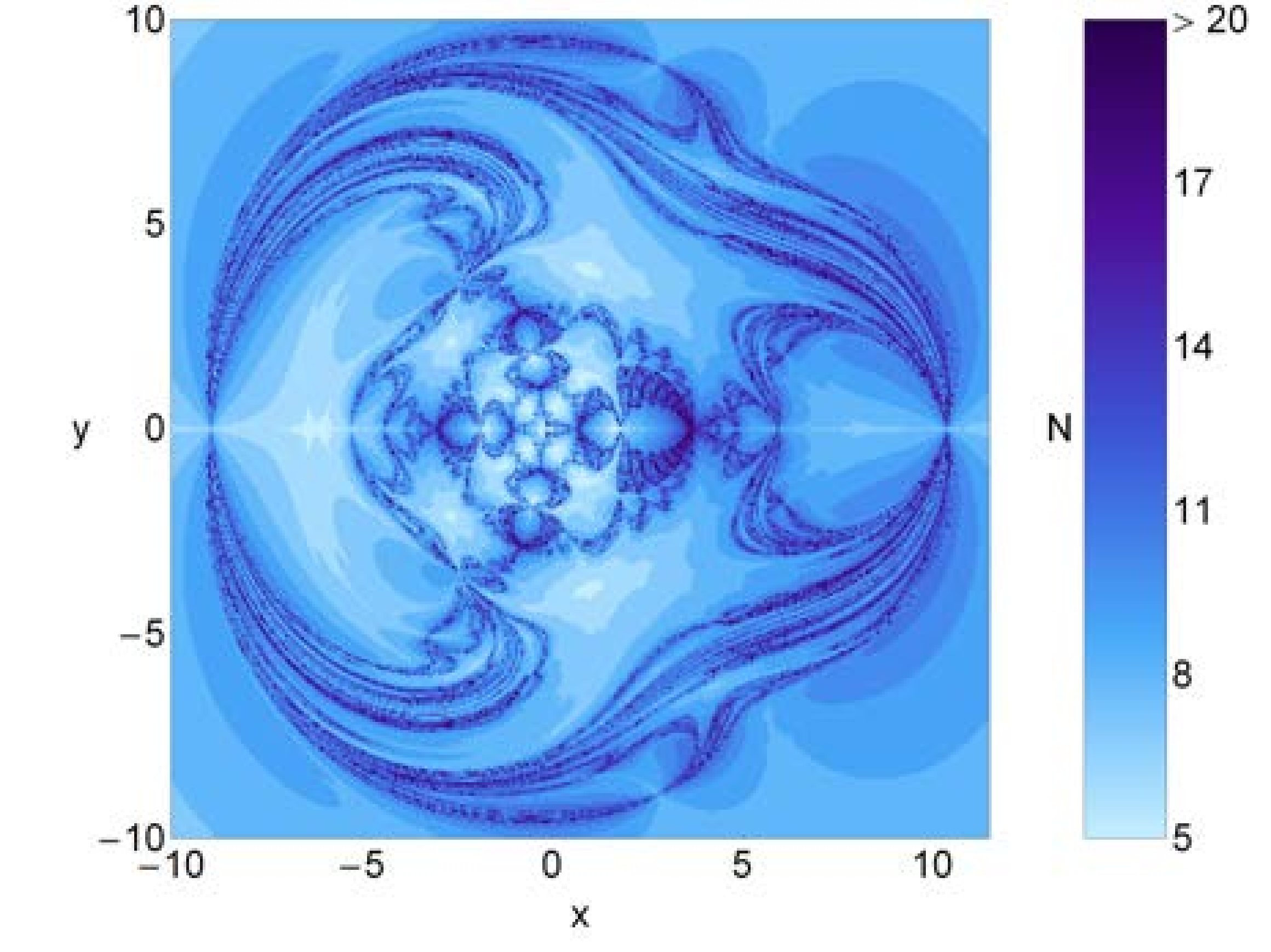}\\
(c)\includegraphics[scale=.35]{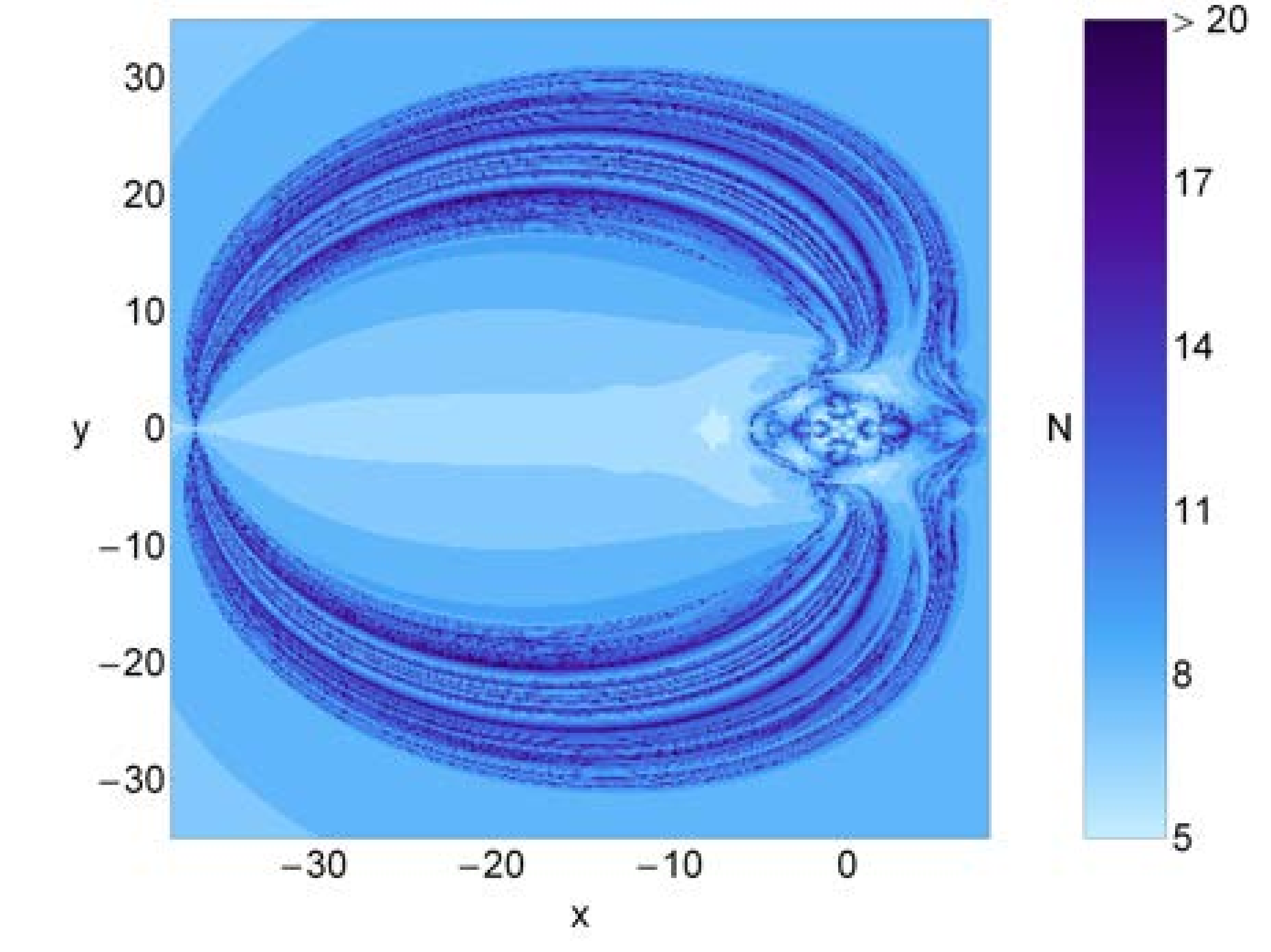}
(d)\includegraphics[scale=.35]{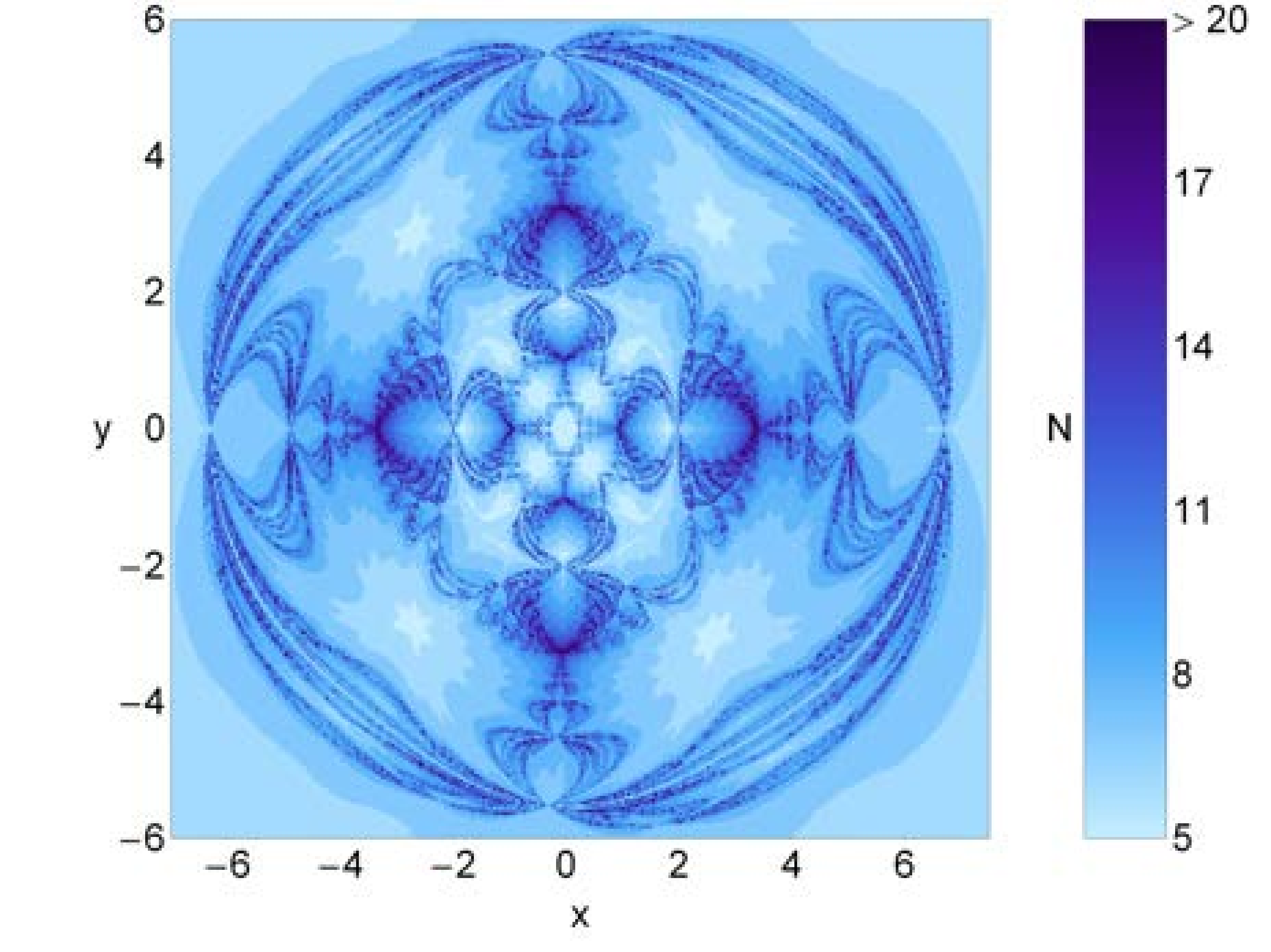}
\caption{The distribution of the corresponding number $N$ of the required iterations for obtaining the Newton-Raphson basins of attraction shown in Fig.\ref{NR_Fig_4}. (Color figure online).}
\label{NR_Fig_4a}
\end{figure*}
\begin{figure*}[!t]
\centering
(a)\includegraphics[scale=.4]{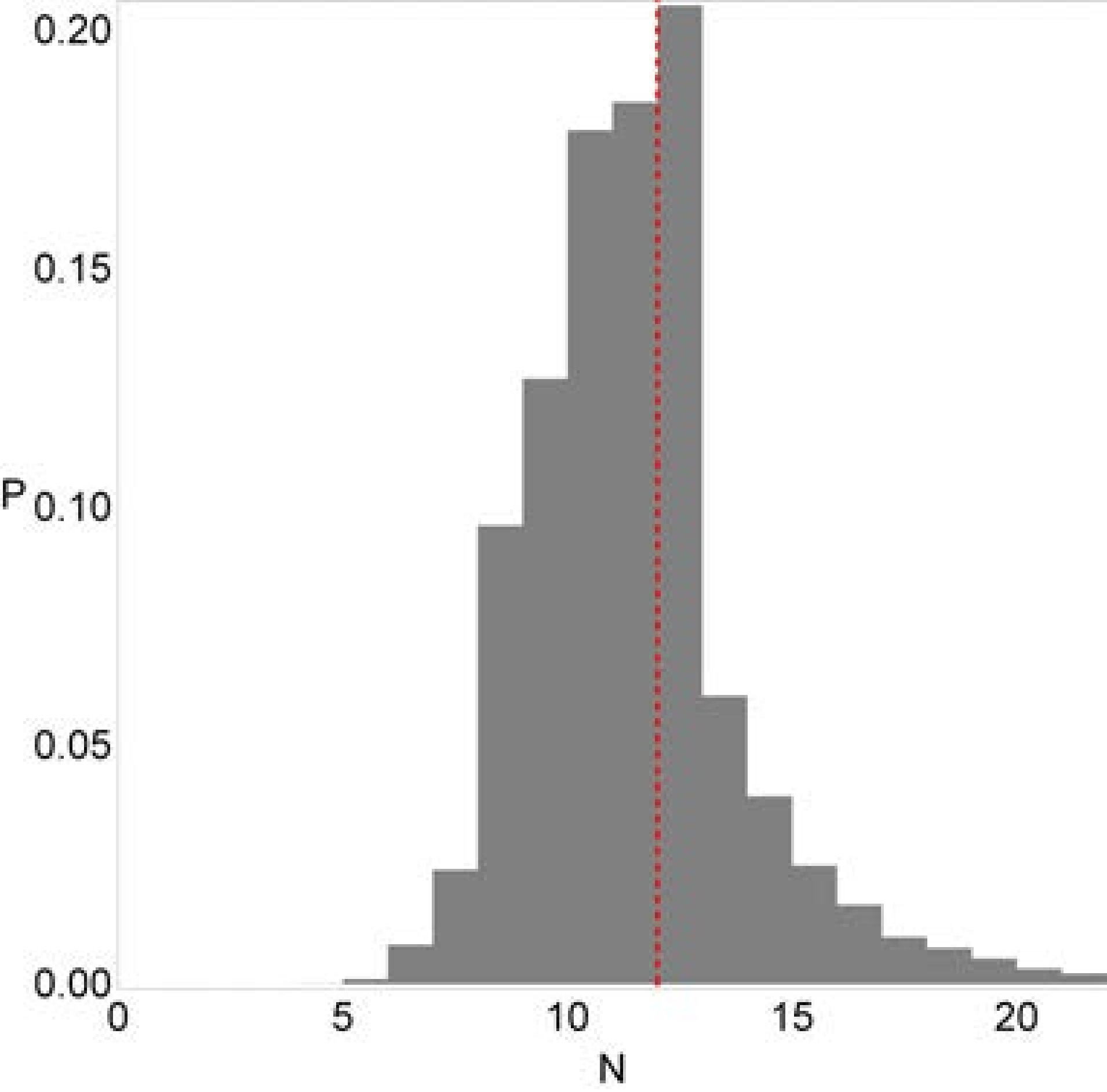}
(b)\includegraphics[scale=.4]{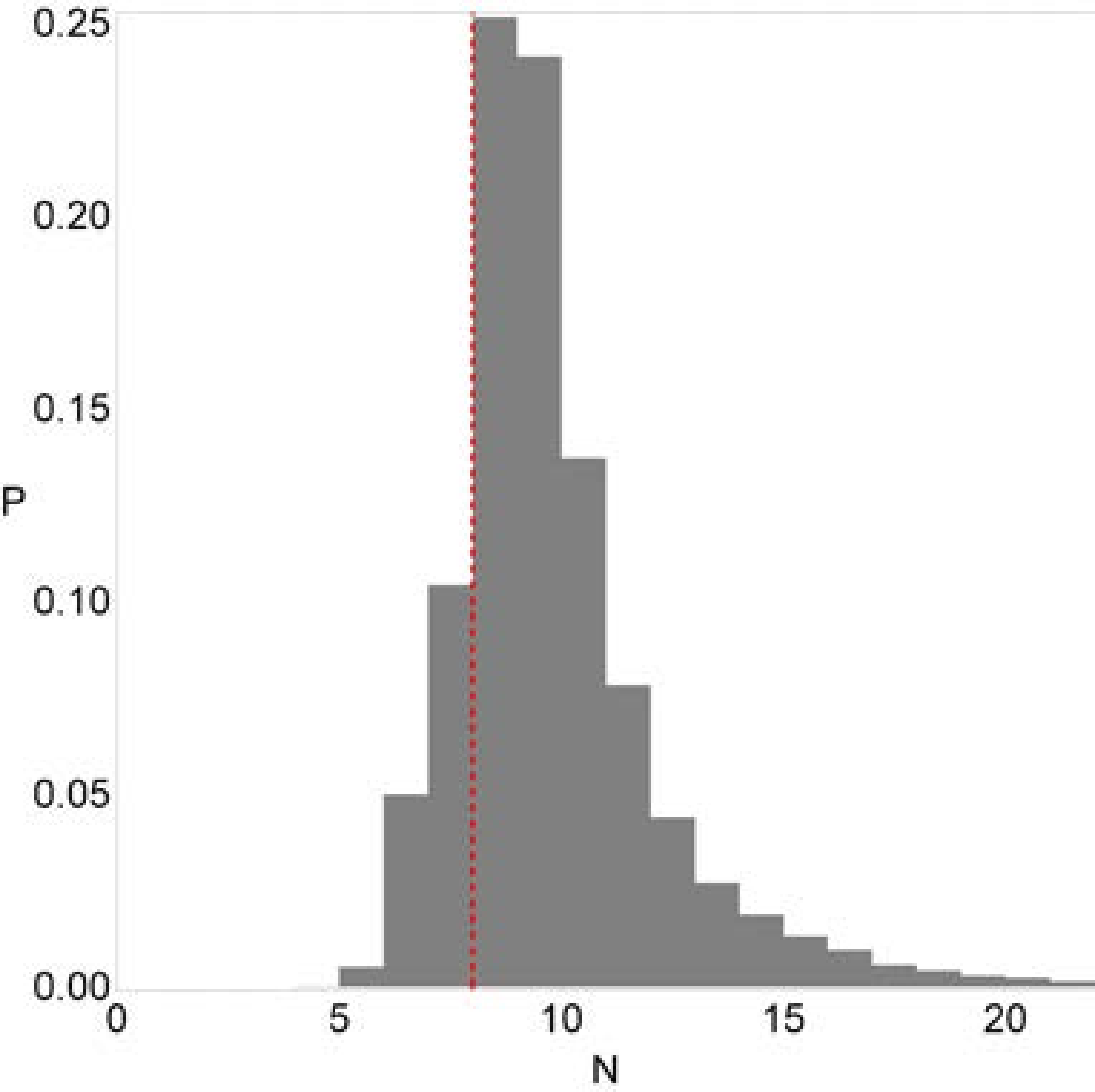}\\
(c)\includegraphics[scale=.4]{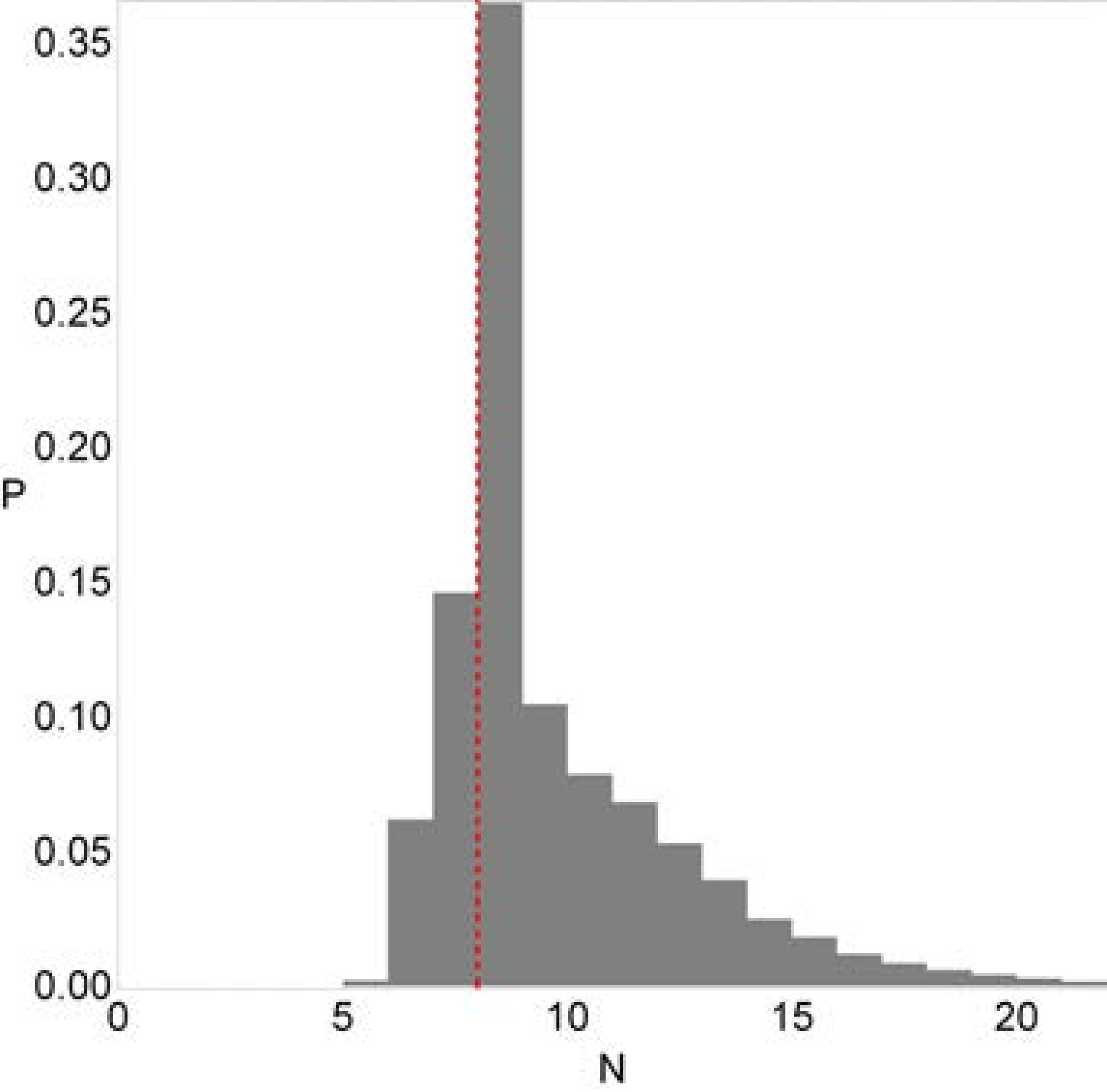}
(d)\includegraphics[scale=.4]{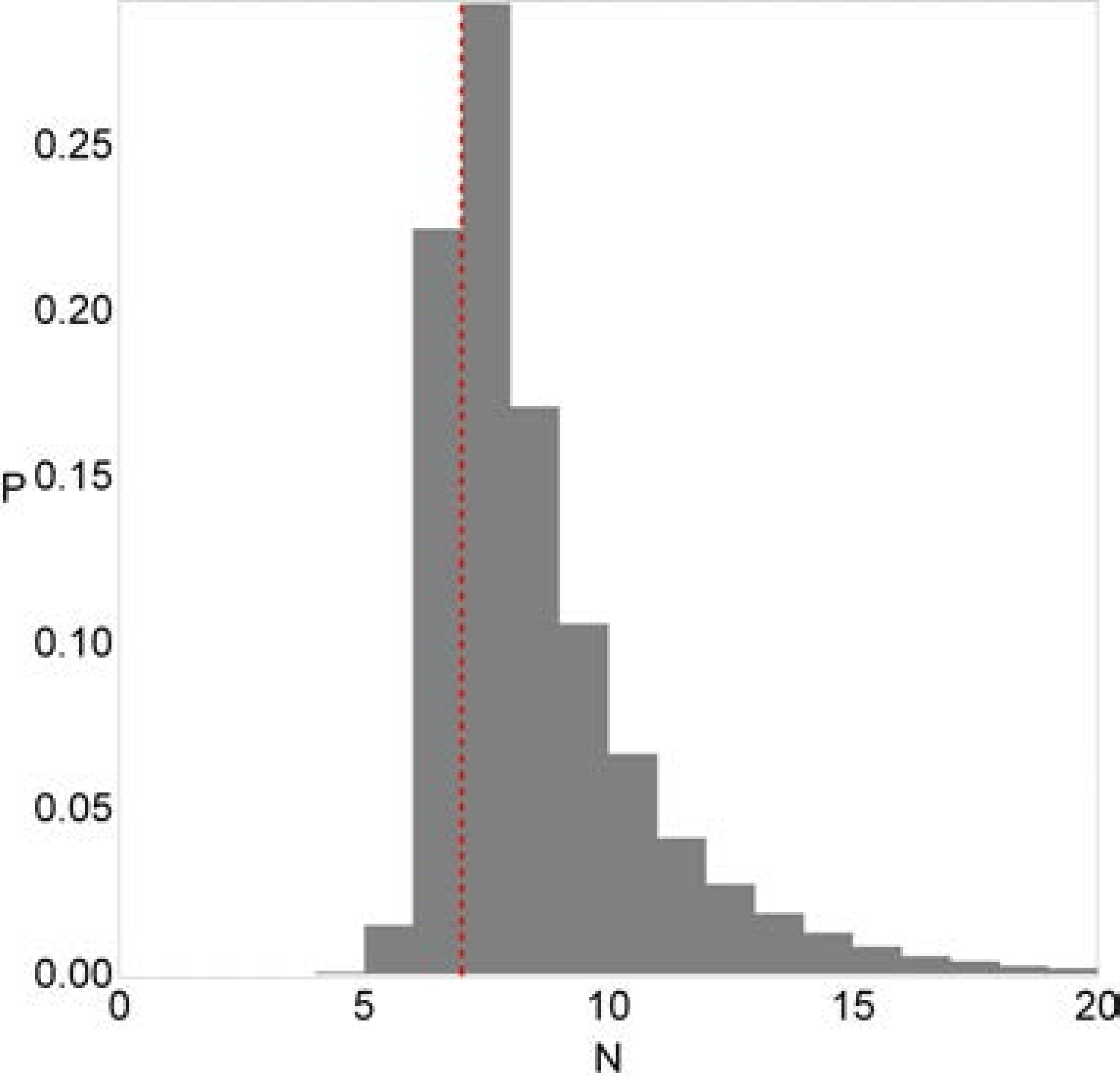}
\caption{The corresponding probability distributions of the required iterations for obtaining the Newton-Raphson basins of attraction shown in Fig. \ref{NR_Fig_4}. (Color figure online).}
\label{NR_Fig_4b}
\end{figure*}
\begin{figure*}[!t]
\centering
(a)\includegraphics[scale=5]{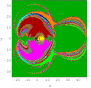}
(b)\includegraphics[scale=5]{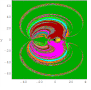}
(c)\includegraphics[scale=3]{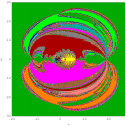}\\
(d)\includegraphics[scale=5]{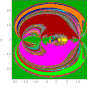}
(e)\includegraphics[scale=5]{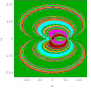}
(f)\includegraphics[scale=5]{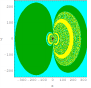}
\caption{The Newton-Raphson basins of attraction on the configuration $(x, y)$ plane for fixed value of  $\alpha=55.5 \degree$ (a) $\beta=25 \degree$, (b)$\beta=26 \degree$, (c) $\beta=28 \degree$, (d) $\beta=37 \degree$, (e) $\beta=45 \degree$, (f) $\beta=50 \degree$, when 13 libration points exist. The color code is same as in  Fig. \ref{NR_Fig_4}. (Color figure online).}
\label{NR_Fig_5}
\end{figure*}
\begin{figure*}[!t]
\centering
(a)\includegraphics[scale=.35]{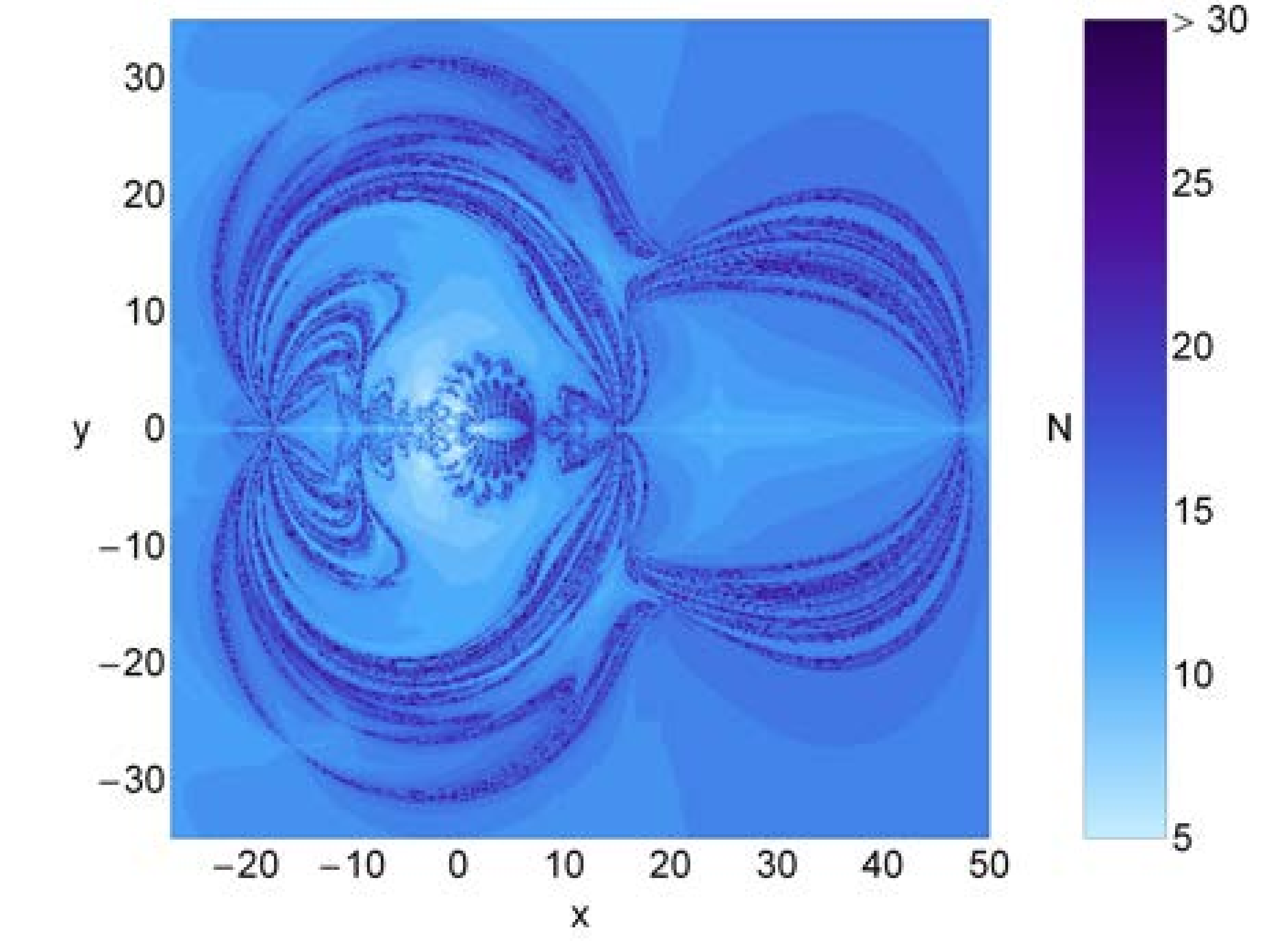}
(b)\includegraphics[scale=.35]{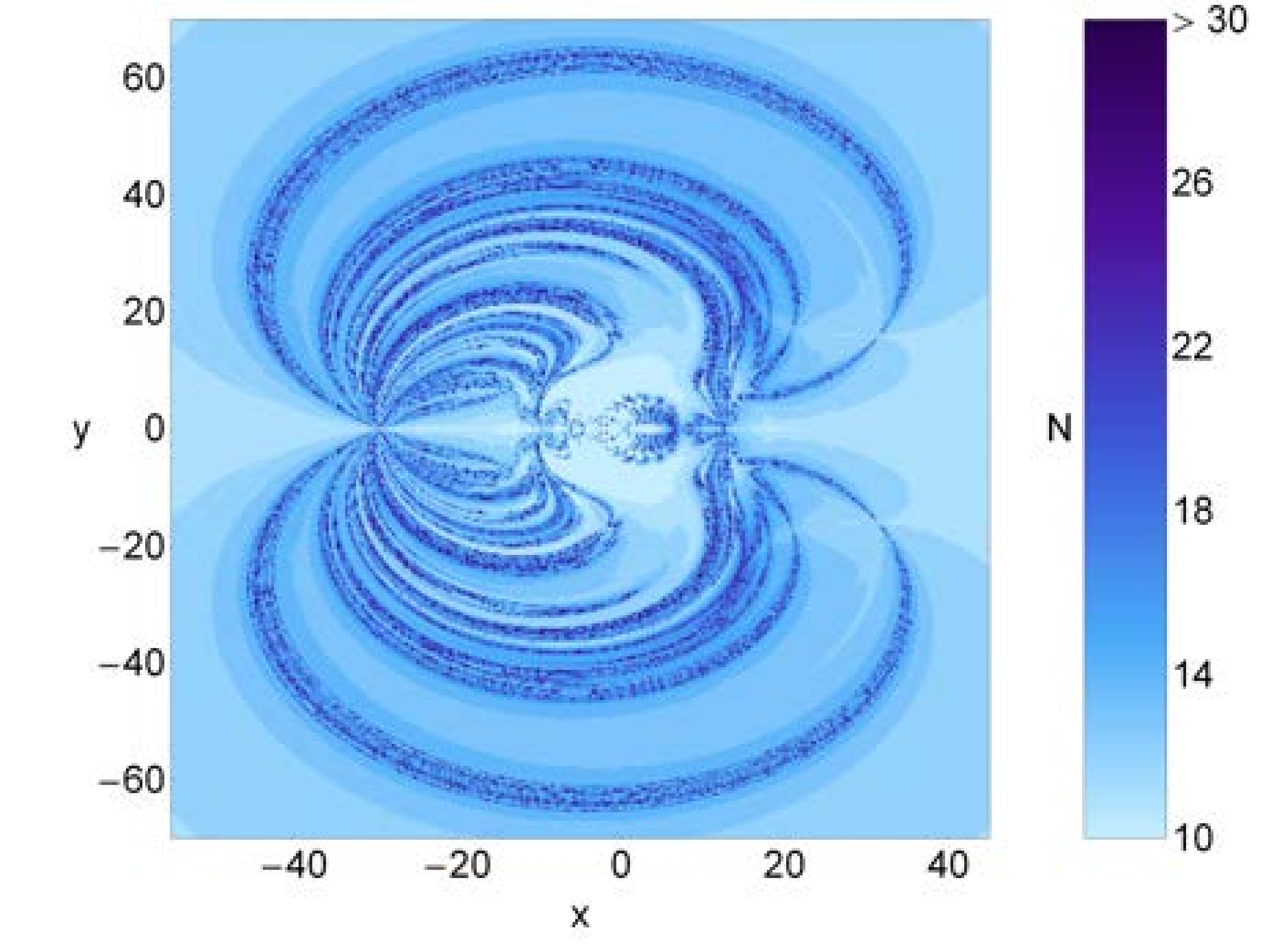}\\
(c)\includegraphics[scale=.35]{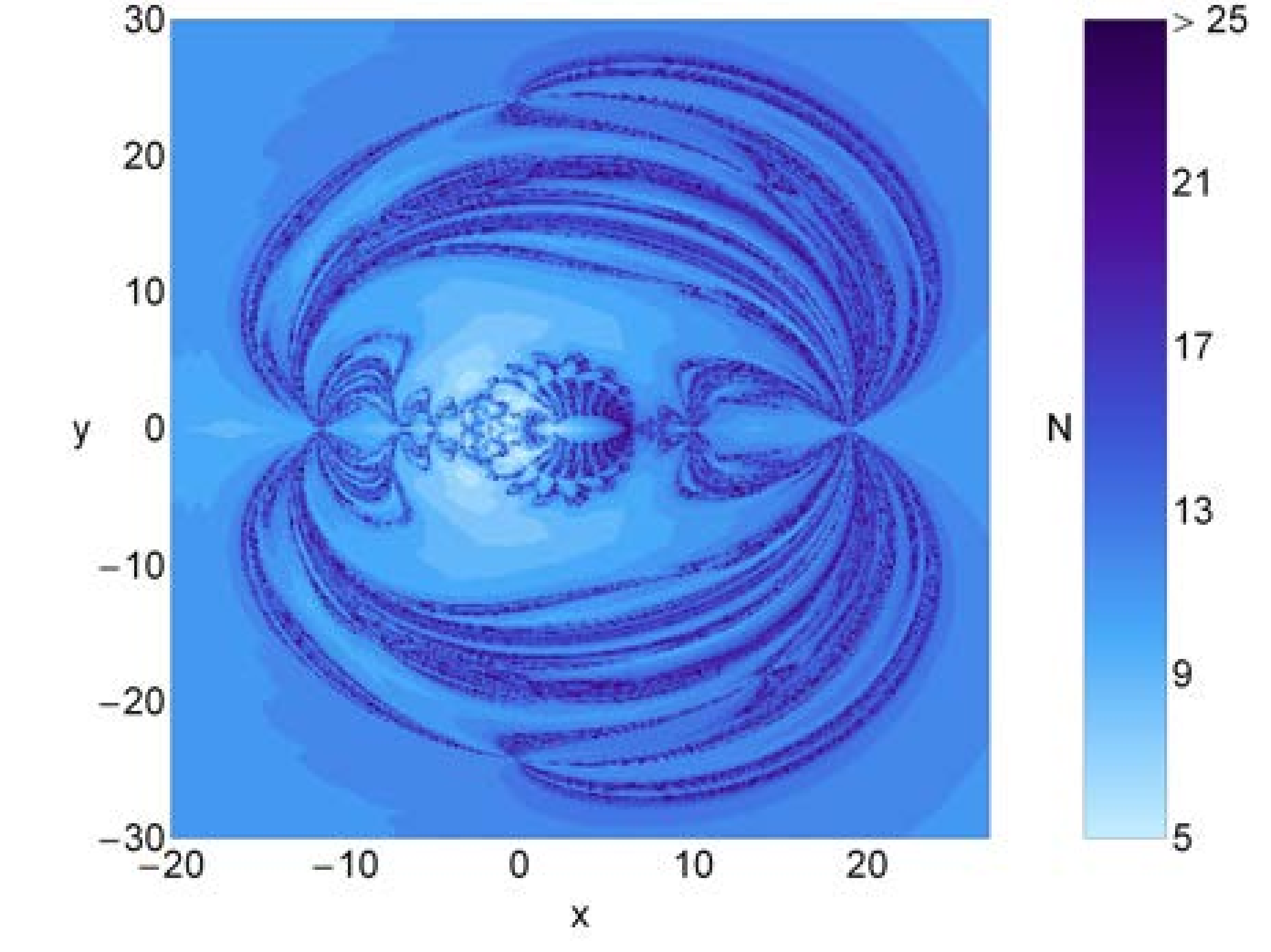}
(d)\includegraphics[scale=.35]{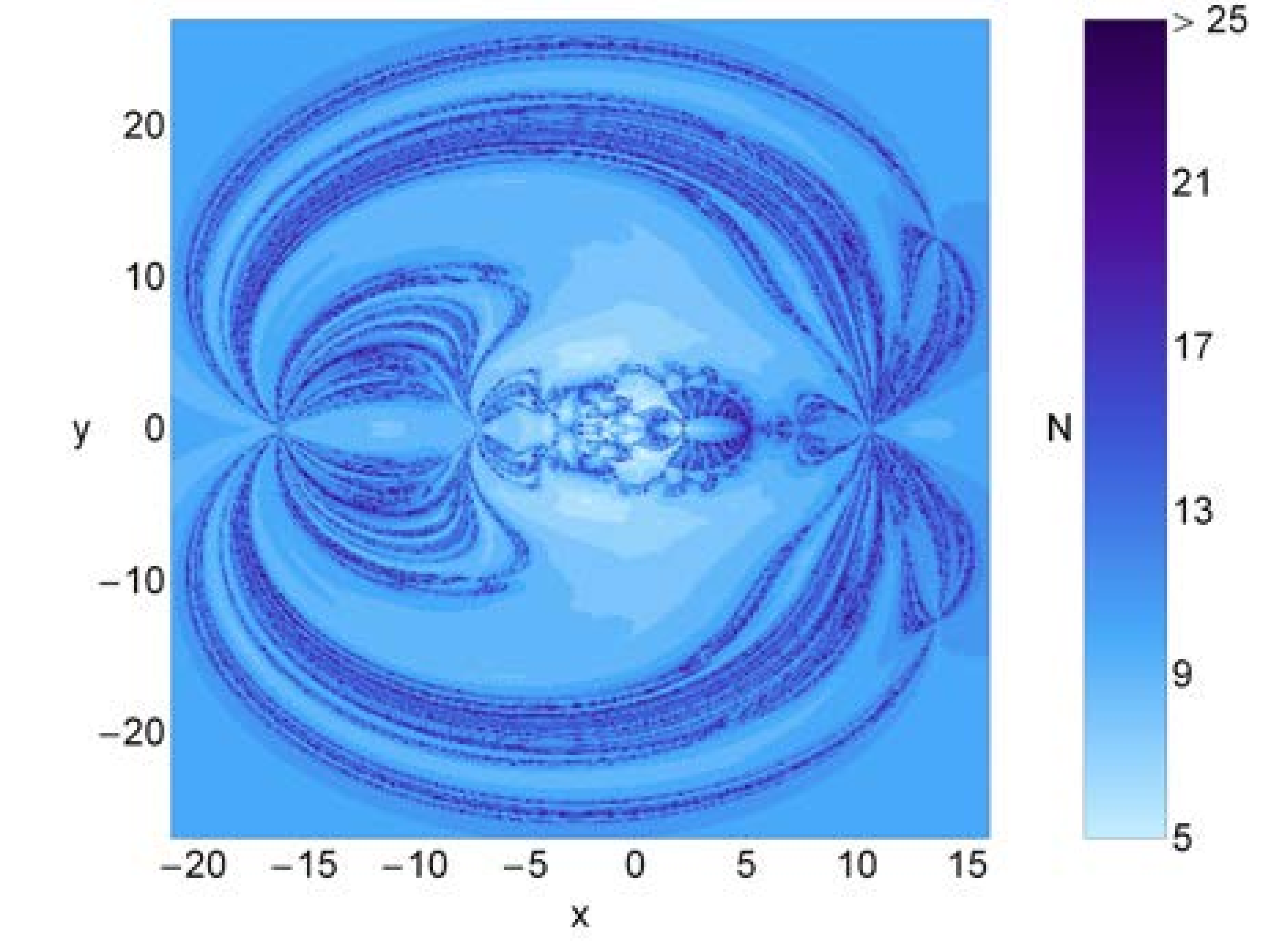}\\
(e)\includegraphics[scale=.35]{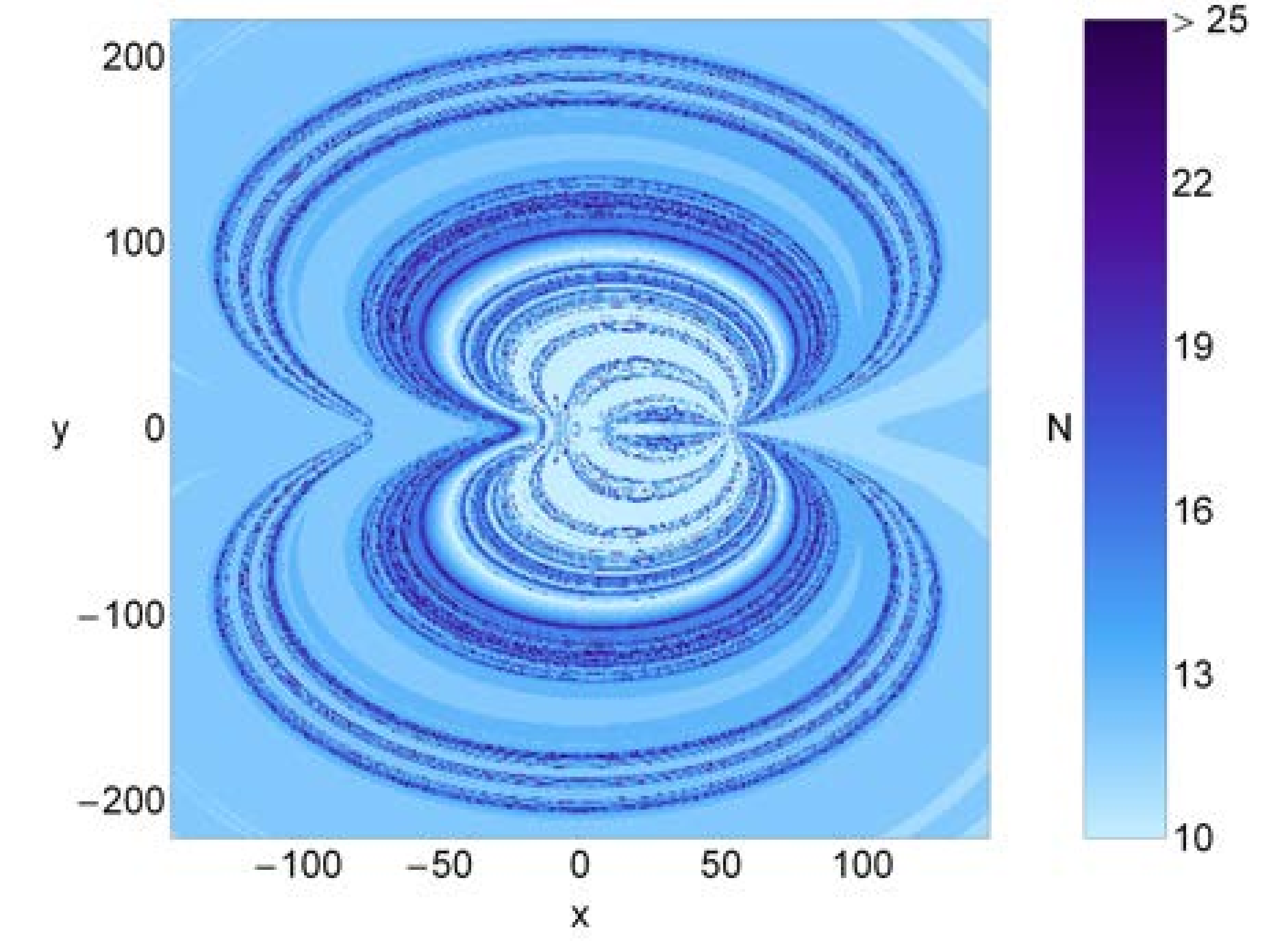}
(f)\includegraphics[scale=.35]{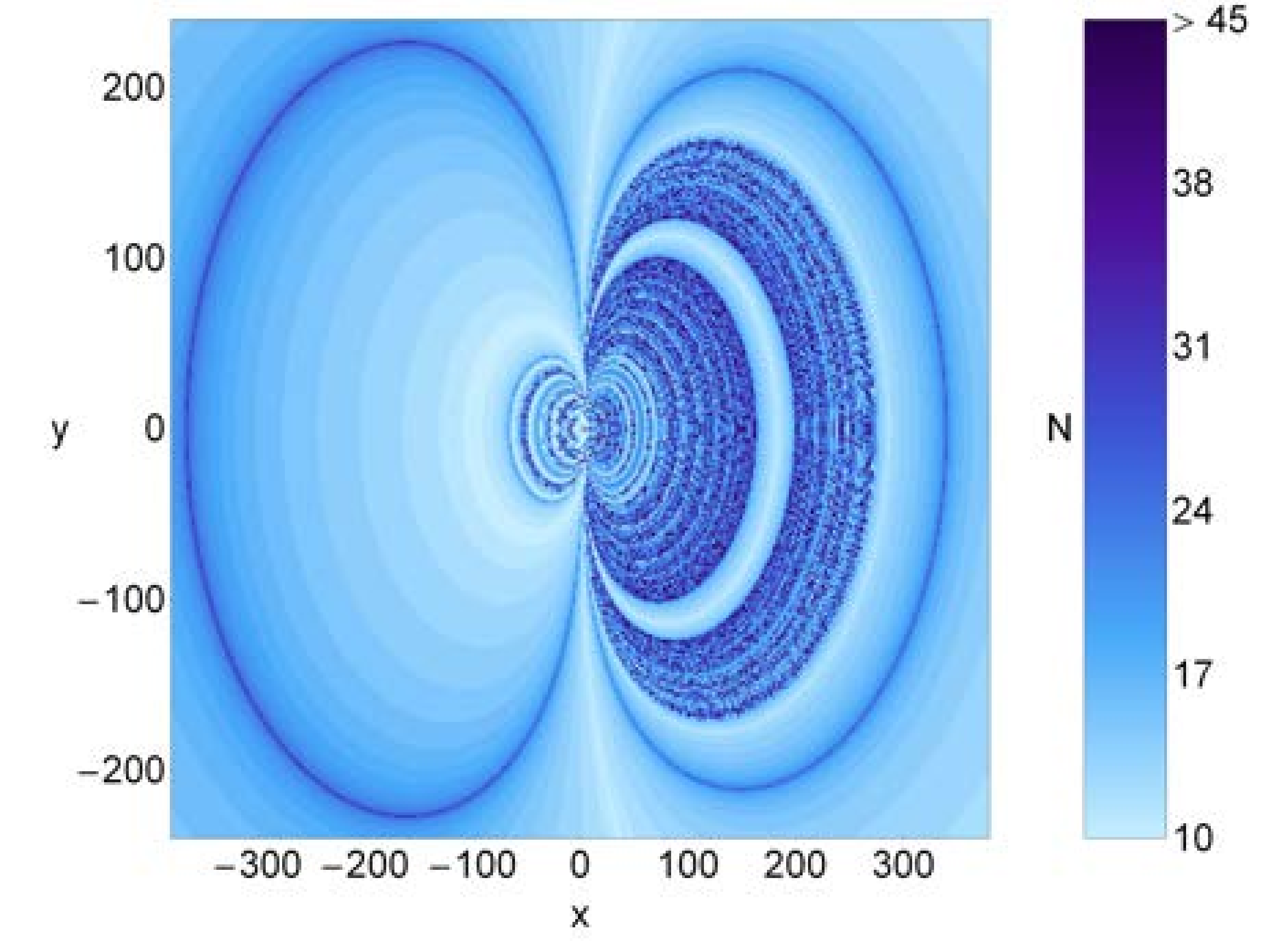}
\caption{The distribution of the corresponding number $N$ of the required iterations for obtaining the Newton-Raphson basins of attraction shown in Fig.\ref{NR_Fig_5}. (Color figure online).}
\label{NR_Fig_5a}
\end{figure*}
\begin{figure*}[!t]
\centering
(a)\includegraphics[scale=.4]{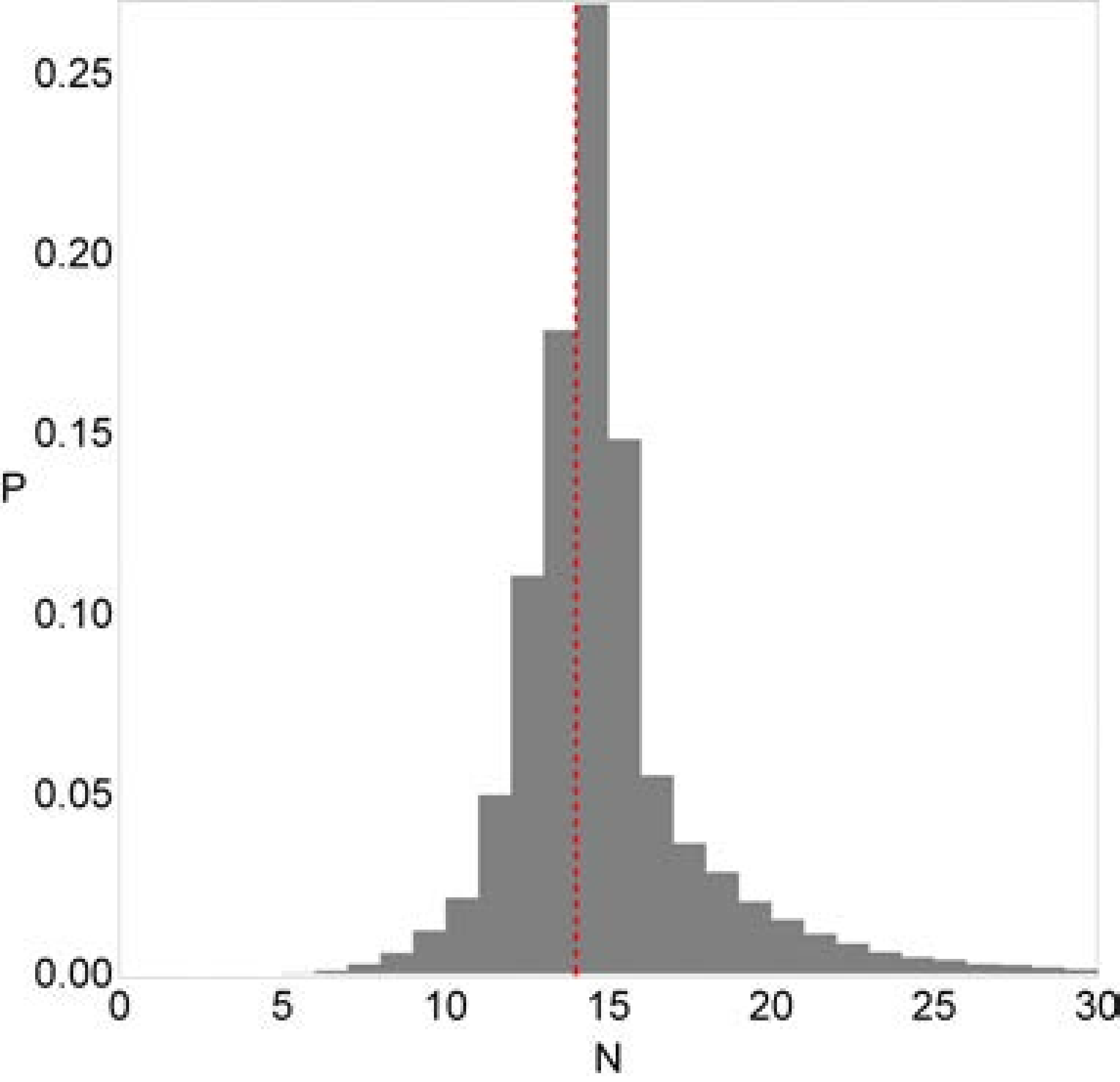}
(b)\includegraphics[scale=.4]{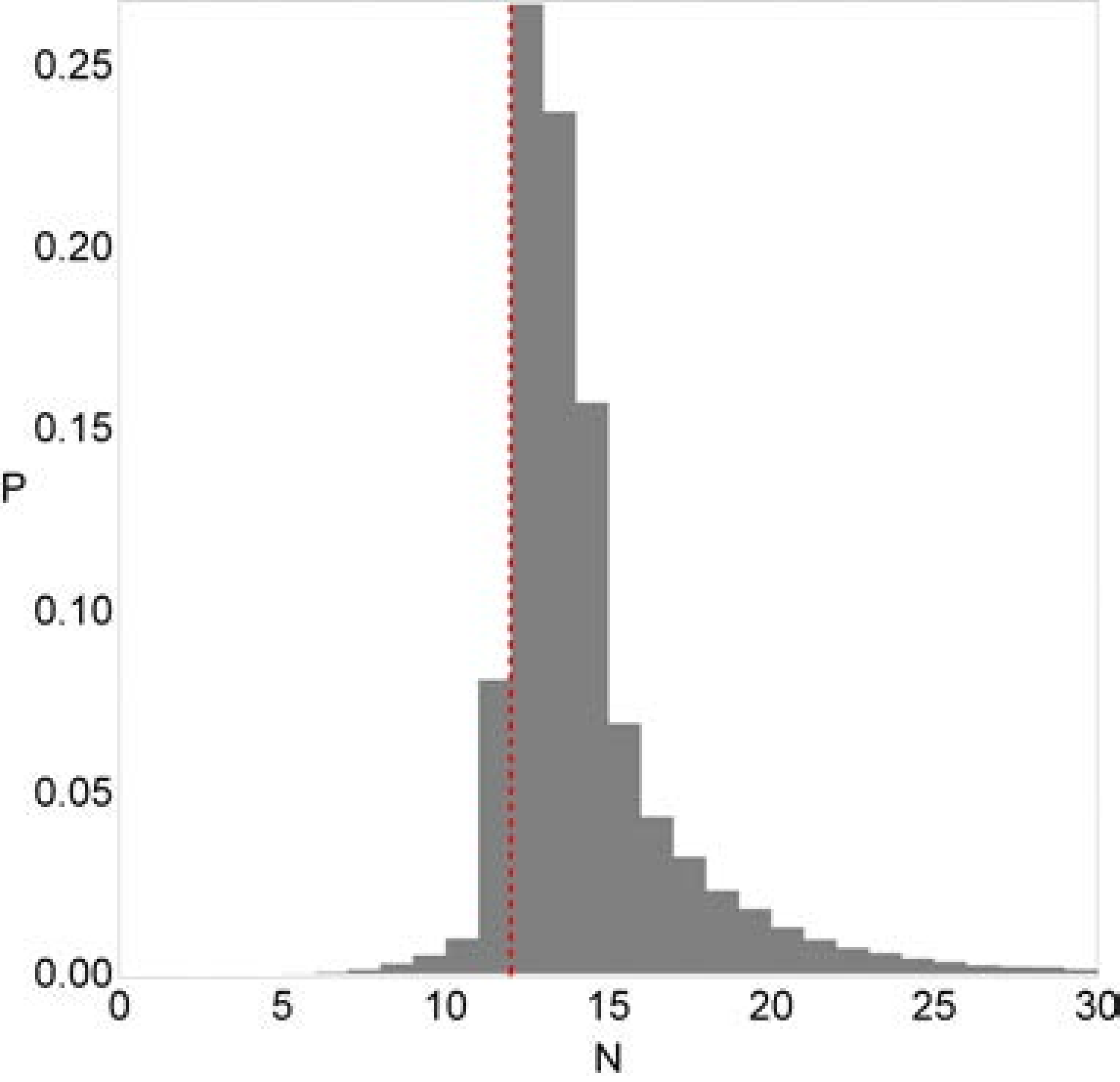}\\
(c)\includegraphics[scale=.4]{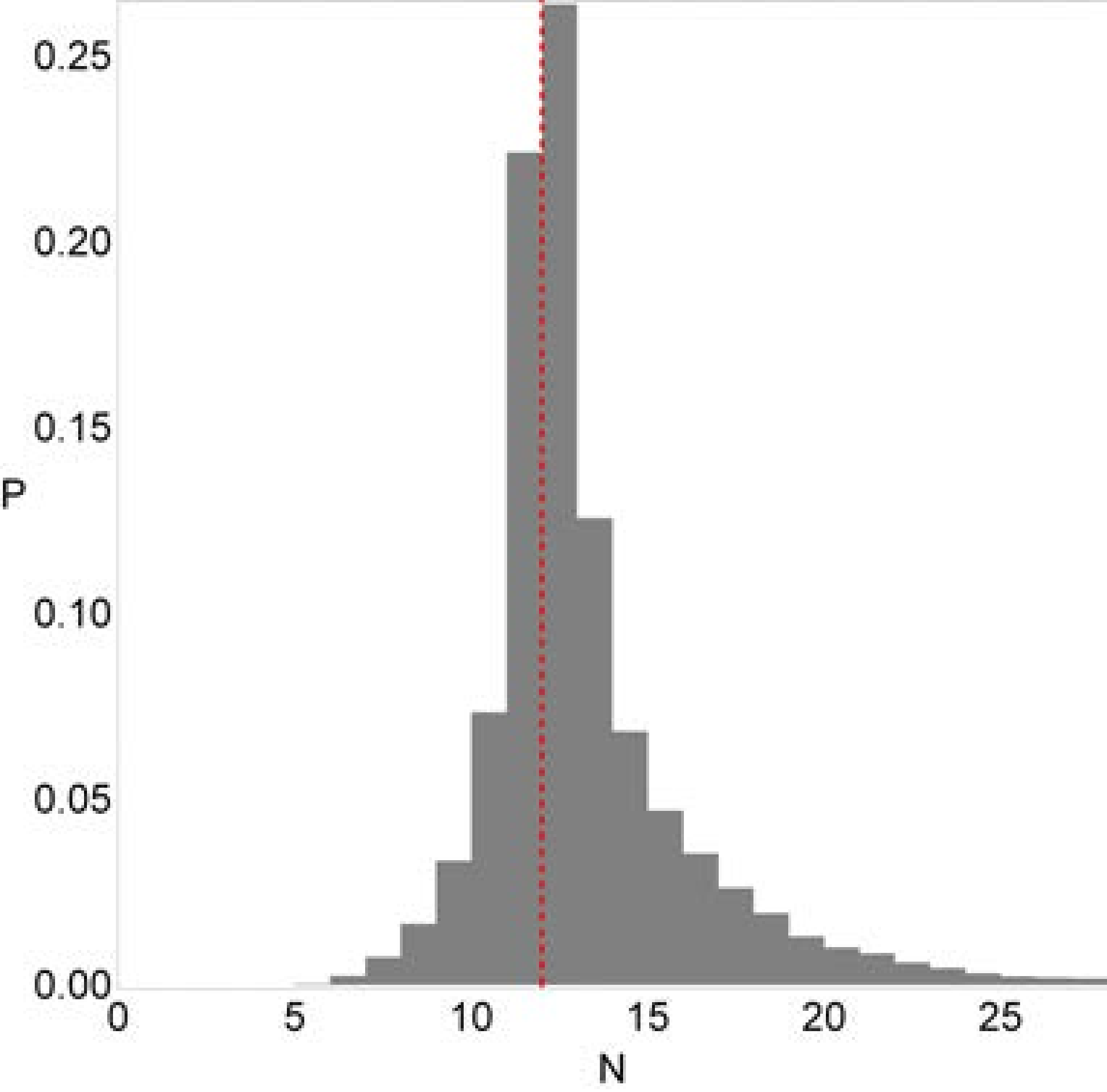}
(d)\includegraphics[scale=.4]{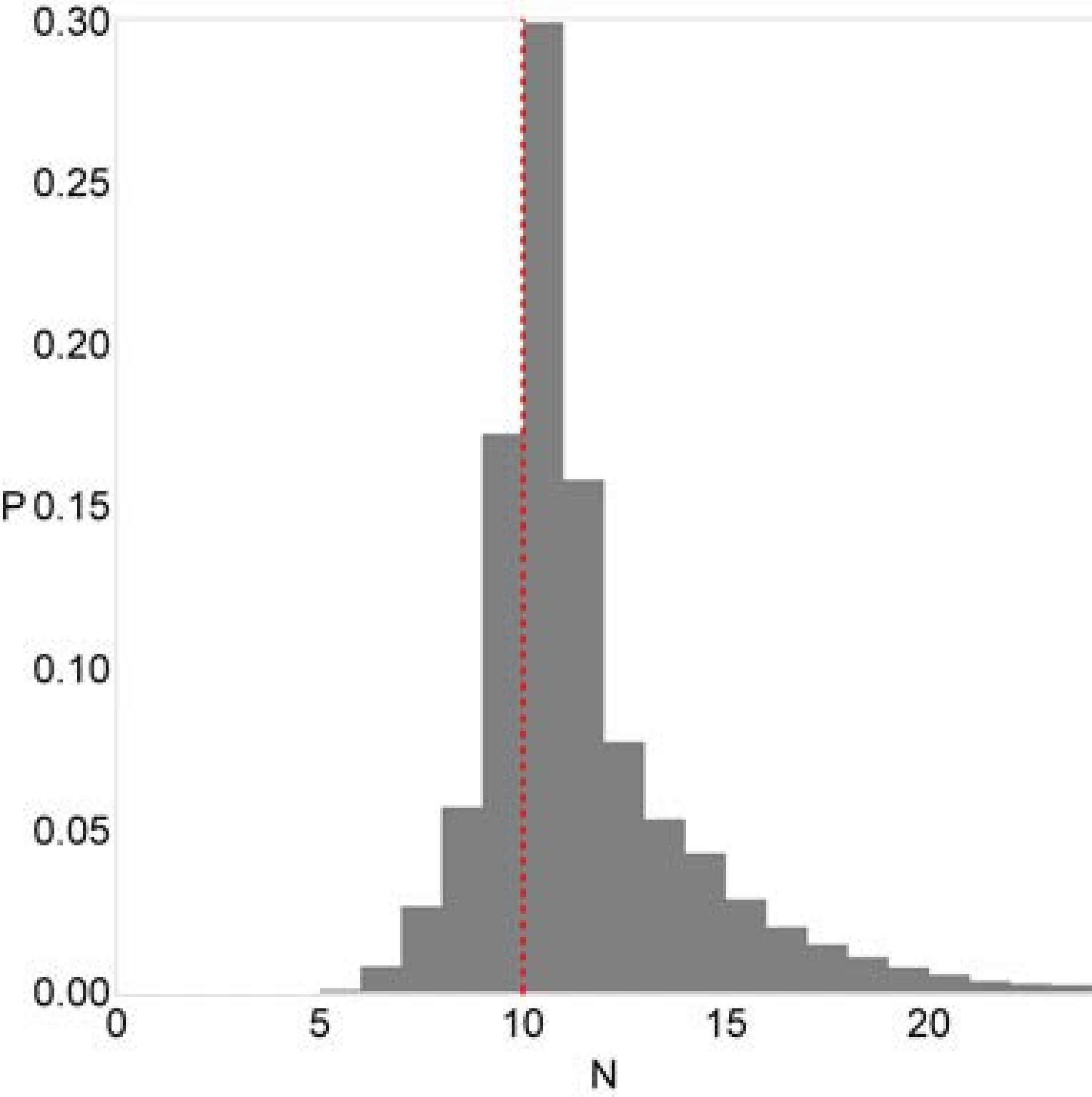}\\
(e)\includegraphics[scale=.4]{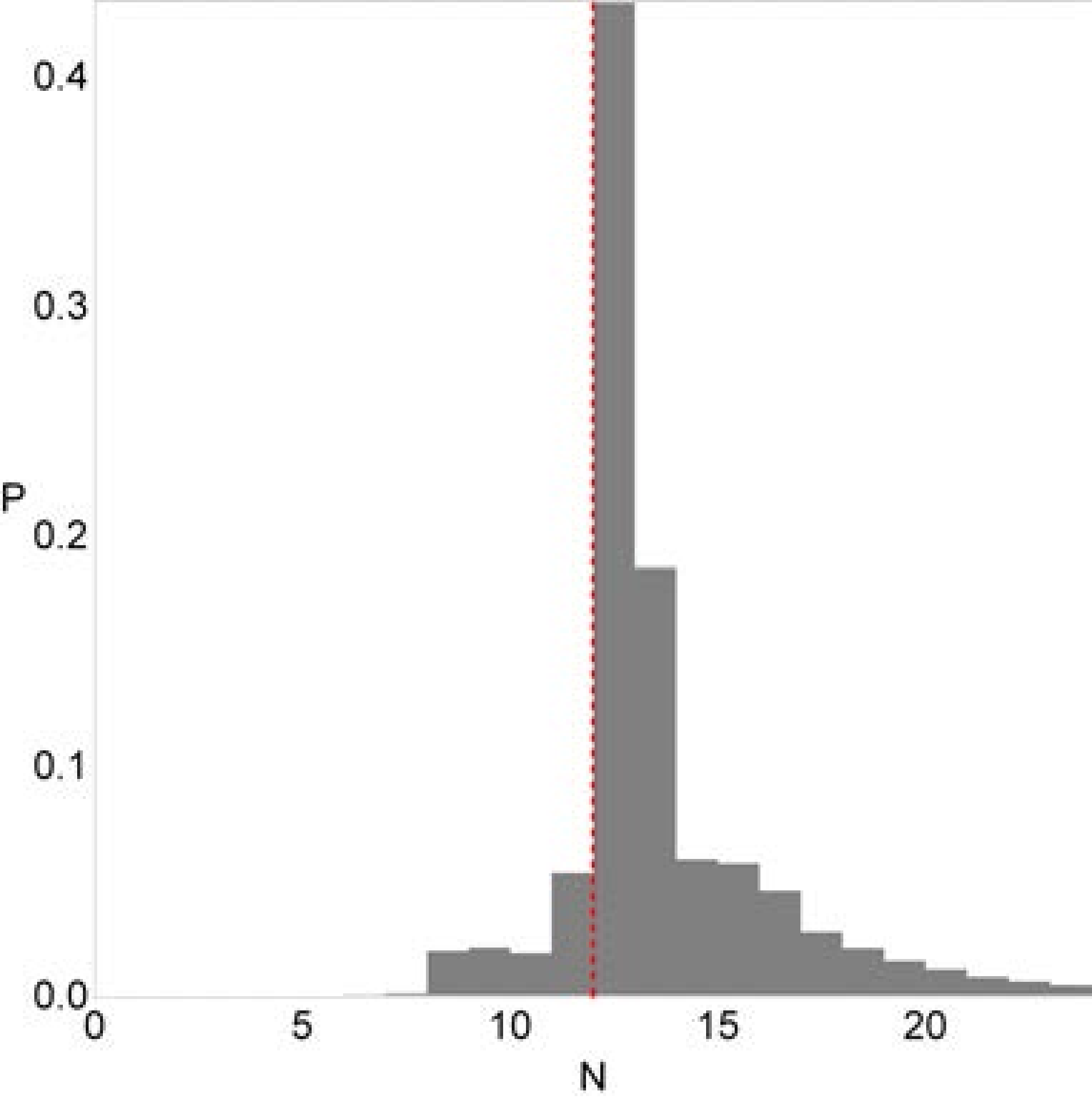}
(f)\includegraphics[scale=.4]{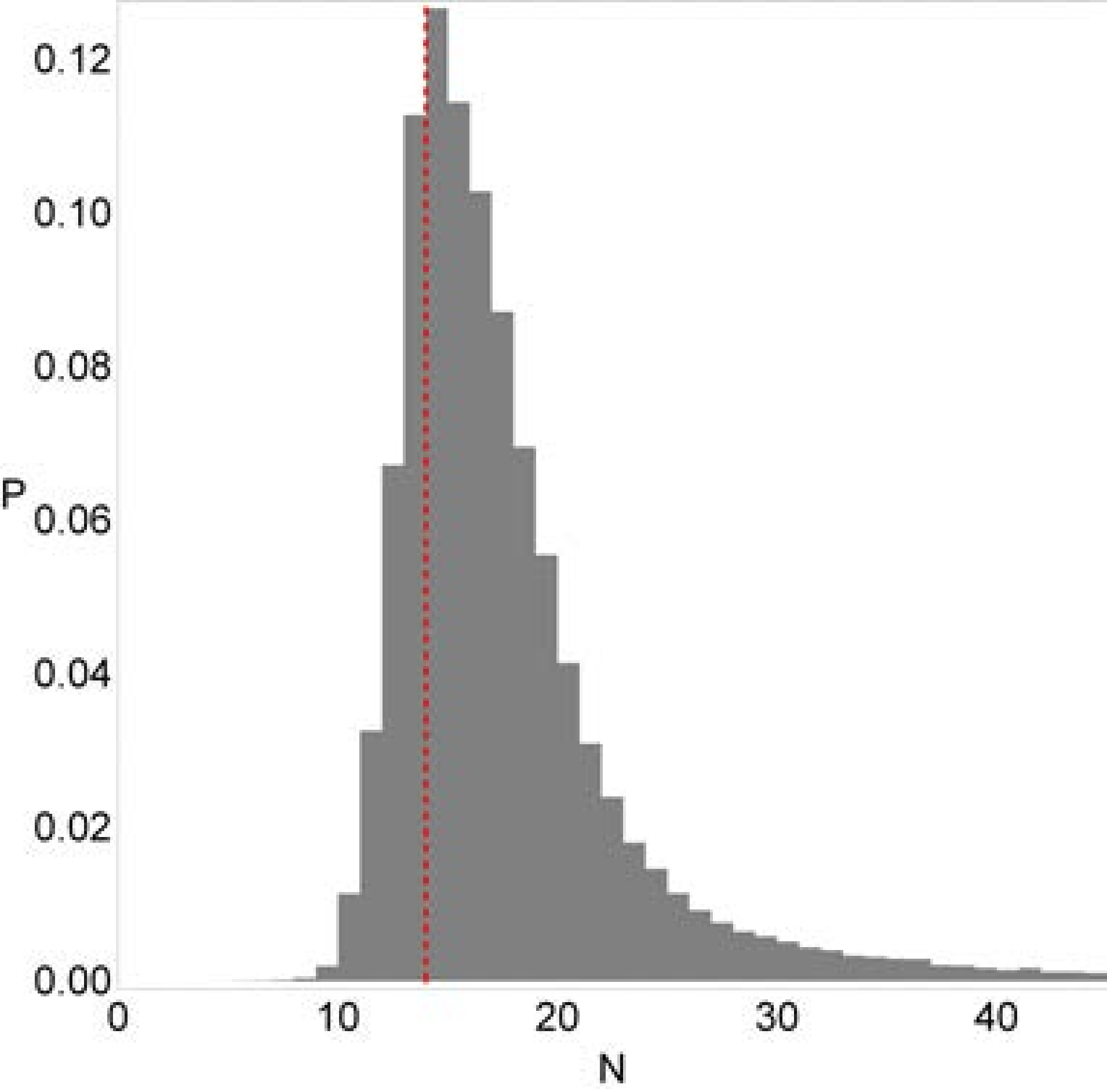}
\caption{The corresponding probability distributions of the required iterations for obtaining the Newton-Raphson basins of attraction shown in Fig. \ref{NR_Fig_5}. (Color figure online).}
\label{NR_Fig_5b}
\end{figure*}
\begin{figure*}[!t]
\centering
(a)\includegraphics[scale=5]{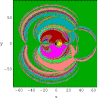}
(b)\includegraphics[scale=5]{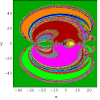}
(c)\includegraphics[scale=6]{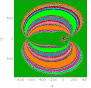}\\
(d)\includegraphics[scale=5]{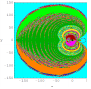}
(e)\includegraphics[scale=5]{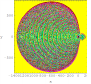}
(f)\includegraphics[scale=5]{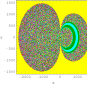}
\caption{The Newton-Raphson basins of attraction on the configuration $(x, y)$ plane for fixed value of $\alpha=59 \degree$ (a) $\beta=22\degree$, (b) $\beta=27 \degree$, (c) $\beta=36 \degree$, (d) $\beta=41 \degree$, (e) $\beta=45 \degree$, (f) $\beta=49 \degree$, when 13 libration points exist. The color is same as in Fig. \ref{NR_Fig_4}.  (Color figure online).}
\label{NR_Fig_6}
\end{figure*}
\begin{figure*}[!t]
\centering
(a)\includegraphics[scale=.35]{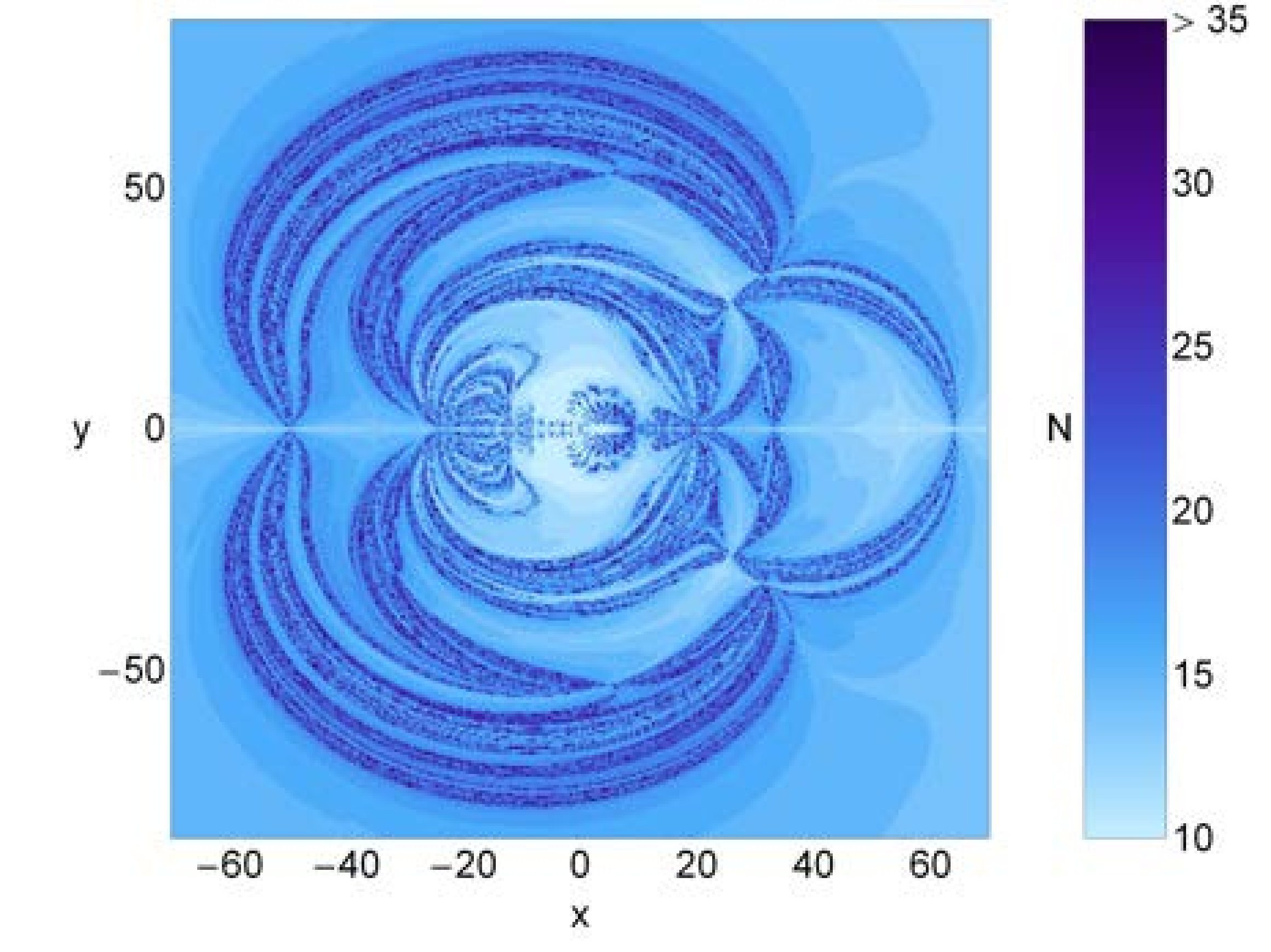}
(b)\includegraphics[scale=.35]{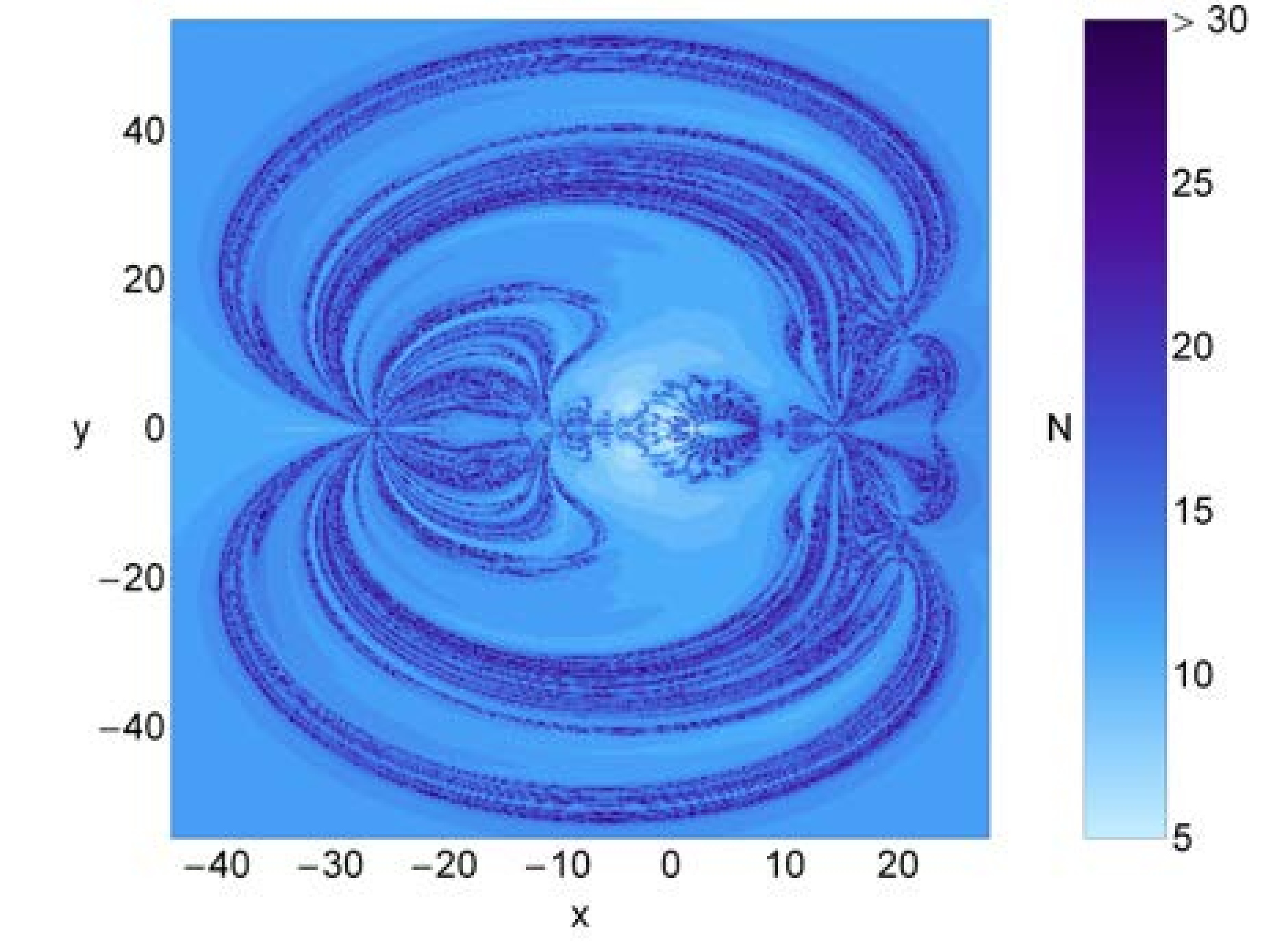}\\
(c)\includegraphics[scale=.35]{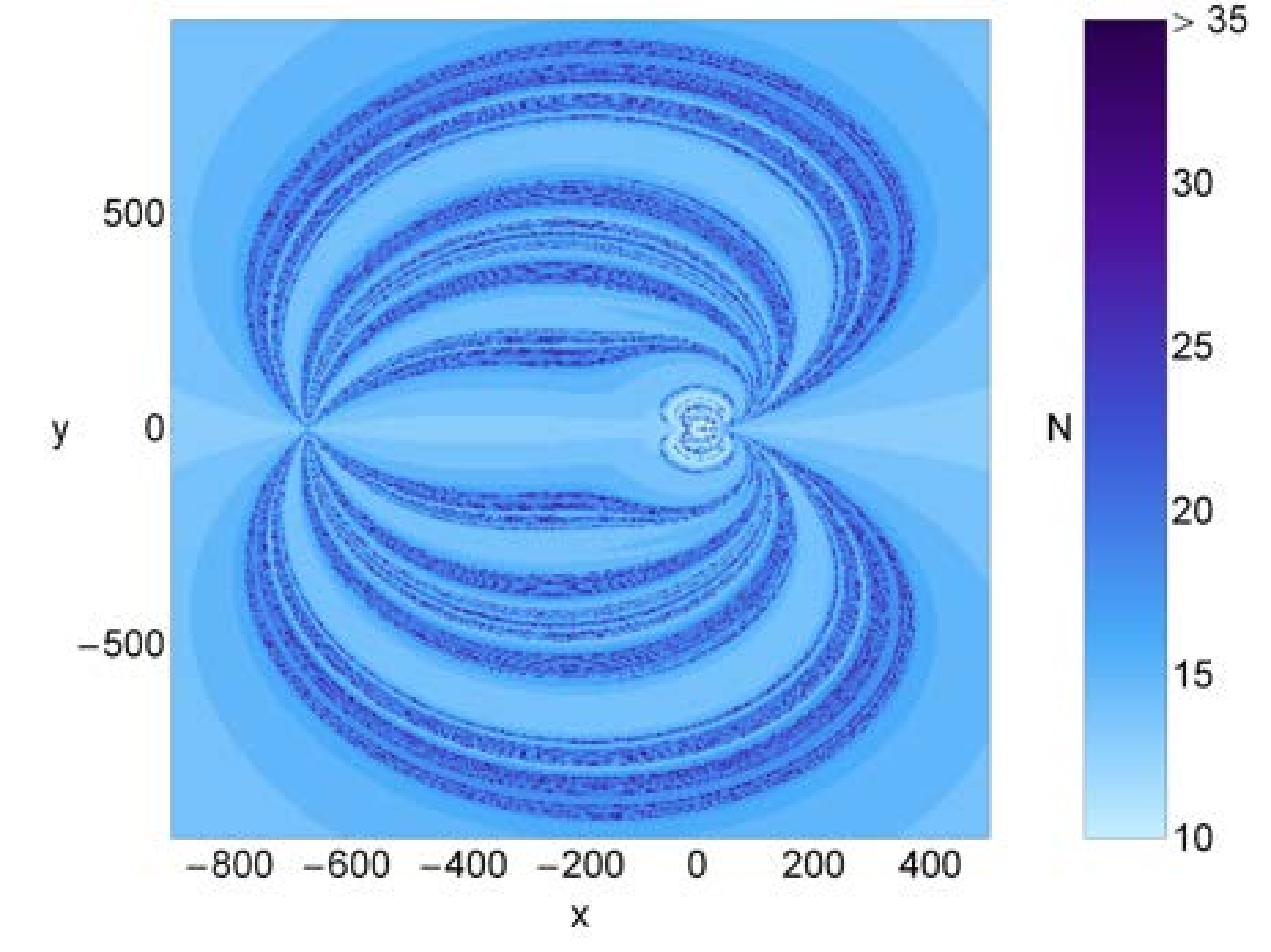}
(d)\includegraphics[scale=.35]{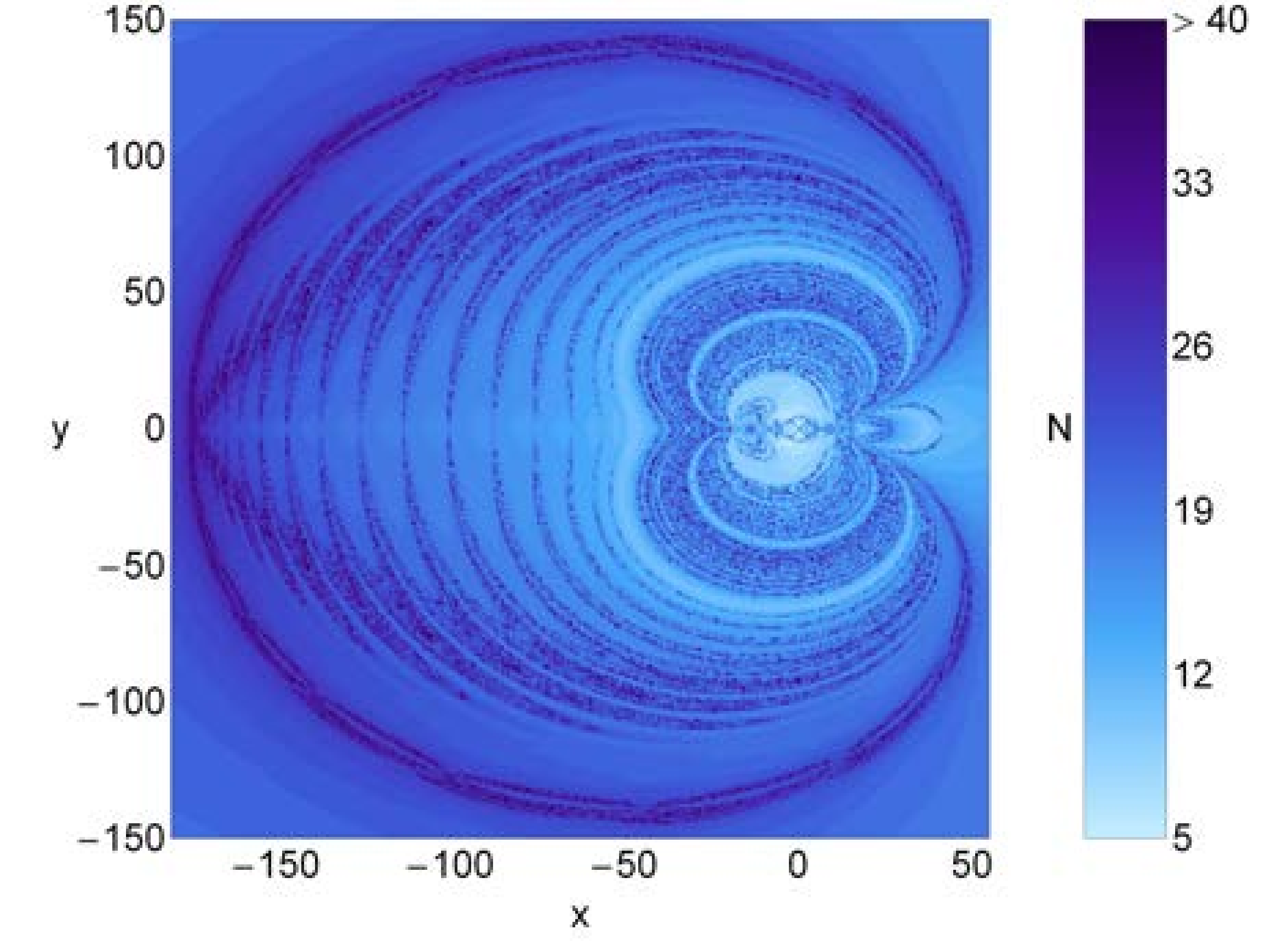}\\
(e)\includegraphics[scale=.35]{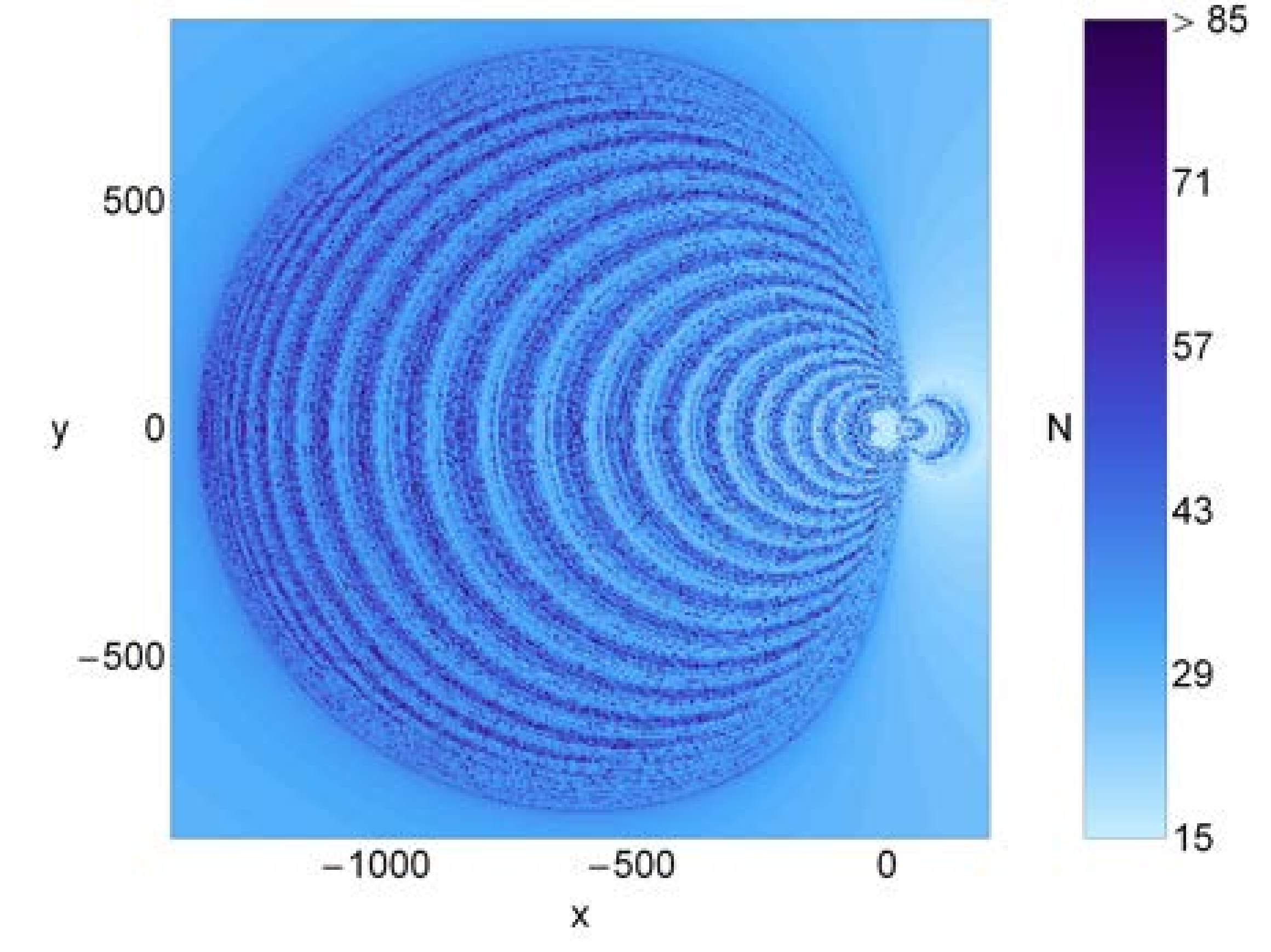}
(f)\includegraphics[scale=.35]{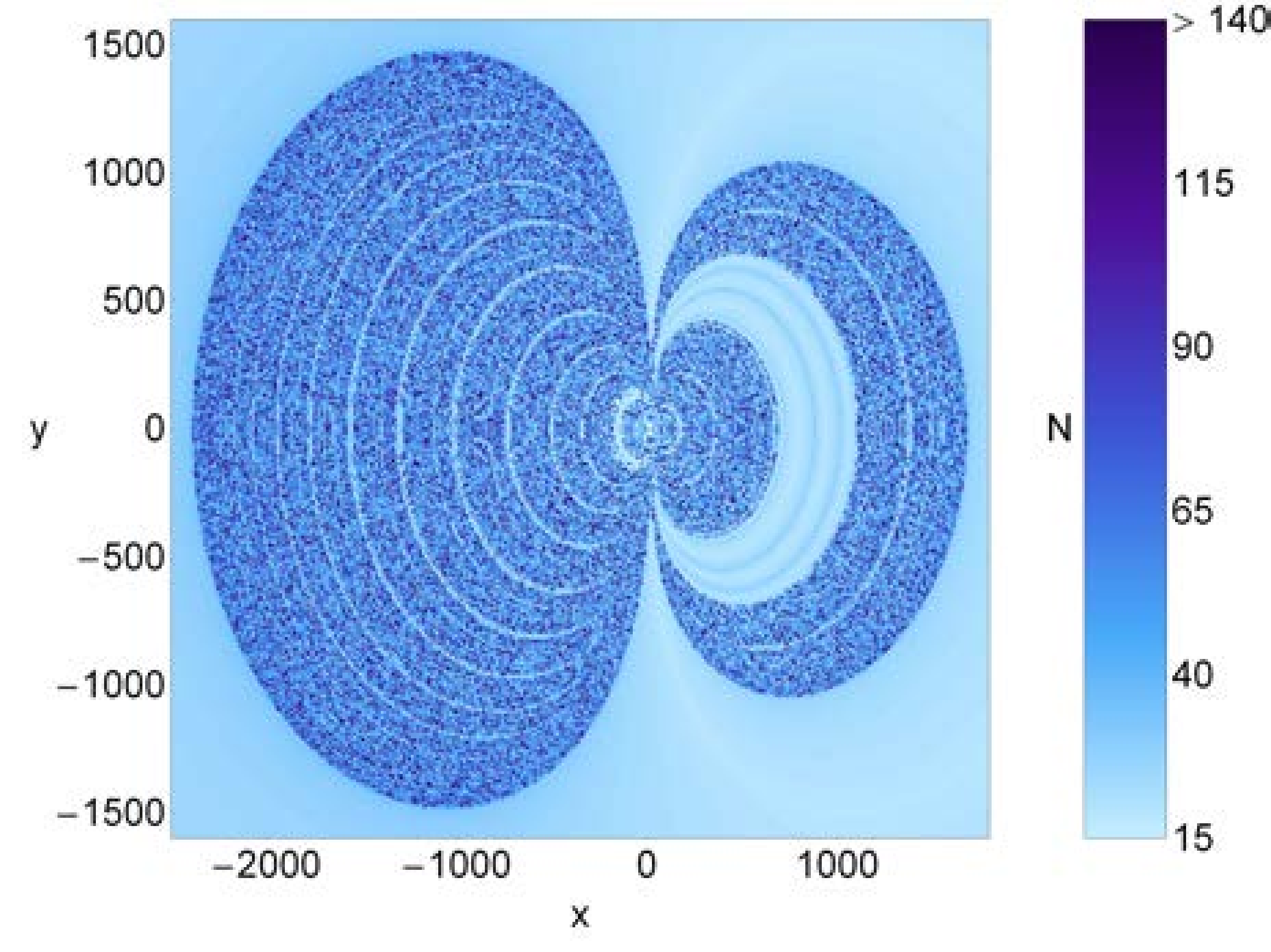}
\caption{The distribution of the corresponding number $N$ of the required iterations for obtaining the Newton-Raphson basins of attraction shown in Fig.\ref{NR_Fig_6}.  (Color figure online).}
\label{NR_Fig_6a}
\end{figure*}
\begin{figure*}[!t]
\centering
(a)\includegraphics[scale=.4]{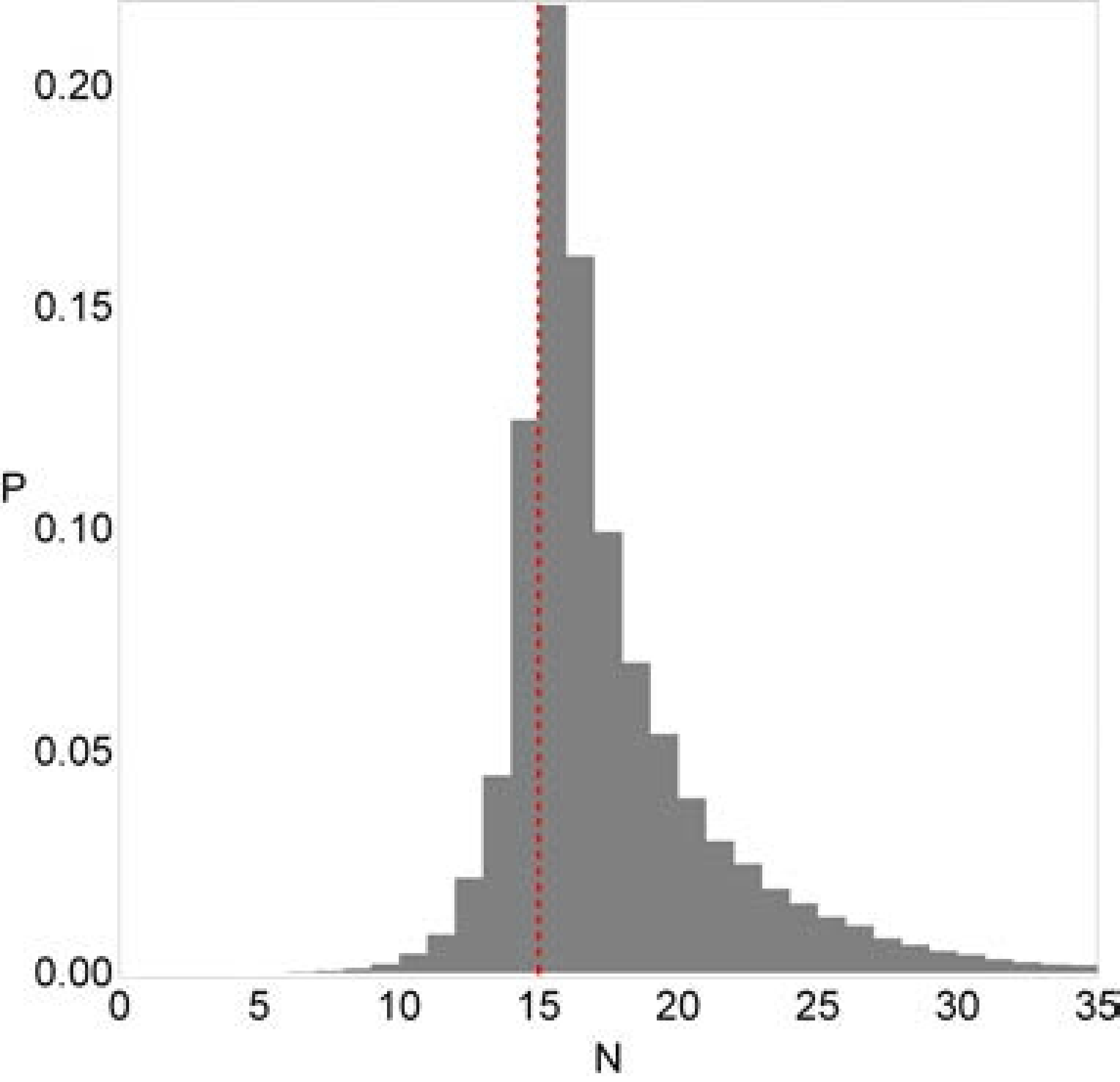}
(b)\includegraphics[scale=.4]{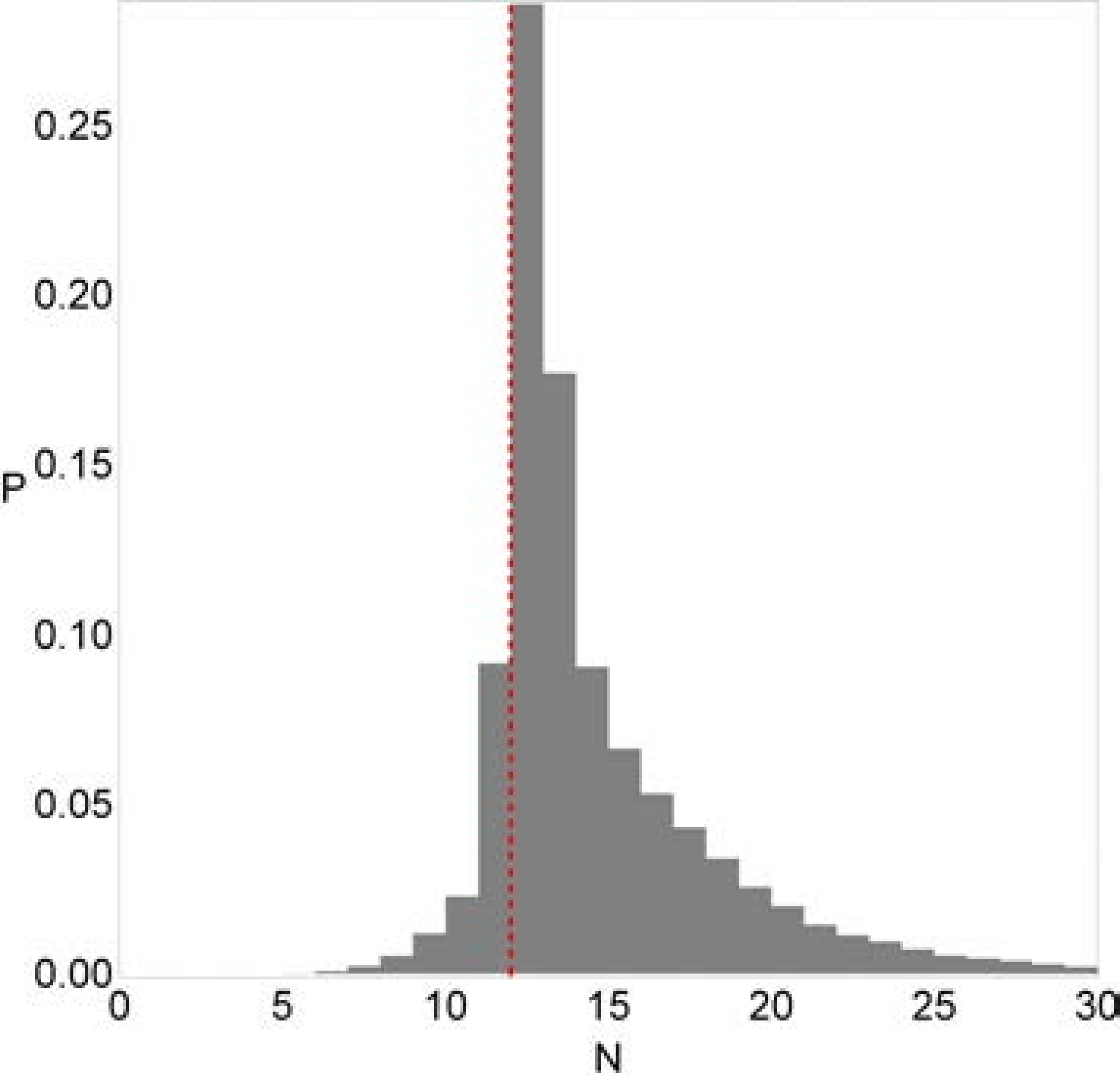}\\
(c)\includegraphics[scale=.4]{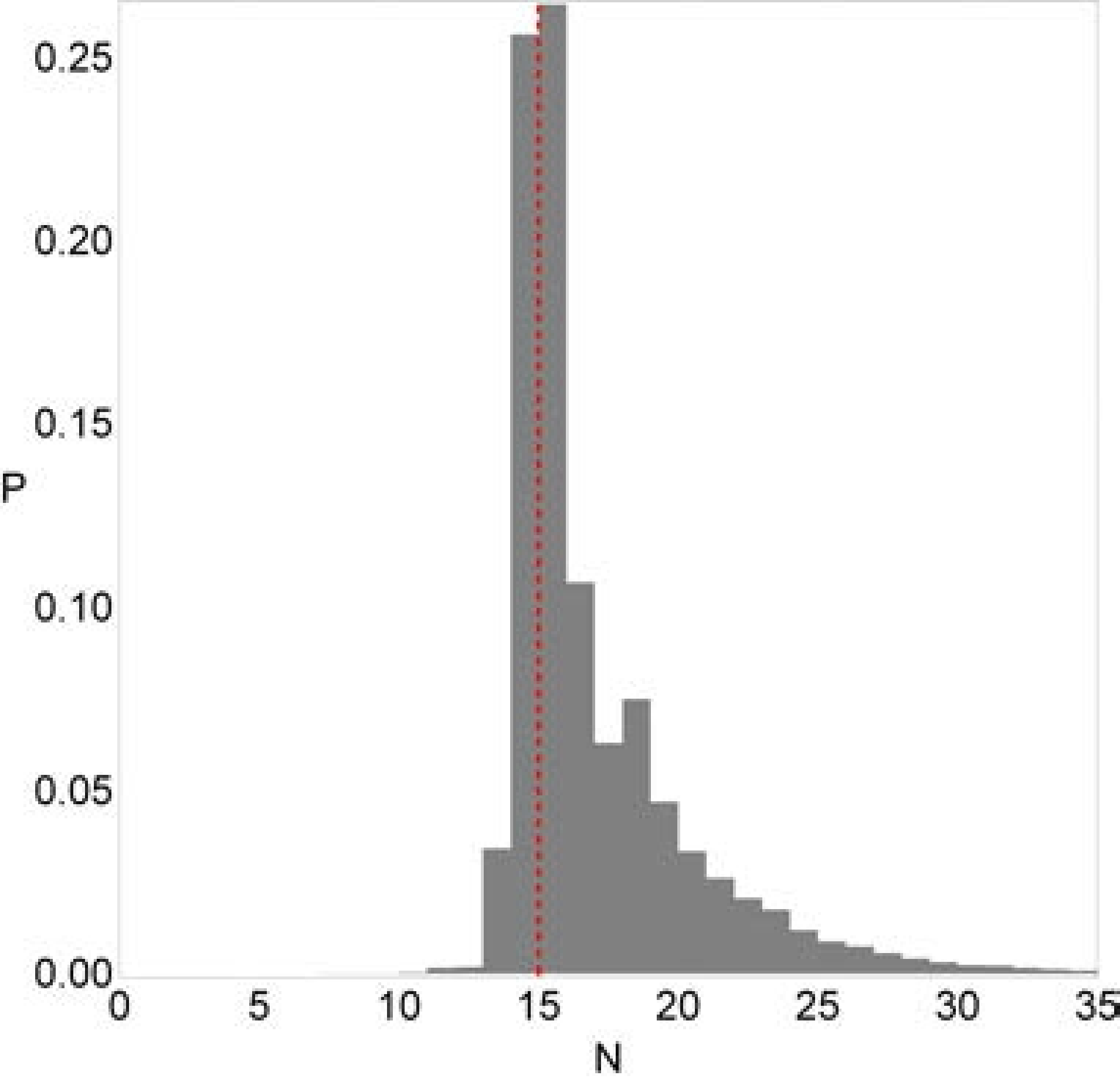}
(d)\includegraphics[scale=.4]{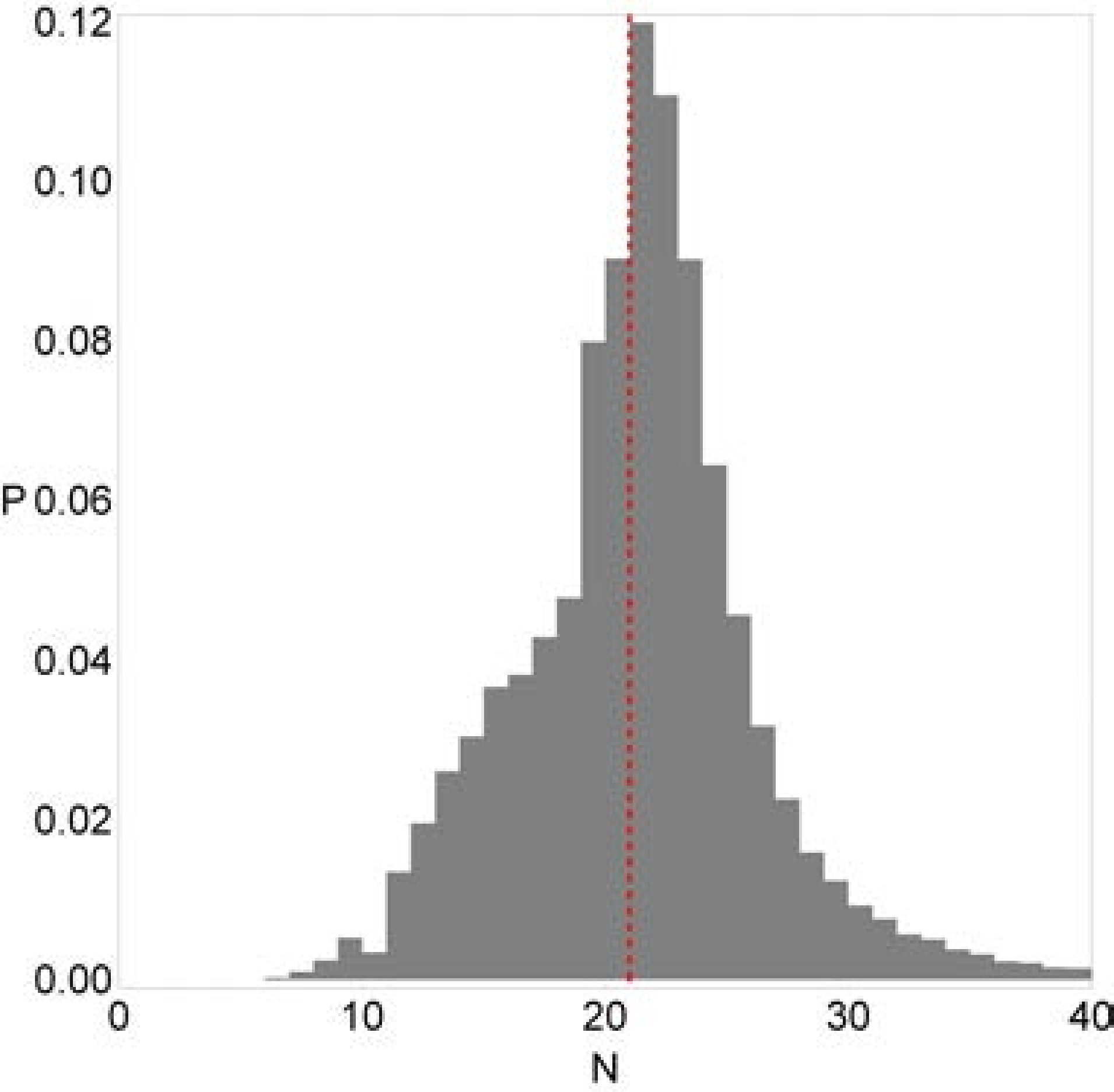}\\
(e)\includegraphics[scale=.4]{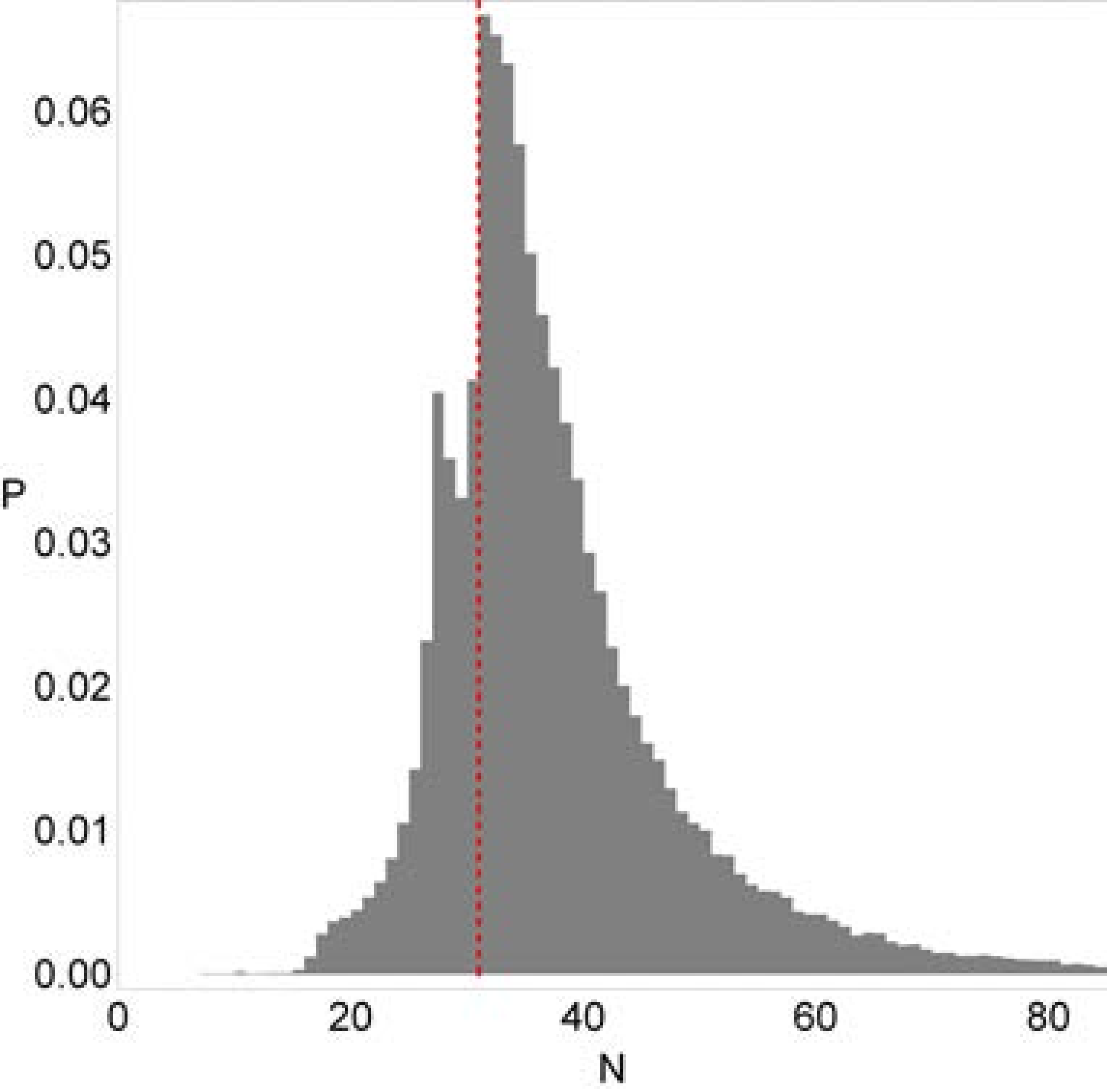}
(f)\includegraphics[scale=.4]{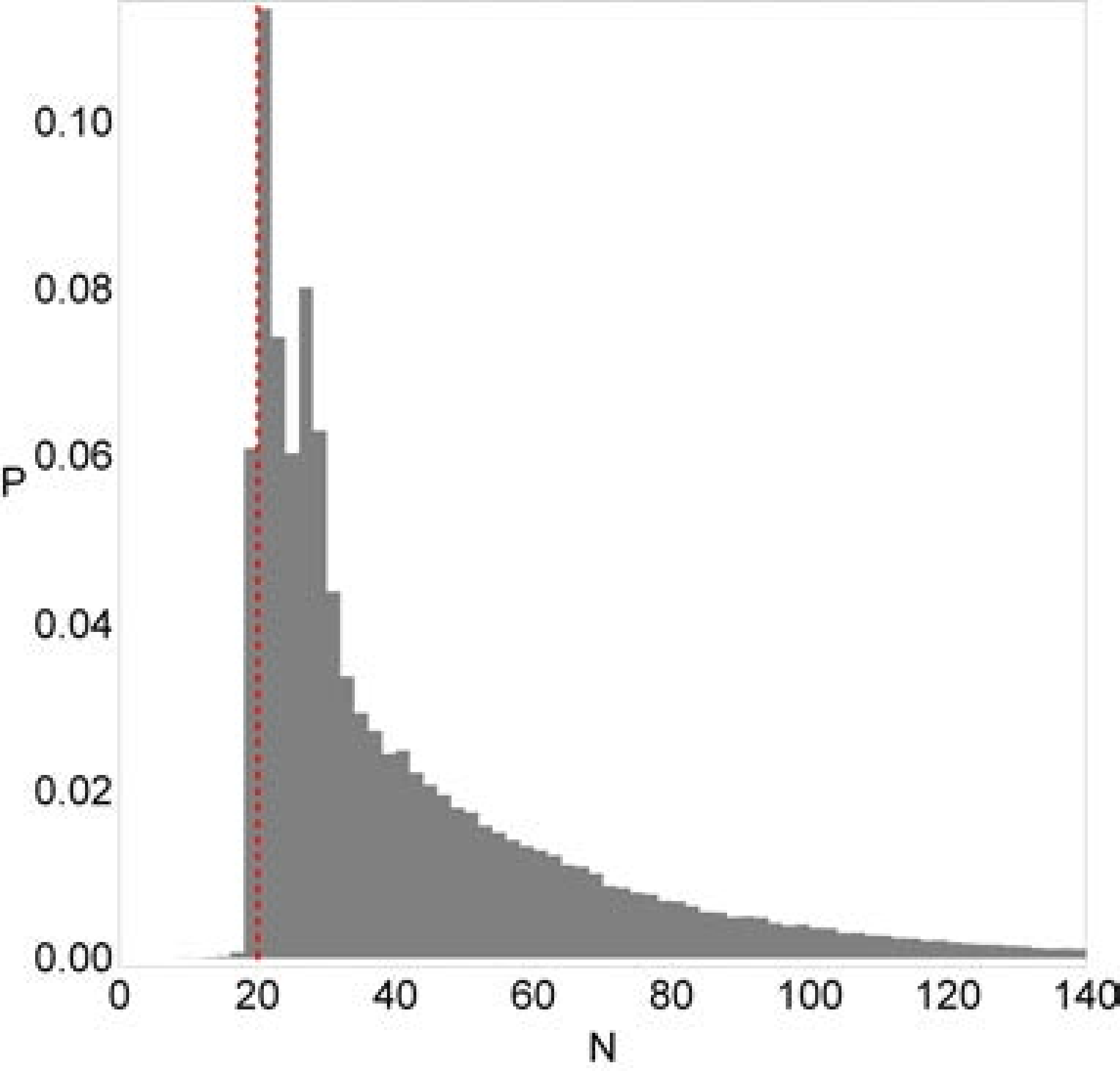}
\caption{The corresponding probability distributions of the required iterations for obtaining the Newton-Raphson basins of attraction shown in Fig. \ref{NR_Fig_6}. (Color figure online).}
\label{NR_Fig_6b}
\end{figure*}
\begin{figure*}[!t]
\centering
(a)\includegraphics[scale=2]{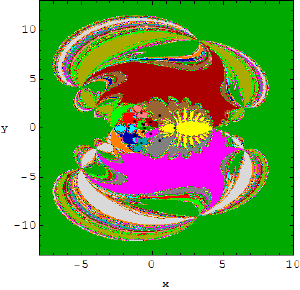}
(b)\includegraphics[scale=2]{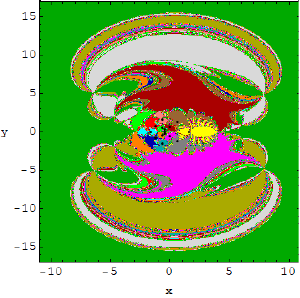}\\
(c)\includegraphics[scale=2]{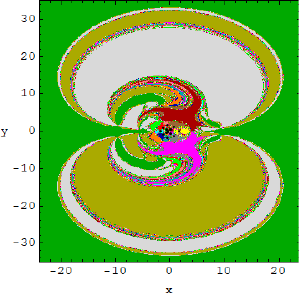}
(d)\includegraphics[scale=0.5]{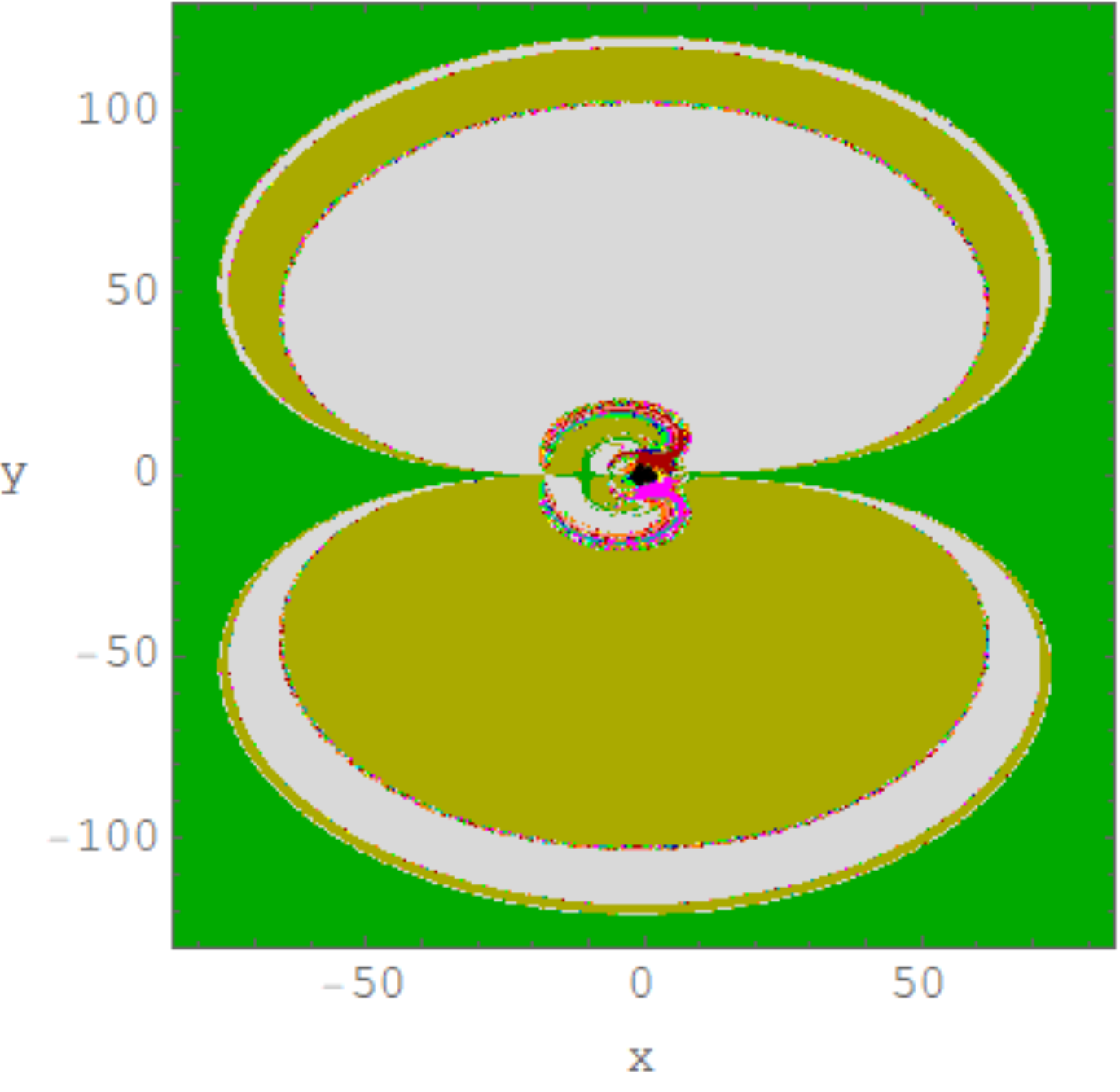}
\caption{The Newton-Raphson basins of attraction on the configuration $(x, y)$ plane for fixed value of  $\alpha=43.5 \degree$ (a) $\beta=30 \degree$, (b) $\beta=32 \degree$, (c) $\beta=34\degree$, (d)$\beta=34.458\degree$, when 15 libration points exist.  The color code for libration points $L_1,...,L_{13}$ is same as in Fig.\ref{NR_Fig_4} while $L_{14}$ (\emph{light grey}), $L_{15}$ (\emph{darker yellow}) and non-converging points (white). (Color figure online).}
\label{NR_Fig_7}
\end{figure*}
The algorithm we follow to determine the Newton-Raphson basins of convergence can be summarized
as: we classify dense, uniform grids of $1024 \times 1024$ initial conditions $(x_0, y_0)$ by applying
a double scan of the configuration $(x, y)$. During the numerical computations, we request an accuracy
of $10^{-15}$ (regarding the positions of the equilibrium points), while the maximum permissible number of iterations is set to $N_{\rm max} = 500$.

The following subsections reveal how the angle parameters influence the topology of the Newton-Raphson basins of convergence, associated with the libration points, in the axisymmetric five-body problem. We consider three cases, regarding the total number of the libration points which act as numerical attractors. In each of the cases the basins of convergence are plotted in different and suitable scale to have a better view of the topology of the basins of convergence.
\subsection{Case I: eleven libration points exist}
\label{Sec:4.1}
We begin with the first case where eleven libration points exist. It can be observed that for $\alpha= 43.5\degree, 55.5\degree,$ and $59\degree$ the range of the intervals of $\beta$ are $(23.25\degree, 29.685\degree]$, $(17.25\degree, 23.640\degree]$, and $(15.5\degree, 21.656\degree]\& [57.948\degree, 59\degree)$, respectively. In Fig. \ref{NR_Fig_1}, \ref{NR_Fig_2}, and \ref{NR_Fig_3}, we discuss the basins of convergence, when eleven libration points exist.

\subsubsection{$\alpha=43.5\degree$}
\label{Sec:4.1.1}
Our numerical analysis starts with the first subcase, i.e., the case when $\alpha=43.5\degree$, where three collinear and eight non-collinear libration points exist. In Fig.\ref{NR_Fig_1}(a-d), we depict the evolution of the Newton-Raphson basins of convergence for four different values of the angle parameter $\beta$ and for a fixed value of $\alpha$. It can be seen that the well formed basins of convergence is spread all over the configuration $(x, y)$ plane. The domain of the basin of convergence, associated with the central libration point $L_2$, has infinite area, whereas the domains of the convergence, associated with all the remaining libration points, are finite. In addition, the neighborhood of the basin boundaries (i.e., the various local areas) are composed by an extremely fractal mixture of initial conditions. It can be noted that the word "fractal" refers to the particular local region which displays a fractal-like geometry.

In fig. \ref{NR_Fig_1}(a-d), it can be observed that the domains of convergence, associated with the libration point $L_1$ (\emph{yellow}), look like exotic bugs with many legs and antennas whereas the domains of convergence, associated with $L_{4,5}$ (\emph{magenta, crimson}), look like butterfly wings. Moreover, the domains of convergence, associated with the libration points $L_{3, 8, 9}$ (\emph{cyan, green, orange}), look like an eight-shaped region, which remains almost unperturbed with the increase in angle parameter $\beta$. On the other hand, as the angle parameter $\beta$ increases, the outer boundaries of the basins of convergence become highly chaotic which leads to the fact that it is almost impossible to predict which initial conditions, falling inside these regions, will converge to which attractors.

In Fig. \ref{NR_Fig_1a}, the distribution of the associated number $N$ of iterations required to reach the desired accuracy are depicted, using tones of blue. We may note that the initial conditions falling inside the fractal regions are very slow converging nodes, while the initial conditions falling inside the attracting domains have relatively very fast rate of convergence. In panel (d) of Fig. \ref{NR_Fig_1a} an interesting phenomenon is observed. For the initial conditions inside the basins of attraction, the required number of iterations is very low but it increases very fast for those initial conditions which lie inside the chaotic region composed of initial conditions. The following Fig. \ref{NR_Fig_1b} illustrates the corresponding probability distribution of the iterations. In every panel, the histograms include almost $98\%$ of the corresponding distributions. The philosophy according which the probability $P$ works is as follows: if $N_0$ initial conditions $(x_0, y_0)$ on the configuration plane converge, after $N$ iterations, to one of the libration points, then $P = \frac{N_0}{N_t}$, where $N_t$ corresponds to total number of initial conditions in each color coded diagram. Moreover, the most probable number of iterations is not constant throughout; it is equal to 10 for all panels (b-d) while for panel-(a), the number of iterations is 15.

\subsubsection{$\alpha=55.5\degree$}
\label{Sec:4.1.2}
We continue our discussion with the case when $\alpha=55.5\degree$, when the range of the angle parameter $\beta\in(17.25\degree, 23.656\degree]$. In Fig. \ref{NR_Fig_2}, we present the Newton-Raphson basins of convergence with $\beta=20\degree, 21\degree, 22\degree, 23\degree$. One can easily observe a very interesting phenomenon associated with the extent of basins of convergence. More precisely, the extent of the basins of convergence associated with the central libration point $L_4$ is infinite where as the extent of the attracting domains of all the other libration points are always finite. As the angle parameter $\beta$ increases, unpredictable changes occur in the domains of the basins of convergence. In panel (a), the the extent of the basins of convergence associated with the libration points $L_6$ and $L_{7}$ are much higher than all the other basins except of course from the domain of the basins of convergence corresponding to the libration point $L_4$, which extends to infinity. Moreover, as we increase the value of the angle parameter $\beta$, the area of the basins of attraction of the libration points $L_6$ and $L_{7}$ slowly decrease and the remaining attractors arrogate the configuration space. In panel (c), when $\beta=22\degree$, the basins of convergence associated with libration points $L_{10}$ and $L_{11}$ are much higher than any other finite domain of convergence. On the other hand in panel (d), almost entire configuration plane which is covered by finite domains of the basins of convergence turn into a chaotic sea, composed  of mixtures of initial conditions.

In Fig. \ref{NR_Fig_2a}(a--d), we depict the distribution of the corresponding number $(N)$ of iterations required for obtaining the predefined accuracy where as the probability distribution of the iterations is illustrated in Fig. \ref{NR_Fig_2b}(a--d). We observe that in all the discussed cases, for more than $95\%$ of the initial conditions on the configuration $(x, y)$ plane, the iterative scheme needs not more than $35$ iterations to converge to one of the attractors, with predefined accuracy (see panels: \ref{NR_Fig_2a}a-c), while for panel (d) it increases slightly. In addition, the average value of the required number $(N)$ of iterations is not constant and it varies in every panels.

\subsubsection{$\alpha=59\degree$}
\label{Sec:4.1.3}

This subsection is devoted to the case for the values of angle parameters $\alpha=59\degree$ and $\beta \in(15.5\degree, 21.680\degree]$. Here, we can notice a tremendous change on the geometry of the basins of convergence with the change in the angle parameter $\beta$ (see Fig. \ref{NR_Fig_3}). In panel (a), where $\beta=16\degree$, most of the area of the finite regions of the basins of convergence is covered by the extent of the basins of convergence associated with the libration points $L_6$ and $L_7$. On the other hand, when $\beta=18\degree$, the area of the basins of attraction of the libration points $L_6$ and $L_7$ decreases drastically and other attractors take over the configuration space. Moreover, in both cases the extent of the basins of convergence associated with the central libration point $L_4$ is infinite, while the extent of the basins of convergence corresponding to other libration points are finite and well formed. In Fig.\ref{NR_Fig_3} (c-d), we have analyzed the distributions of the corresponding number $(N)$ of iterations required to obtain the predefined accuracy, while the probability distributions of iterations are depicted in Fig.\ref{NR_Fig_3} (e-f). It can be easily observed that for more than $95\%$ of the initial conditions on the configuration $(x, y)$ plane, the iterative scheme needs at least 35 iterations to obtain the coveted accuracy. Moreover, the average value of required number $(N)$ of iterations for panel-e is 20 while for panel-f is 16 which shows that it is not constant throughout.
\subsection{Case II: thirteen libration points exist}
\label{Sec:4.2}
We continue our study with the case when thirteen libration points exist in which three are collinear, while ten are non-collinear. The Newton-Raphson basins of attraction for three different values of the angle parameter $\alpha$ are presented in three different subcases.

\subsubsection{$\alpha=43.5\degree$}
\label{Sec:4.2.1}
In this case for $\alpha=43.5\degree$, the basins of convergence are illustrated in Fig. \ref{NR_Fig_4} for four values of $\beta$. We observe that shape as well as the geometry of the basins of convergence change drastically even with a slight change on $\beta$. Indeed, the extent of the basins of convergence associated with the central libration point $L_2$ is infinite, while for all the remaining libration points their domains of convergence are finite. As the value of the angle parameter $\beta$ increases, the following phenomenon take place on the configuration $(x, y)$ plane.
\begin{description}
  \item[-] The domain of the basins of convergence associated with the collinear libration points $L_{1, 3}$ and non-collinear libration points $L_{8, 9}$ look like exotic bugs with many legs and antennas.
  \item[-] Two antenna shaped regions exist in the left and right sides of the well shaped basins of convergence and they are composed by a mixture of initial conditions which converge to non-collinear libration points only. When $\beta$ increases the antenna shaped regions exist in left side increase, while in right side  they are decreasing (see Fig. \ref{NR_Fig_4}a-c). In panel (d), it can be observed that the antenna shaped region in the left and the right side are now equal and look like heart shaped region.
  \item[-] In Fig.\ref{NR_Fig_4}a-c, the geometry of the basins of convergence is symmetrical about $x-$axis while in Fig.\ref{NR_Fig_4}d, the topology of the basins of convergence looks symmetrical about both the axes.
  \item[-]There exists no evidence in numerical calculations of non-converging initial conditions, whatsoever.
\end{description}

The associated number $(N)$ of the required iterations to obtain the predefined accuracy is depicted in Fig. \ref{NR_Fig_4a}a--d, where as the probability distributions of the iterations are illustrated in Fig. \ref{NR_Fig_4b}a--d. We can observe that more than $95\%$ of the initial conditions on the configuration plane converges to one the attractors within the first 20 iterations. The initial conditions falling inside the region where the basins boundaries separated need much iterations, in comparison to those initial conditions which fall inside the regular domain of convergence.

\subsubsection{$\alpha=55.5\degree$}
\label{Sec:4.2.2}
In Fig. \ref{NR_Fig_5}a-f, we provide the Newton-Raphson basins of convergence corresponding to the libration points for $\alpha=55.5\degree$ and six different values of angle parameter $\beta$. The color-coded convergence diagrams presented in Fig. \ref{NR_Fig_5}a-f have substantial influence due to  change in the angle parameter $\beta$. The most notable changes can be summarized as follows:
\begin{figure*}[!t]
\centering
(a)\includegraphics[scale=.35]{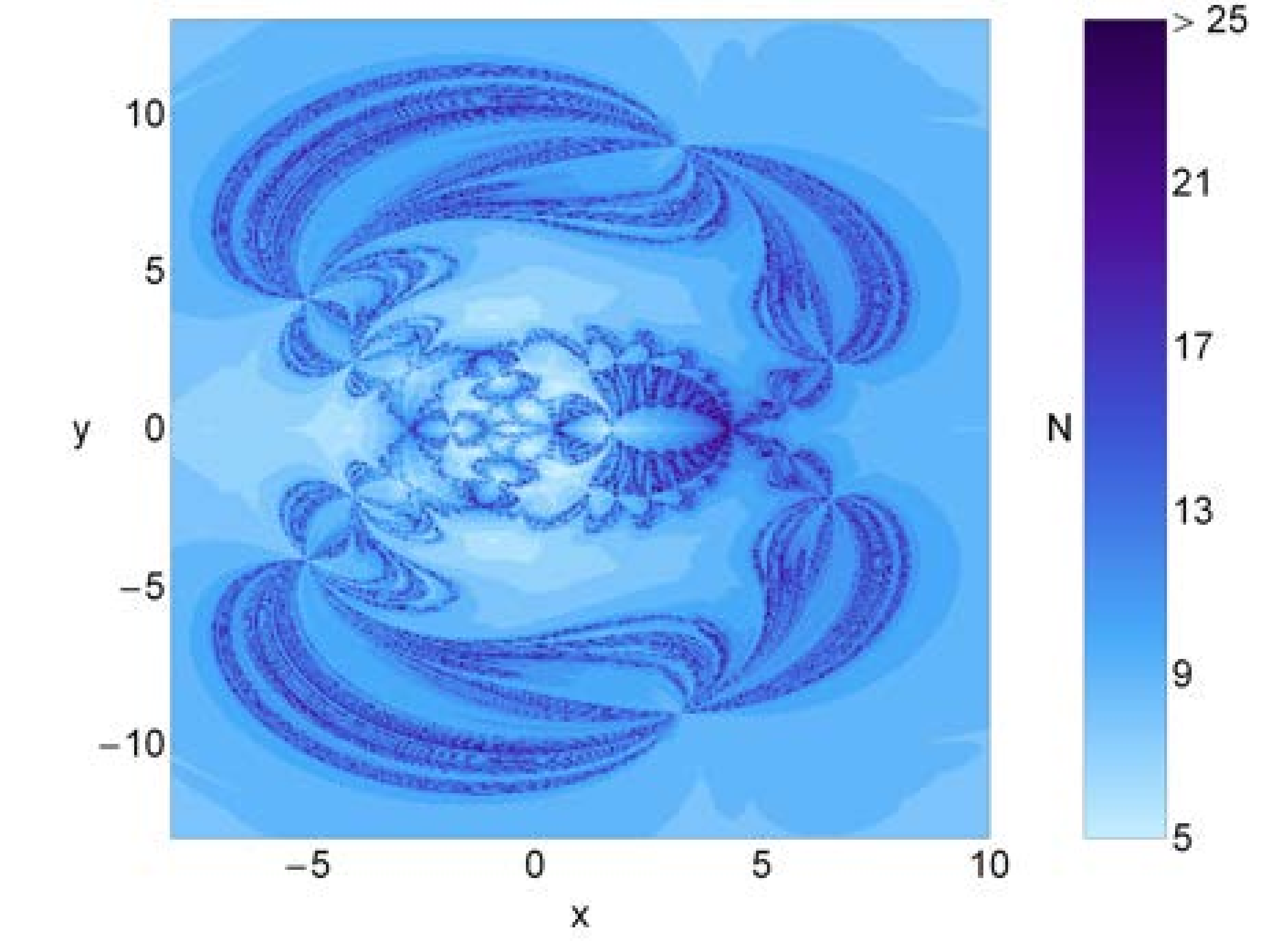}
(b)\includegraphics[scale=.35]{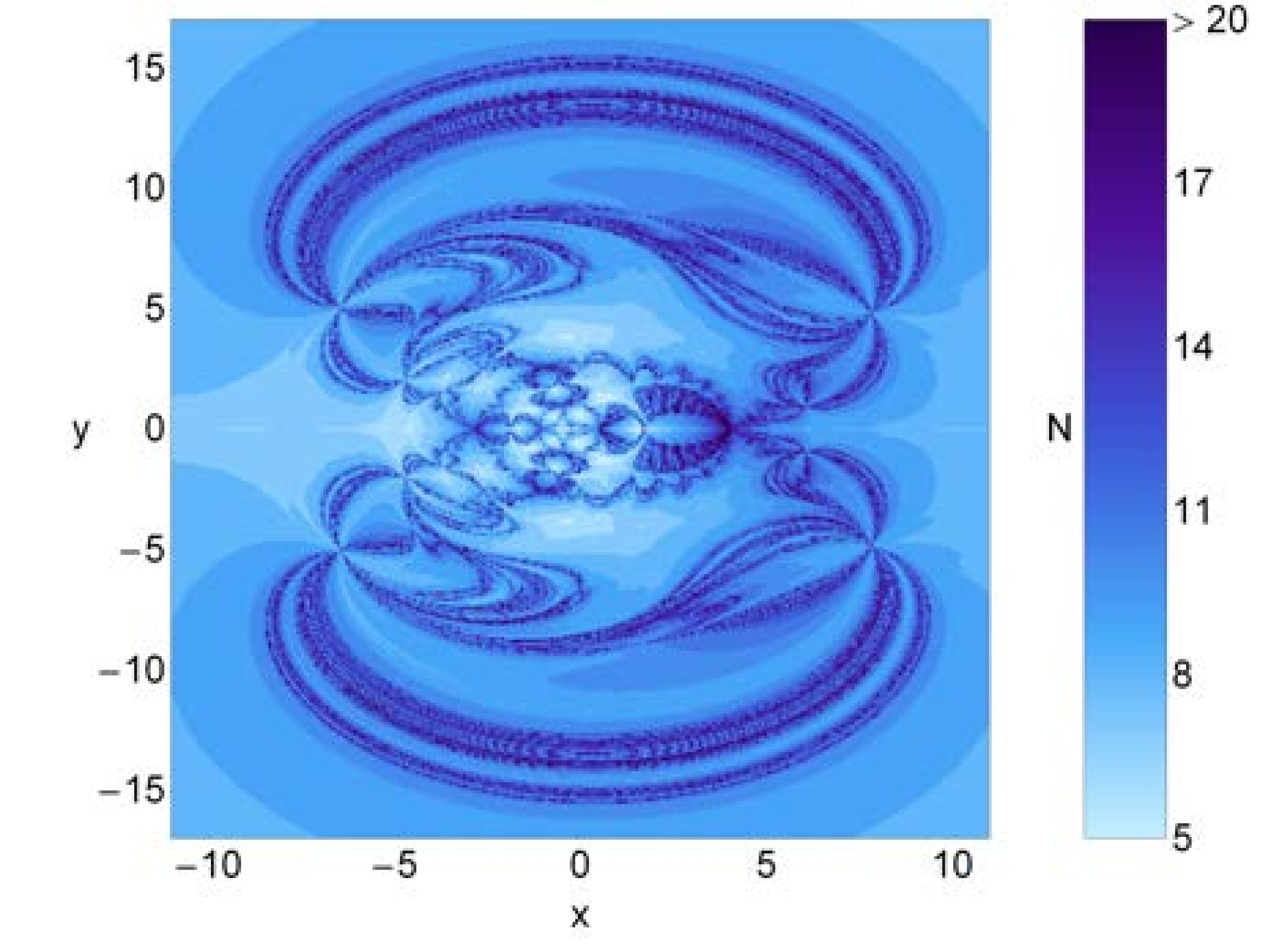}\\
(c)\includegraphics[scale=.35]{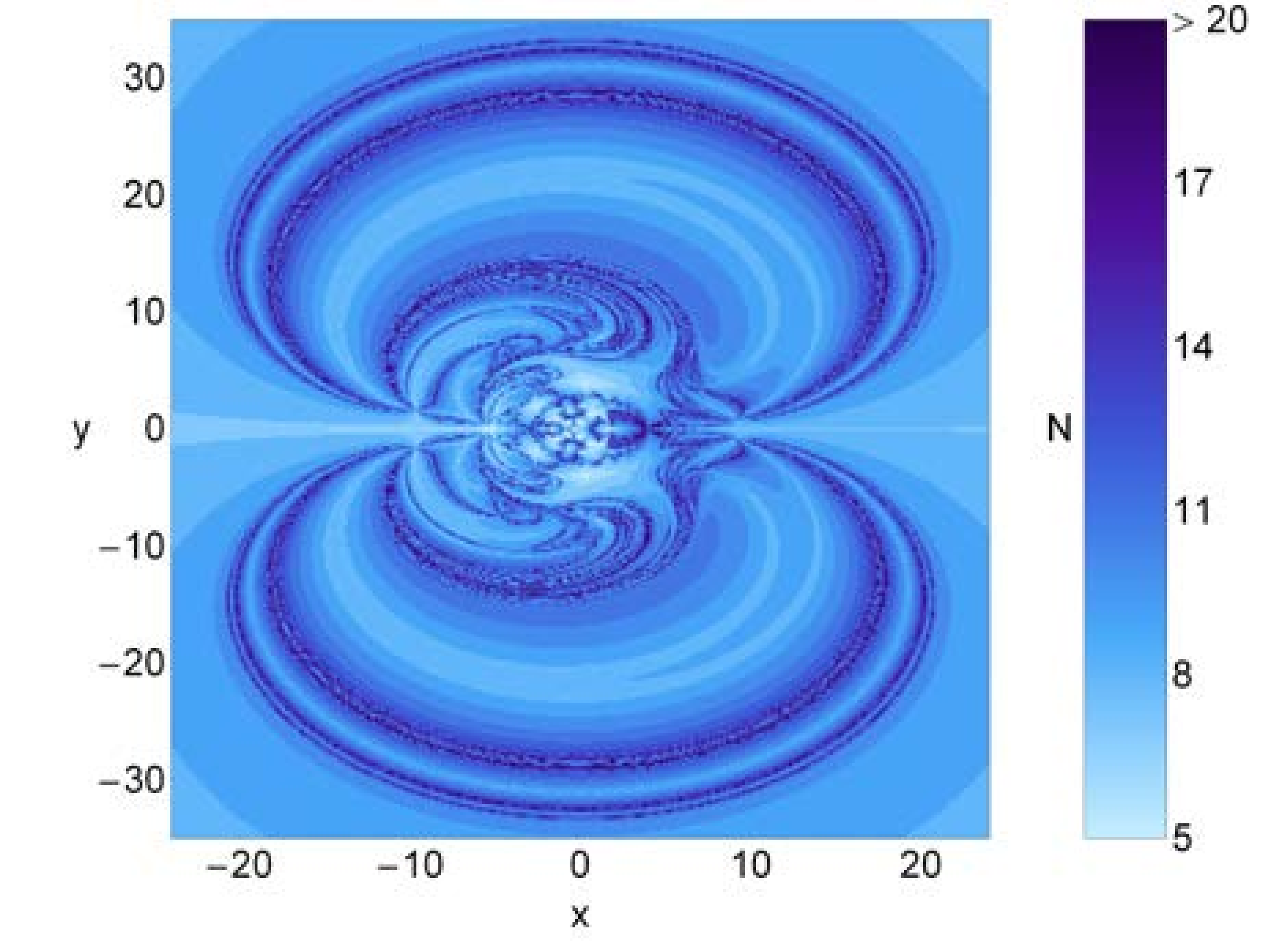}
(d)\includegraphics[scale=.35]{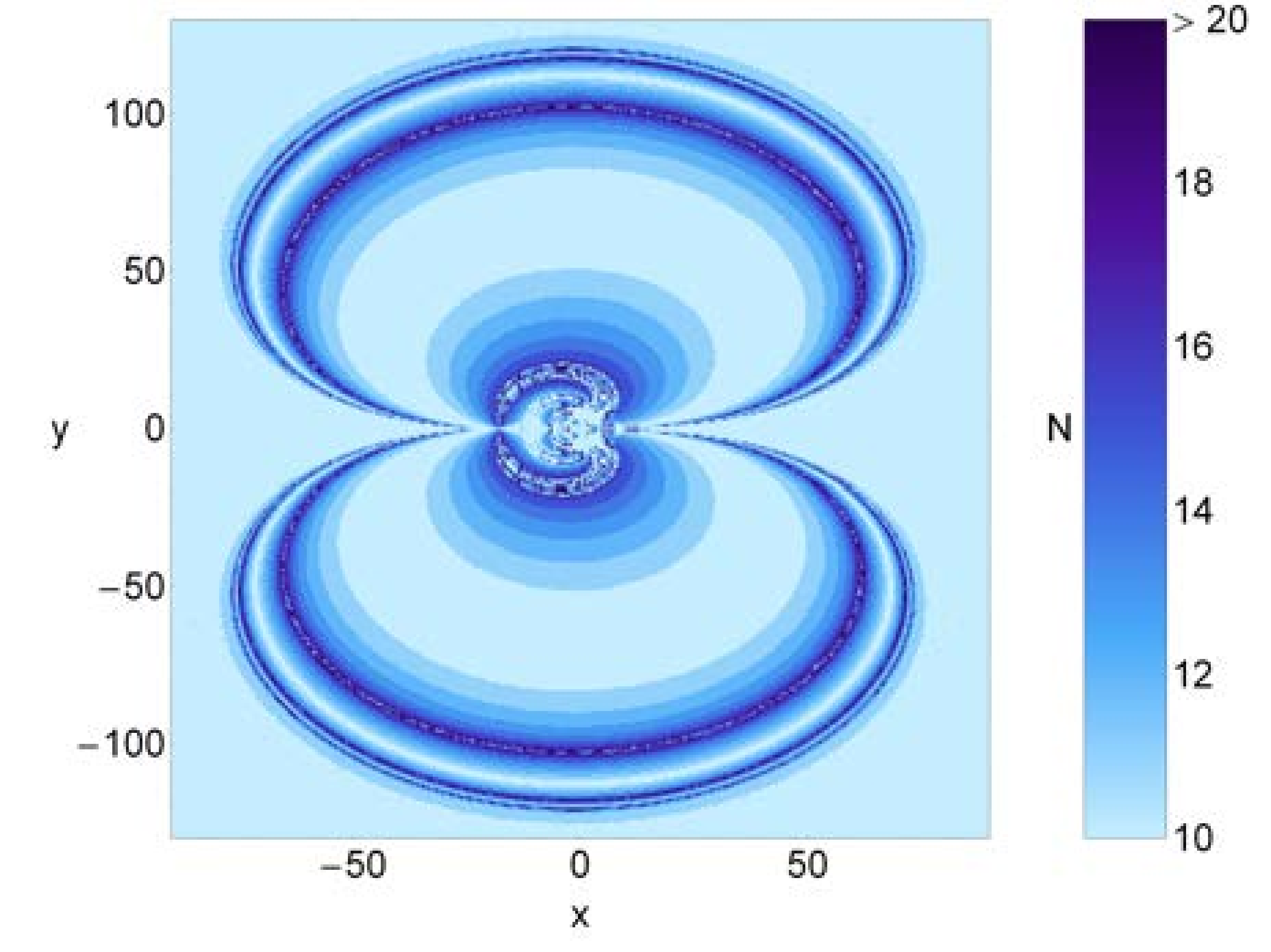}
\caption{The distribution of the corresponding number $N$ of the required iterations for obtaining the Newton-Raphson basins of attraction shown in Fig.\ref{NR_Fig_7}.  (Color figure online).}
\label{NR_Fig_7a}
\end{figure*}
\begin{figure*}[!t]
\centering
(a)\includegraphics[scale=.4]{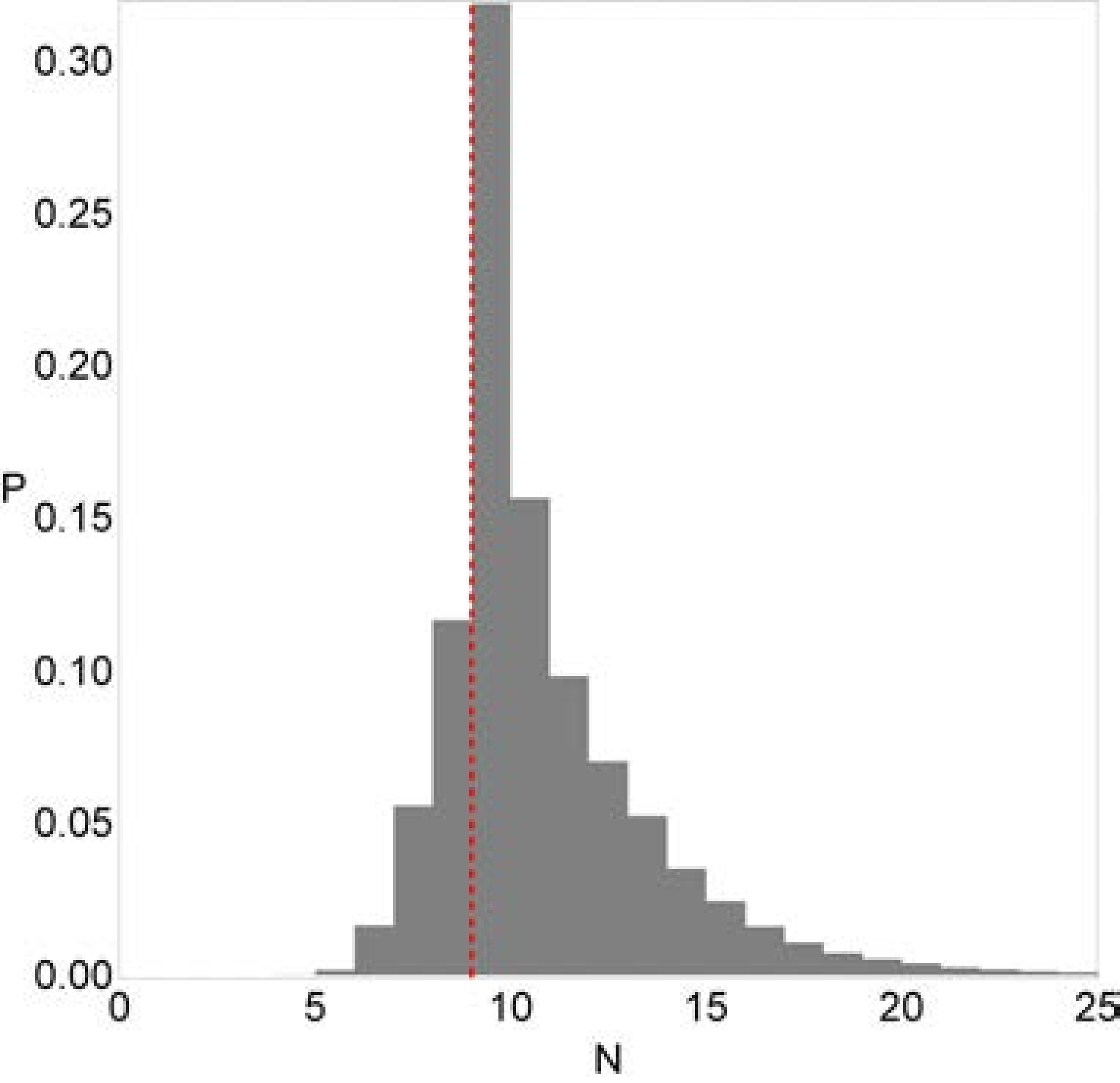}
(b)\includegraphics[scale=.4]{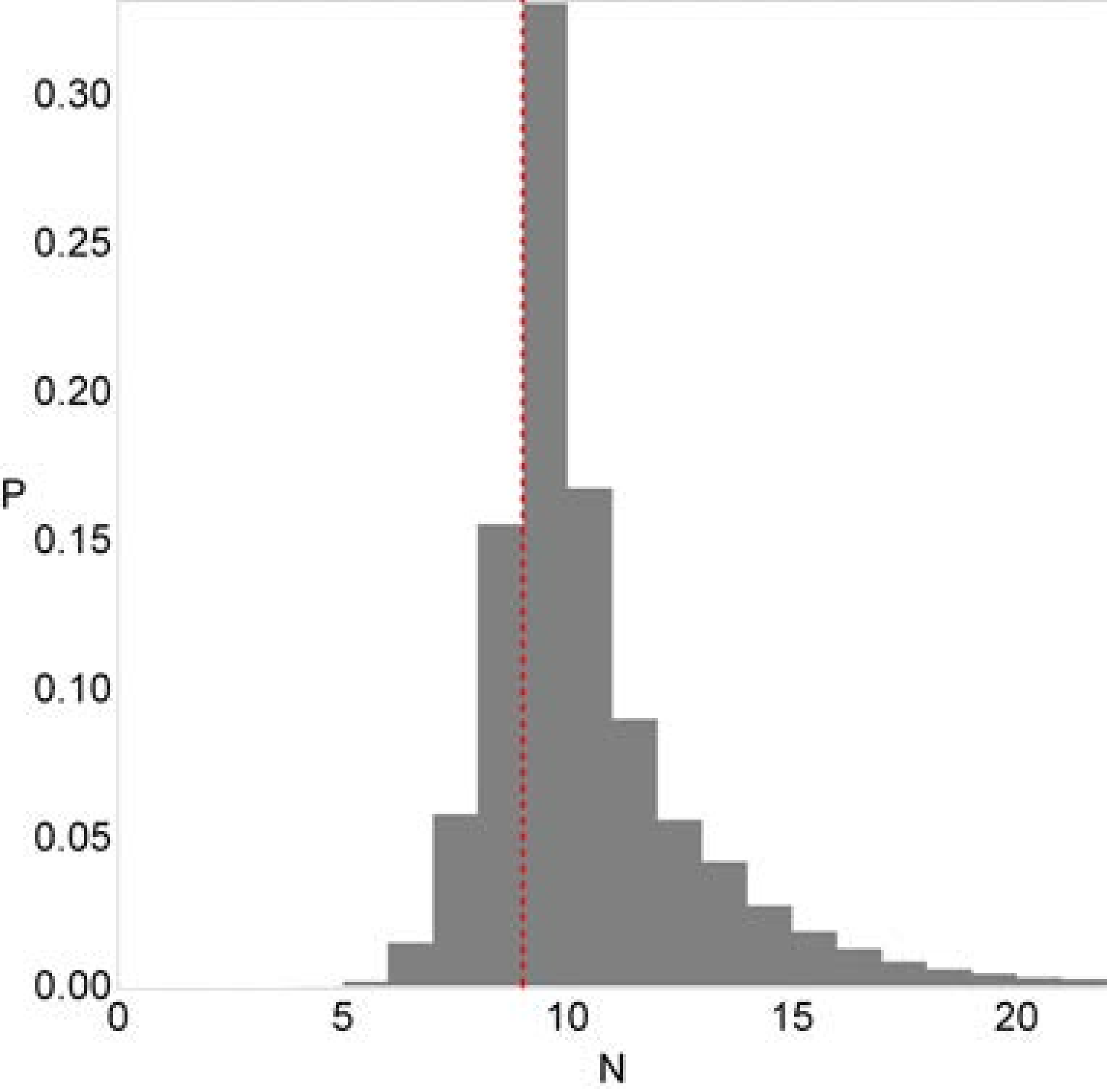}\\
(c)\includegraphics[scale=.4]{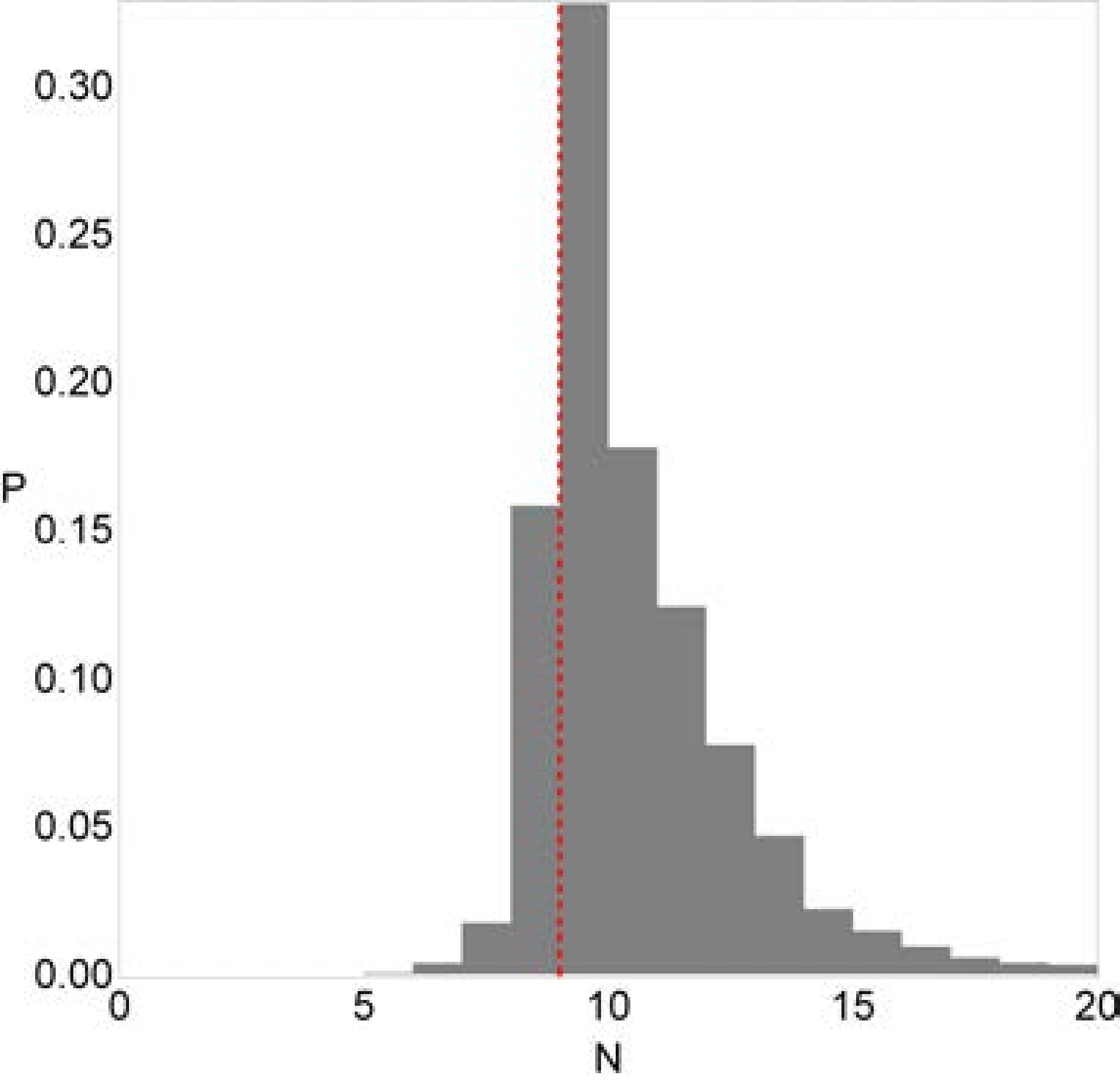}
(d)\includegraphics[scale=.4]{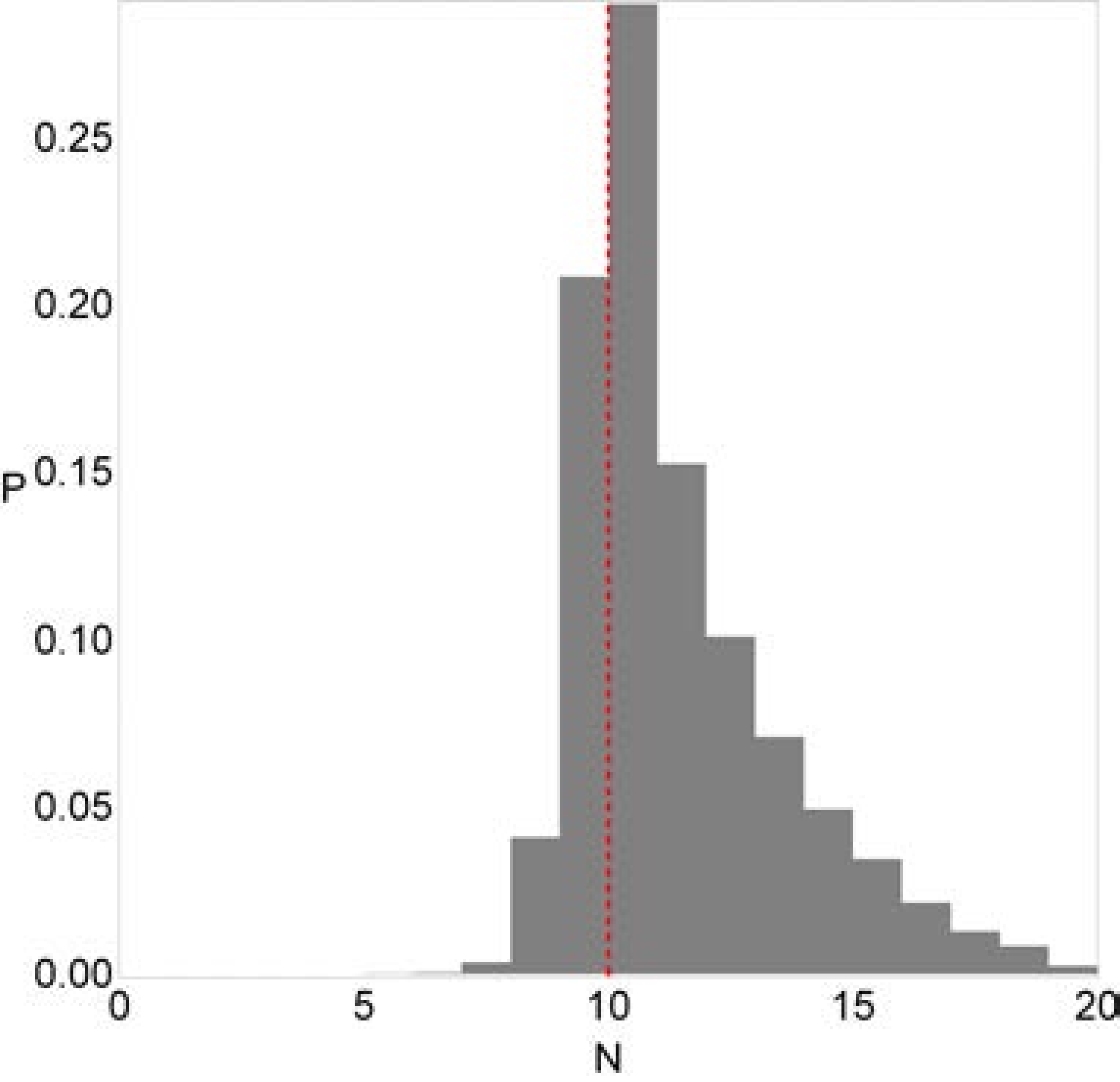}
\caption{The corresponding probability distributions of the required iterations for obtaining the  Newton-Raphson basins of attraction shown in Fig. \ref{NR_Fig_7}. (Color figure online).}
\label{NR_Fig_7b}
\end{figure*}
\begin{figure*}[!t]
\centering
(a)\includegraphics[scale=5]{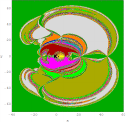}
(b)\includegraphics[scale=7]{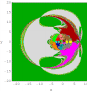}\\
(c)\includegraphics[scale=5]{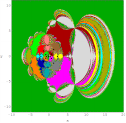}
(d)\includegraphics[scale=7]{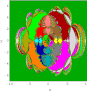}
\caption{The Newton-Raphson basins of attraction on the configuration $(x, y)$ plane for fixed value of  $\alpha=55.5 \degree$ (a) $\beta=24 \degree$, (b) $\beta=51 \degree$,  (c)  $\beta=53 \degree$, (d) $\beta=55 \degree$ when 15 libration points exist. The color code for libration points $L_1,...,L_{13}$ is same as in Fig.\ref{NR_Fig_7}. (Color figure online).}
\label{NR_Fig_8}
\end{figure*}
\begin{figure*}[!t]
\centering
(a)\includegraphics[scale=.35]{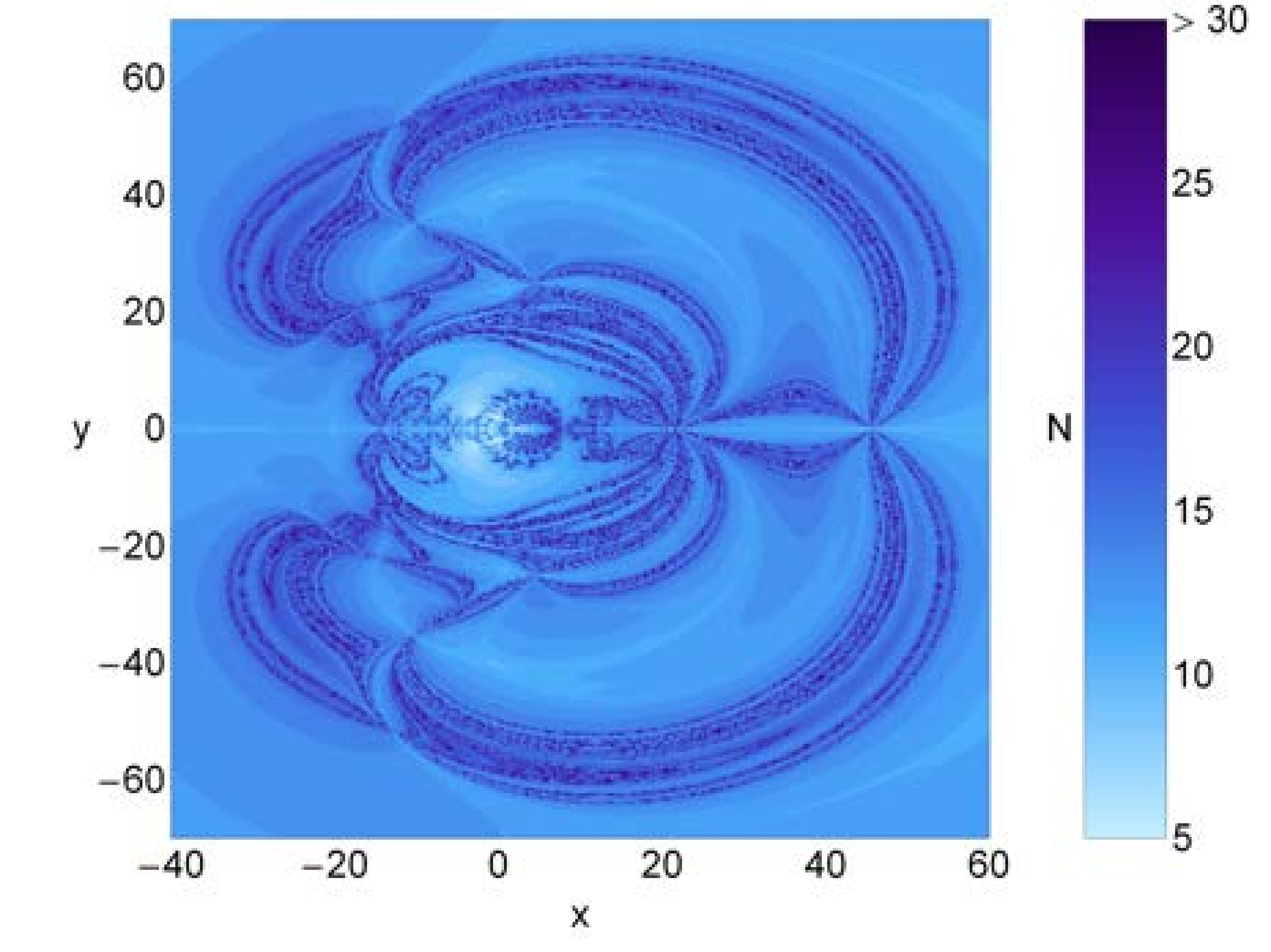}
(b)\includegraphics[scale=.35]{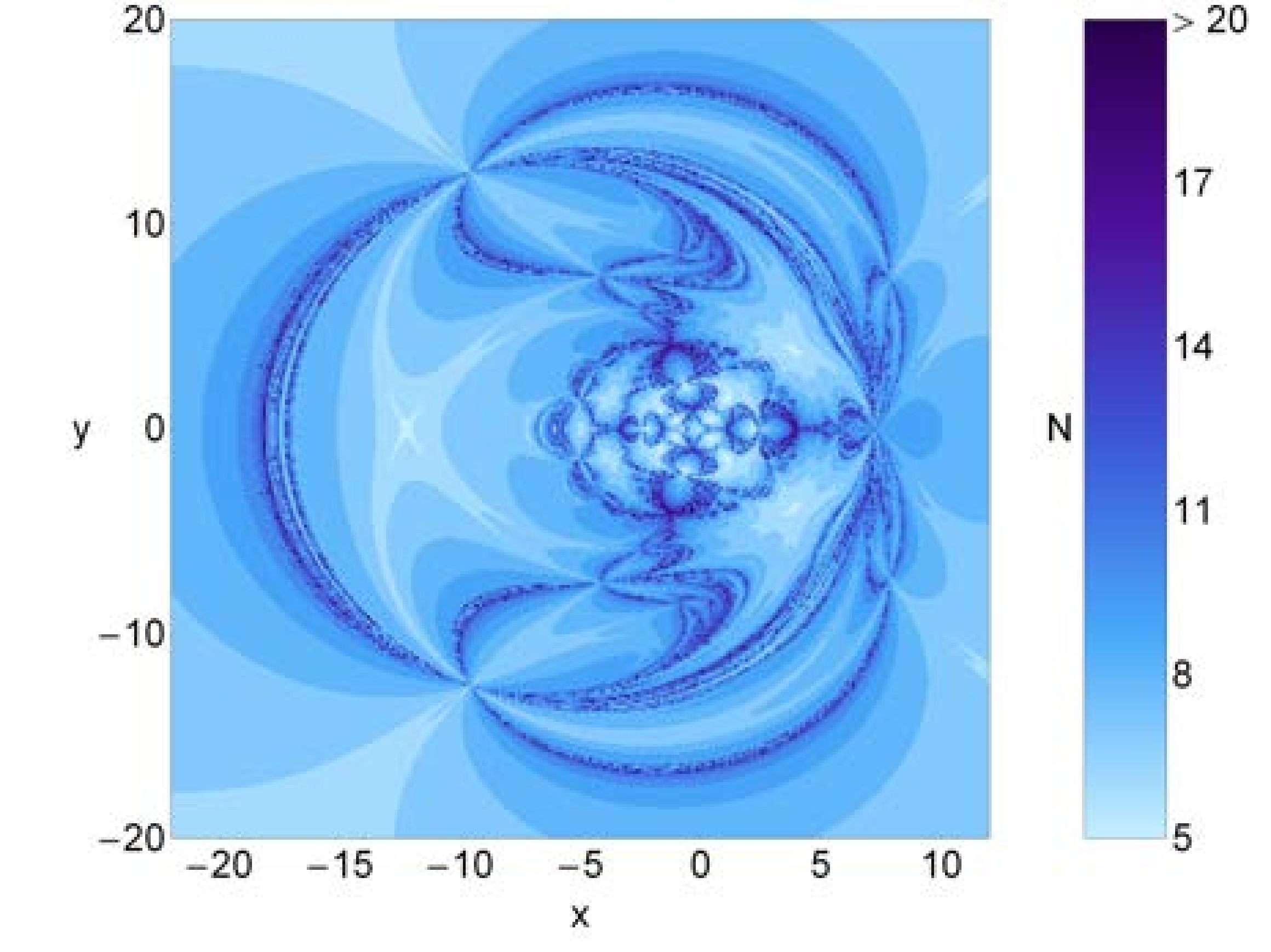}\\
(c)\includegraphics[scale=.35]{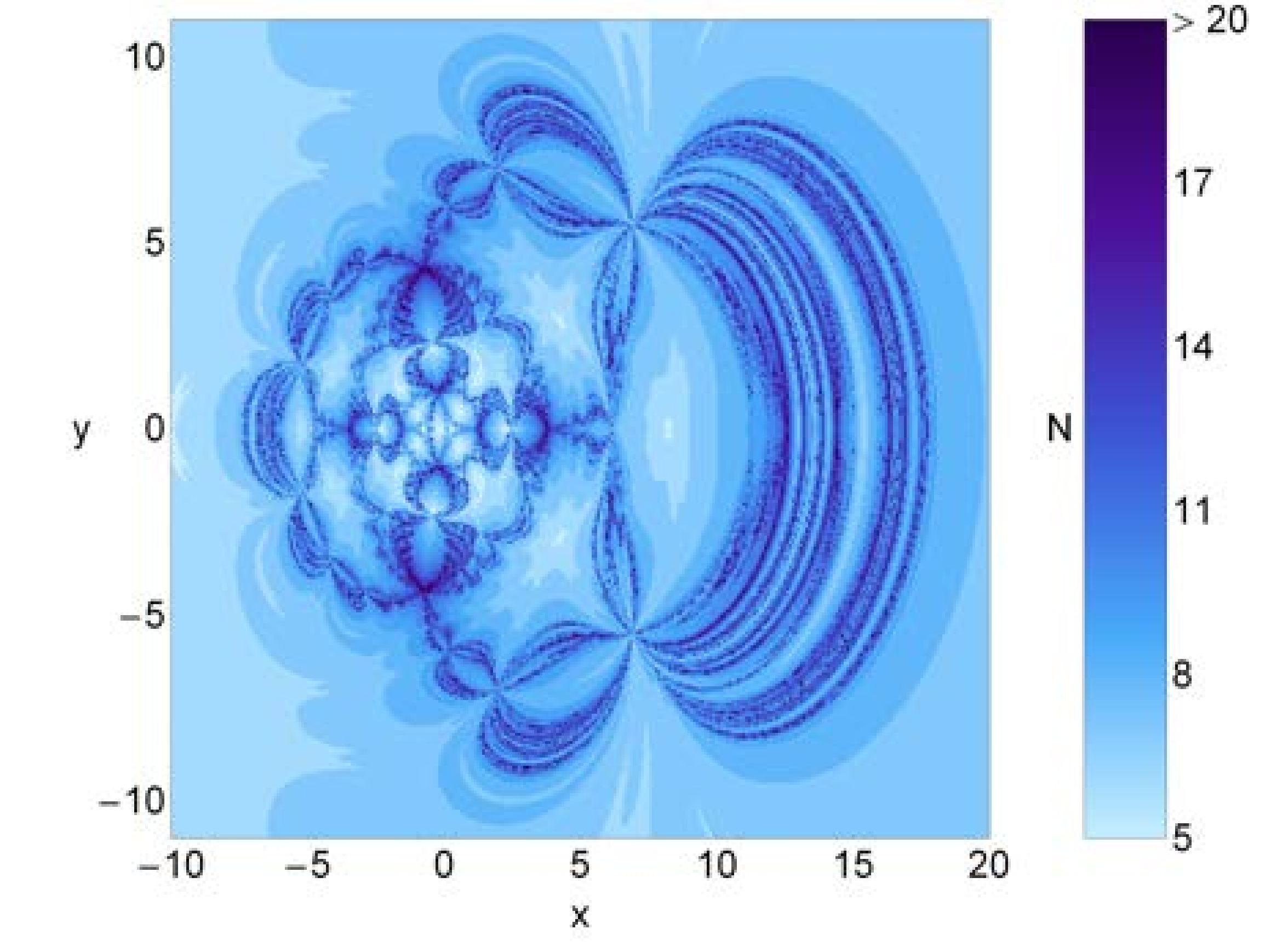}
(d)\includegraphics[scale=.35]{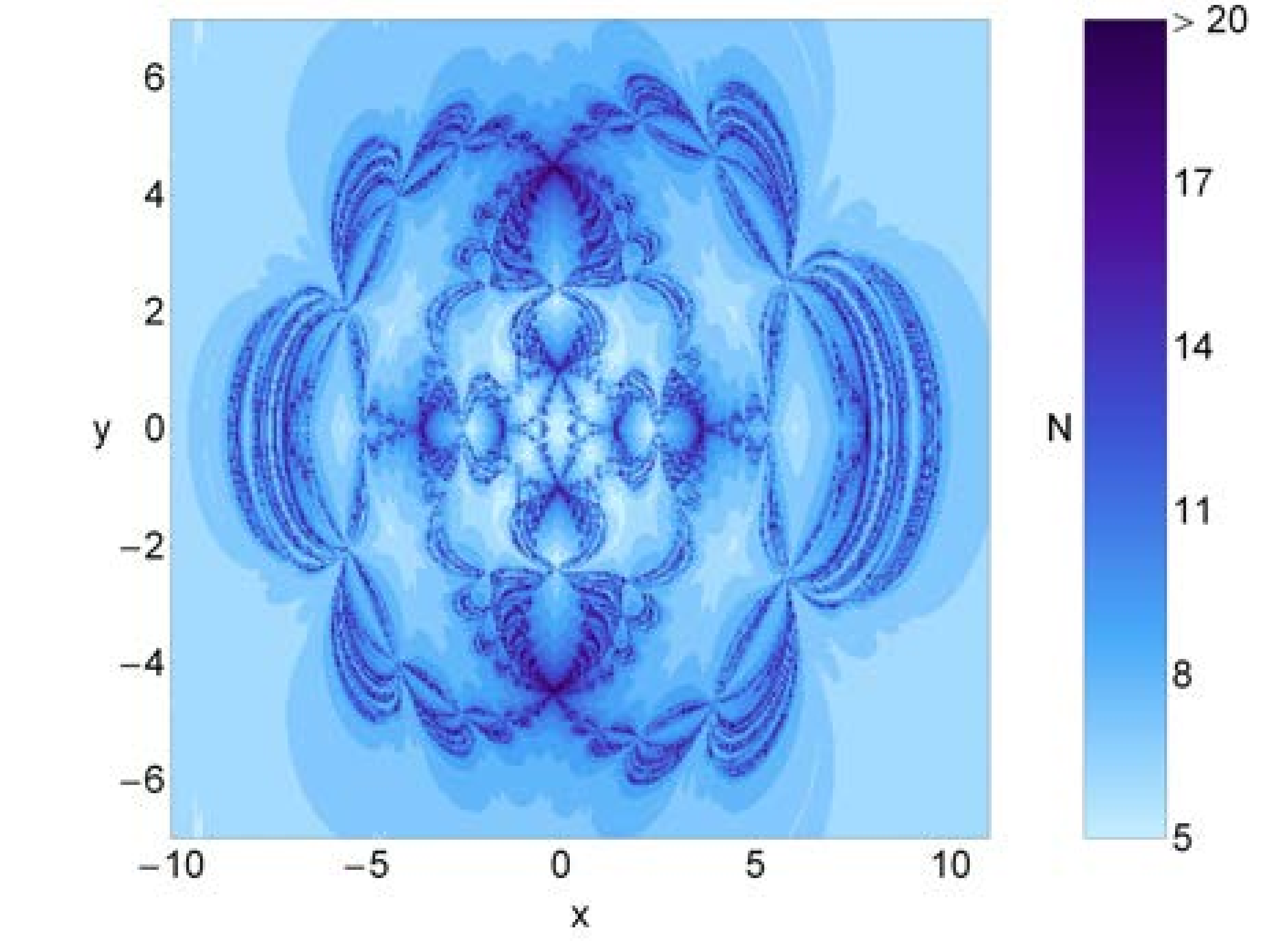}
\caption{The distribution of the corresponding number $N$ of the required iterations for obtaining the Newton-Raphson basins of attraction shown in Fig. \ref{NR_Fig_8}. (Color figure online).}
\label{NR_Fig_8a}
\end{figure*}
\begin{description}
  \item[-] In \ref{NR_Fig_5}a-e, the extent of the basins of convergence corresponding to the central libration point $L_2$ is infinite while in \ref{NR_Fig_5}f, the extent of basins of convergence associated with the libration point $L_3$ is infinite.
  \item[-] In \ref{NR_Fig_5}f, it can be easily seen that most of the configuration plane is covered by basins of convergence corresponding to central libration point $L_2$, of course, which is less than the basins of convergence corresponding to $L_3$ which has infinite extent. Moreover, the other regions on the configuration plane are highly chaotic mixture composed of the initial conditions mainly converges to the collinear libration points $L_1$, $L_2$ and $L_3$.
  \item[-] It can be observed that the basins of convergence associated with the libration point $L_1$ look like exotic bugs with many legs and antenna. If we see the basins of convergence associated with the libration points whose extent is finite, look like exotic bugs with two antennas. These two antennas decrease as the value of $\beta$ increases.
  \item[-] Looking at all the panels, we could claim that there exist unpredictable changes in the shape of basins of convergence associated with the libration points when $\beta$ varies.
  \item[-] In particular, when we analyze the basins of convergence in panel (c), we observe that the finite basins of convergence corresponding to each libration point look like "magnetic field line" shape. Where as the boundaries separating two domain of basins of convergence are highly chaotic and composed of the initial conditions on configuration $(x, y)$ plane.
\end{description}
In Fig. \ref{NR_Fig_5a}a-f, we have plotted the associated number $(N)$ of required iterations for obtaining the predefined  accuracy, while in Fig. \ref{NR_Fig_5b}a--f, we have illustrated the probability distributions of iterations. It is seen that $95\%$ of the initial conditions on the configuration plane converges to one of the attractors within the first 20 iterations. In panel (f), it can be observed that the number of iterations required to converge the initial conditions to any attractor is much higher, i.e., $N=45$ while for those initial conditions which lie in the domain of the basins of convergence associated with the libration point $L_2$ is relatively very low i.e., $N<8$. Moreover, the probability distributions unveil  that the average value of required number of iterations is unpredictable as it neither increases nor decreases if $\beta$ varies.

\subsubsection{$\alpha=59\degree$}
\label{Sec:4.2.3}
In this case, we have depicted the basins of convergence associated with the libration points when $\beta=22\degree, 27\degree, 36\degree$, $41\degree, 45\degree$ and $49\degree$ in six different panels of Fig. \ref{NR_Fig_6}. Looking at these panels, we observe a very interesting phenomenon associated with the extent of the basins of convergence. Indeed, the extent of the basins of convergence associated with the collinear libration point $L_2$ (panels: a-c), $L_3$ (panel: d), and  $L_1$ (panel: e,f) is always infinite, whereas on the other hand the extent of the basins of convergence associated with the non-collinear libration points are always finite. For further increase in the value of $\beta$, the finite domain of the basins of convergence are now surrounded by highly chaotic mixture composed of initial conditions. The chaotic mixture in Fig-\ref{NR_Fig_6}f is mainly composed of five types of libration points: (i) the initial conditions attracted by $L_5$; (ii) the initial conditions attracted by $ L_4$; (iii) the initial conditions attracted by $L_1$; (iv) the initial conditions attracted by $L_2$; (v) the initial conditions attracted by $ L_3$; while in Fig-\ref{NR_Fig_6}-(d, e) the chaotic mixture is composed of initial conditions attracted by almost each of the libration points. Moreover, we do not find any initial conditions for which the multivariate version of the Newton-Raphson iterative scheme does not converge to any of the attractors.

In Fig-\ref{NR_Fig_6a}(a-f), the distributions of the corresponding number $(N)$ of iterations necessary to obtain the predefined accuracy are illustrated, using tones of blue. It is very clear from Fig. \ref{NR_Fig_6a}(a-c) that the initial conditions lying inside the basins of attraction converge relatively fast $(N< 25)$ while the initial conditions lying in the vicinity of the basins boundaries have slowest converging points $(N>35)$. It is interesting to note that the number of iterations required to converge to at least one of the attractors increases (i.e.,  $N>85$) (see Fig-\ref{NR_Fig_6a}e) while it is $N>140$ (see Fig-\ref{NR_Fig_6a}f) for the initial conditions falling inside the chaotic regions.

The corresponding probability distributions of the iterations are given in Fig. \ref{NR_Fig_6b}. One can easily observe that the most probable number $(N^*)$ of iteration is not constant, i.e.,  it varies in each panel. When the angle parameter $\beta$ varies from $22\degree$ to $49\degree$ in Fig.-\ref{NR_Fig_6b}(a-f) the most probable number  $(N^*)$ of iteration varies from 13 (see panel-b, when $\beta=27$) to 32 (see panel-e, when $\beta=45$). Therefore, it is not possible to predict for any value of angle parameter $\beta$ what will be the most probable number $(N^*)$ of iteration.

\begin{figure*}[!t]
\centering
(a)\includegraphics[scale=.4]{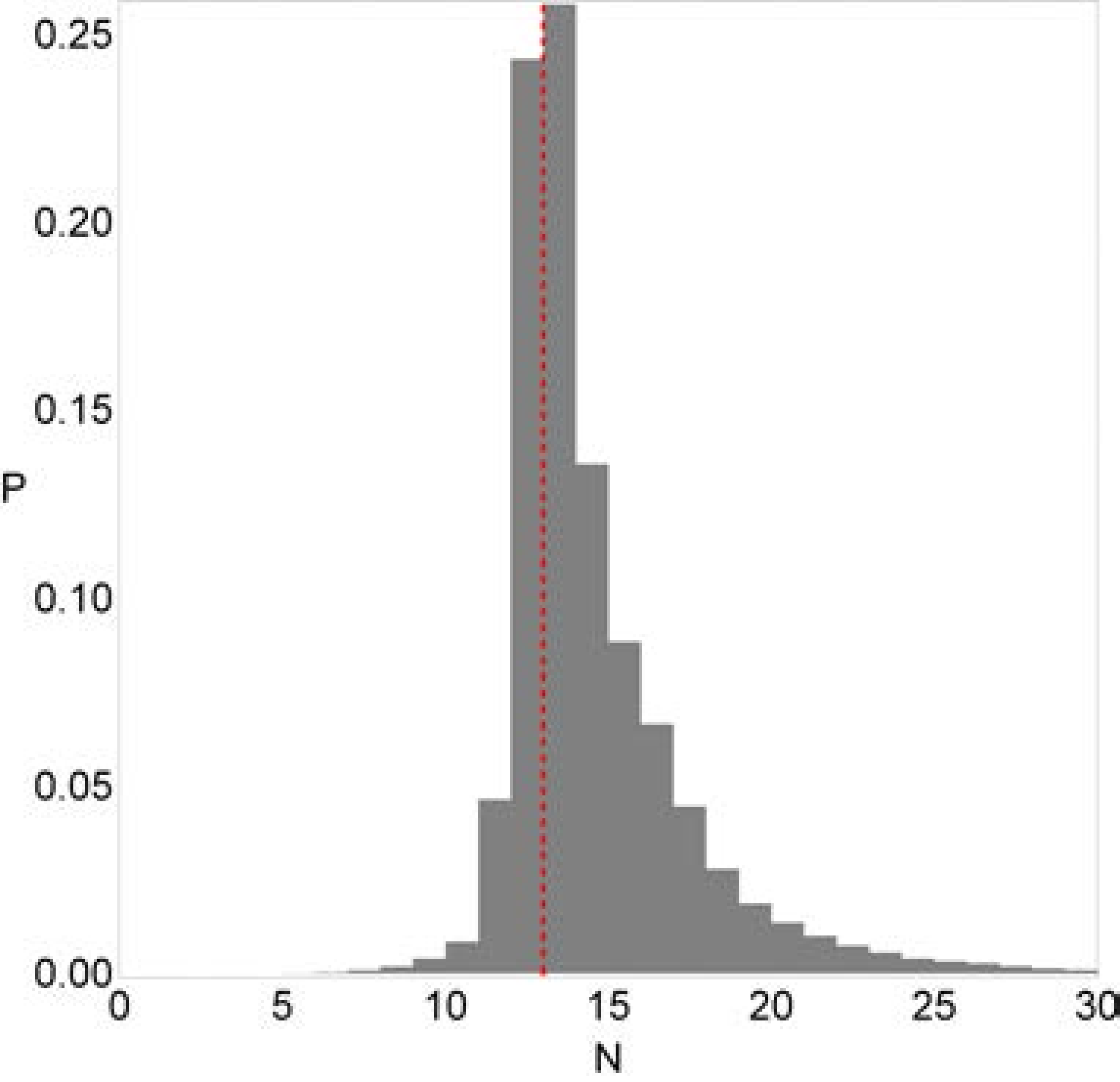}
(b)\includegraphics[scale=.4]{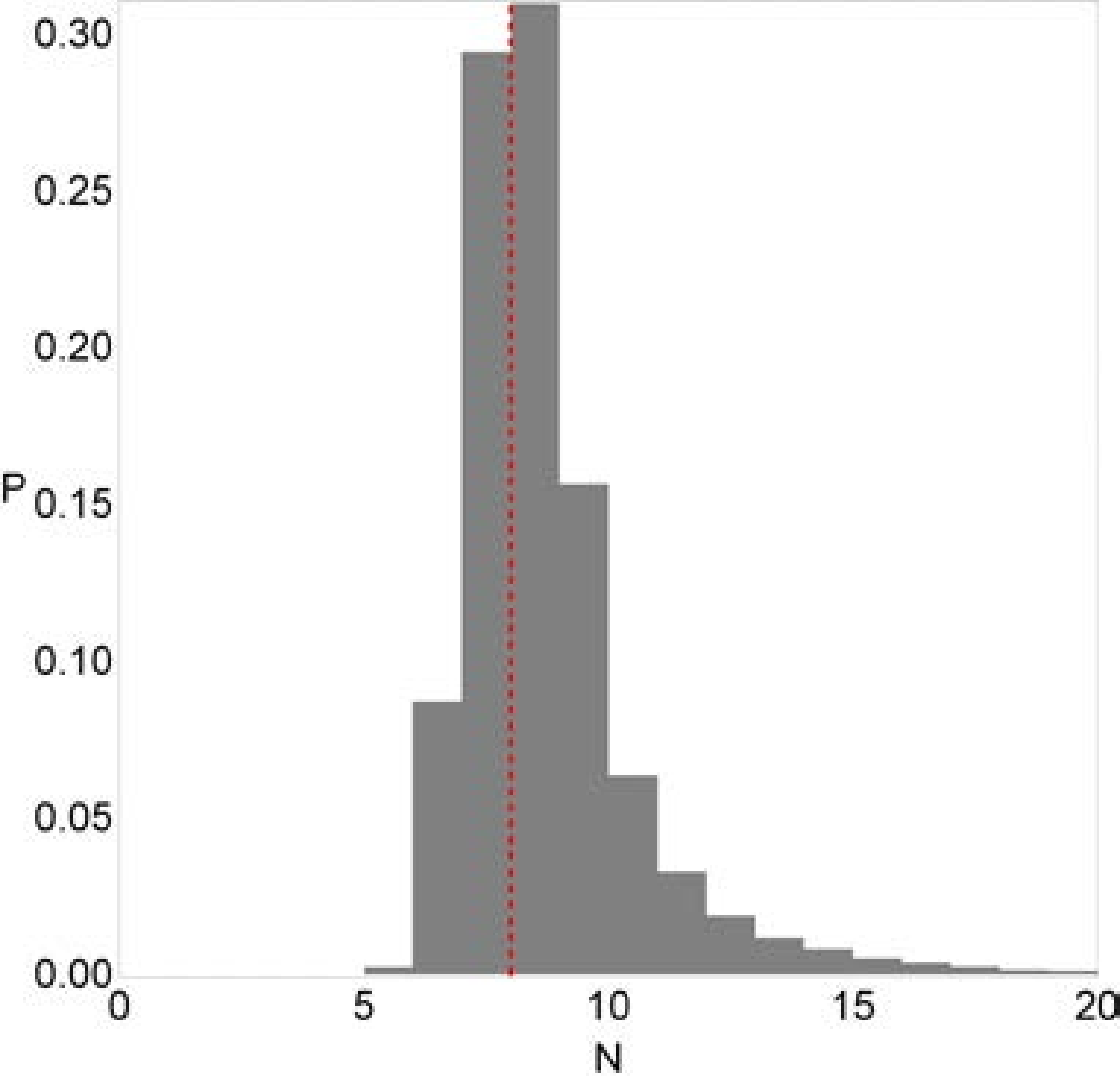}\\
(c)\includegraphics[scale=.4]{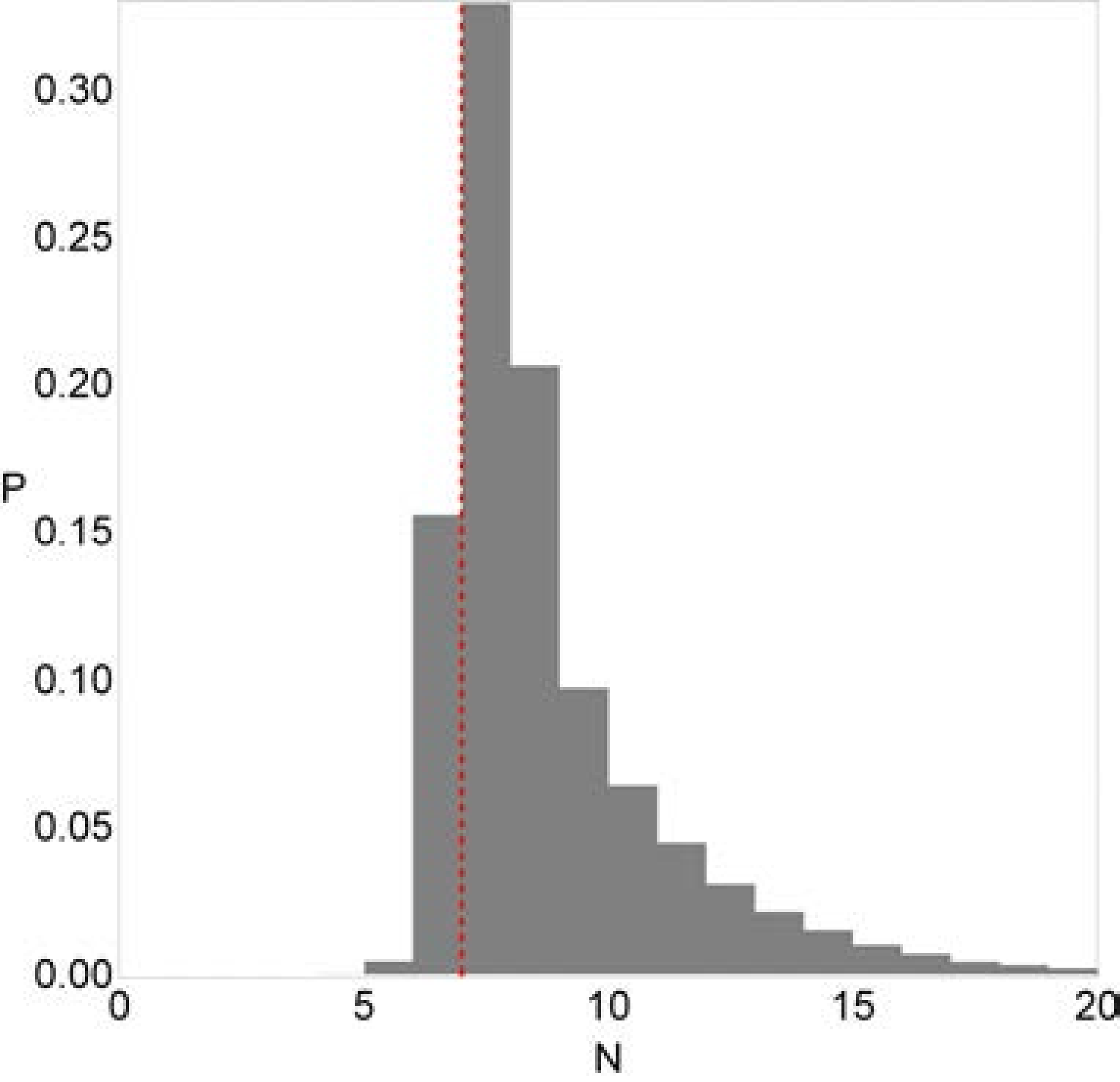}
(d)\includegraphics[scale=.4]{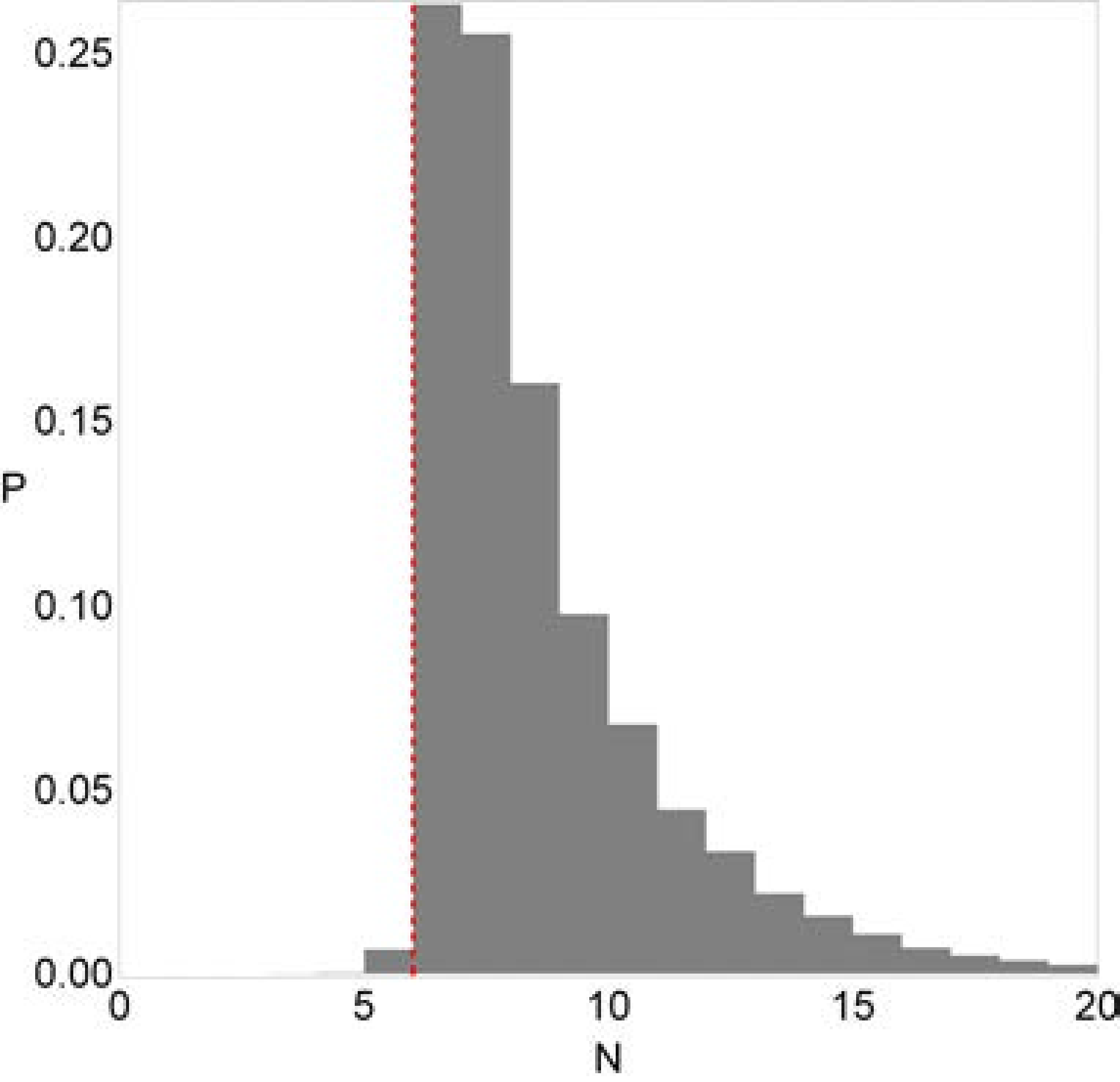}
\caption{The corresponding probability distributions of the required iterations for obtaining the Newton-Raphson basins of attraction shown in Fig. \ref{NR_Fig_8}. (Color figure online).}
\label{NR_Fig_8b}
\end{figure*}

\subsection{Case III: fifteen libration points exist}
\label{Sec:4.3}
The last case under consideration concerns the scenario where the axisymmetric five-body problem has 15 libration points.  We have considered three subcases based on the value of the angle parameter $\alpha$.
\subsubsection{$\alpha=43.5\degree$}
\label{Sec:4.3.1}
Our analysis deals with the value of angle parameter $\alpha=43.5\degree$ in this subcase. In Fig. \ref{NR_Fig_7}(a--d), we have illustrated the basins of convergence for four different values of the angle parameter $\beta$, by using the multivariate version of Newton-Raphson iterative scheme. We observe that in this range of values of angle parameter $\beta$, the changes on the topology of the basins of convergence on the configuration $(x, y)$ plane, due to the variation of this parameter, are effected substantially.

We can observe that in Fig. \ref{NR_Fig_7}(a-d), the extent of the basins of convergence associated with the central libration point $L_2$ is always infinite. On the other hand, the basins of convergence corresponding to other libration points are always finite and well formed. It can be also observed that as the value of angle parameter $\beta$ increases, the range of the basins of convergence corresponding to finite extent increase. It is clear that the basins of convergence corresponding to libration points $L_{14, 15}$ increase relatively fast in comparison of the finite domains corresponding to other libration points. As it is seen in Fig. \ref{NR_Fig_7}(c-d) for the angle parameter $\beta= 34\degree, 34.458\degree$, respectively, the most of finite domain of the basins of convergence is covered by the domain of convergence associated with the libration points $L_{14, 15}$.

The corresponding number $(N)$ of the required iterations for the desired accuracy is illustrated in Fig. \ref{NR_Fig_7a}(a-d), whereas the probability distributions of iterations is depicted in Fig. \ref{NR_Fig_7b}(a-d). It is clear that for obtaining the coveted accuracy, the iterative scheme needs not more than 20 iterations for more than 95 \% of the initial conditions which are examined. Furthermore, in this case the average value of needed number $(N)$ of iterations in Fig. \ref{NR_Fig_7b}(a-c) remains constant, $N^*=9$  whereas it is 10 for Fig. \ref{NR_Fig_7b}d, for all  examined values of $\beta$.
\subsubsection{$\alpha=55.5\degree$}
\label{Sec:4.3.2}
We continue with the case where we consider the value of $\alpha=55.5\degree$, and for four values of angle parameter $\beta$ are depicted in Fig. \ref{NR_Fig_8}. It is necessary to note that in Fig. \ref{NR_Fig_8}a, where $\beta=24\degree$, there exist three collinear and 12 non-collinear libration points whereas Fig. \ref{NR_Fig_8}(b-d) corresponds to the case where five collinear and ten non-collinear libration points exist. We can observe that as the value of $\beta$ increases, the finite regions of the basins of convergence are reduced rapidly and of course the extent of the basins of convergence associated with the libration point $L_2$ increases.

In Fig. \ref{NR_Fig_8a},  the corresponding number $(N) $ of required iterations is depicted, using tones of blue. Moreover, the associated probability distribution of iterations is  presented in Fig. \ref{NR_Fig_8b}(a-d). It is observed that the most probable number $N^*$ is 14 for $\beta=24\degree$, then it decreases to 8 when $\beta$ increases to $51\degree$ and it decreases until it reaches $N^*=6$, the lowest observed value for $\beta=55\degree$.

\subsubsection{$\alpha=59\degree$}
\label{Sec:4.3.3}
\begin{figure*}[!t]
\centering
(a)\includegraphics[scale=5]{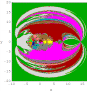}
(b)\includegraphics[scale=5]{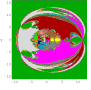}\\
(c)\includegraphics[scale=5]{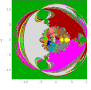}
(d)\includegraphics[scale=5]{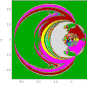}\\
(e)\includegraphics[scale=.65]{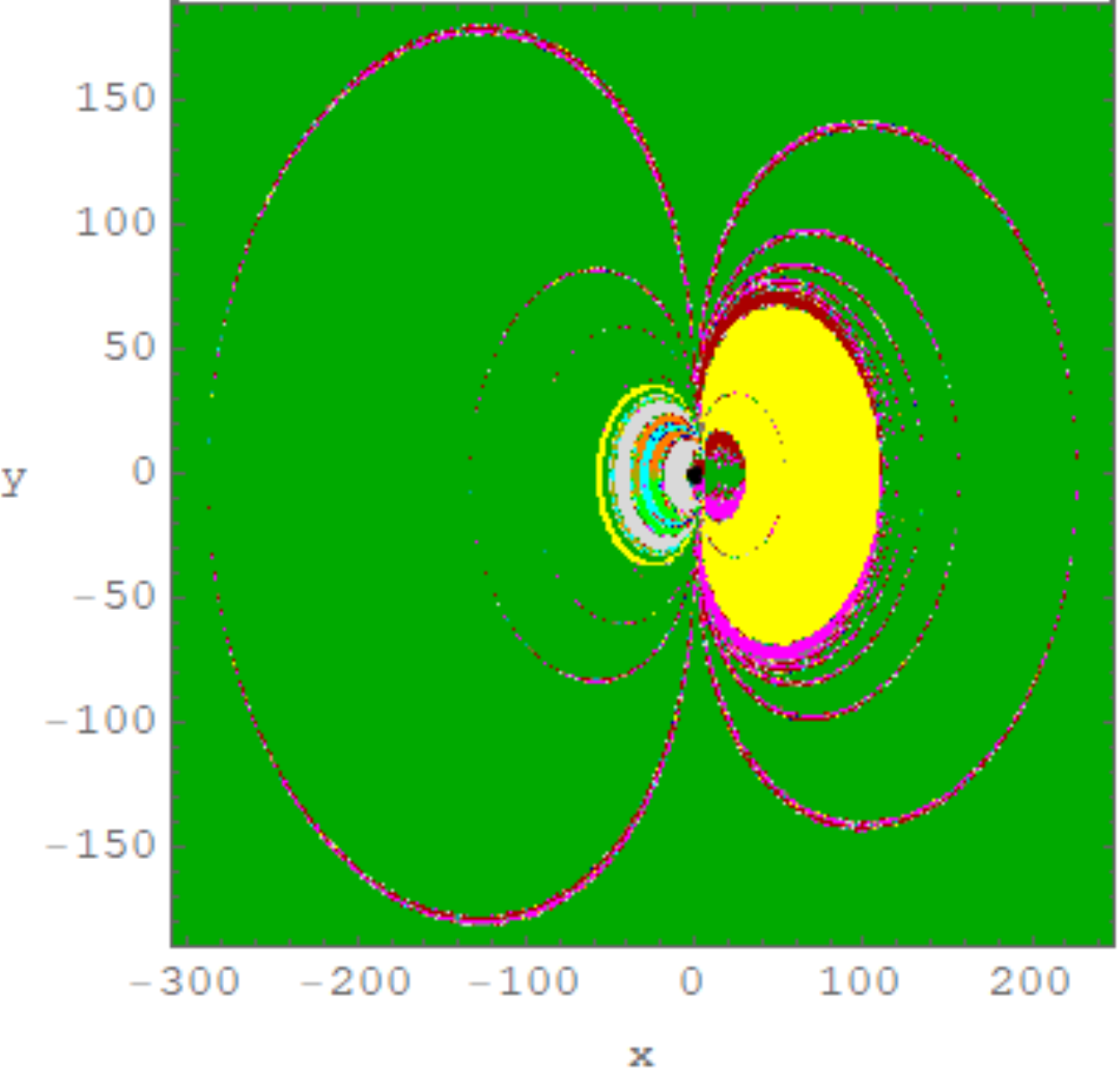}
\caption{The Newton-Raphson basins of attraction on the configuration $(x, y)$ plane for fixed value of $\alpha=59 \degree$ (a) $\beta=50\degree$, (b) $\beta=51\degree$, (c) $\beta=52\degree$, (d) $\beta=53\degree$,(e) $\beta=54 \degree$, when 15 libration points exist. The color code for the attractors is same as in Fig. \ref{NR_Fig_7}. (Color figure online).}
\label{NR_Fig_9}
\end{figure*}

\begin{figure*}[!t]
\centering
(a)\includegraphics[scale=.35]{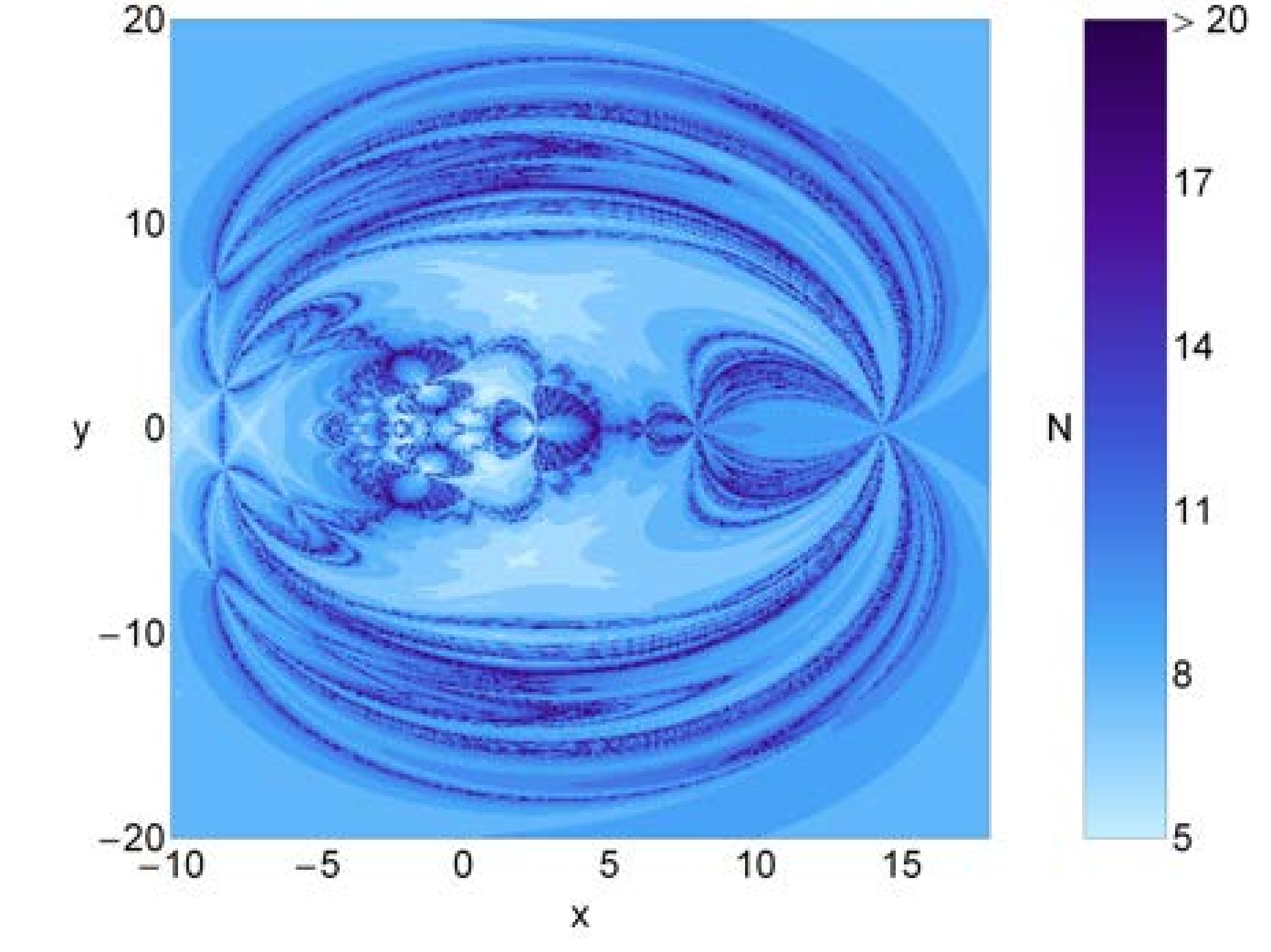}
(b)\includegraphics[scale=.35]{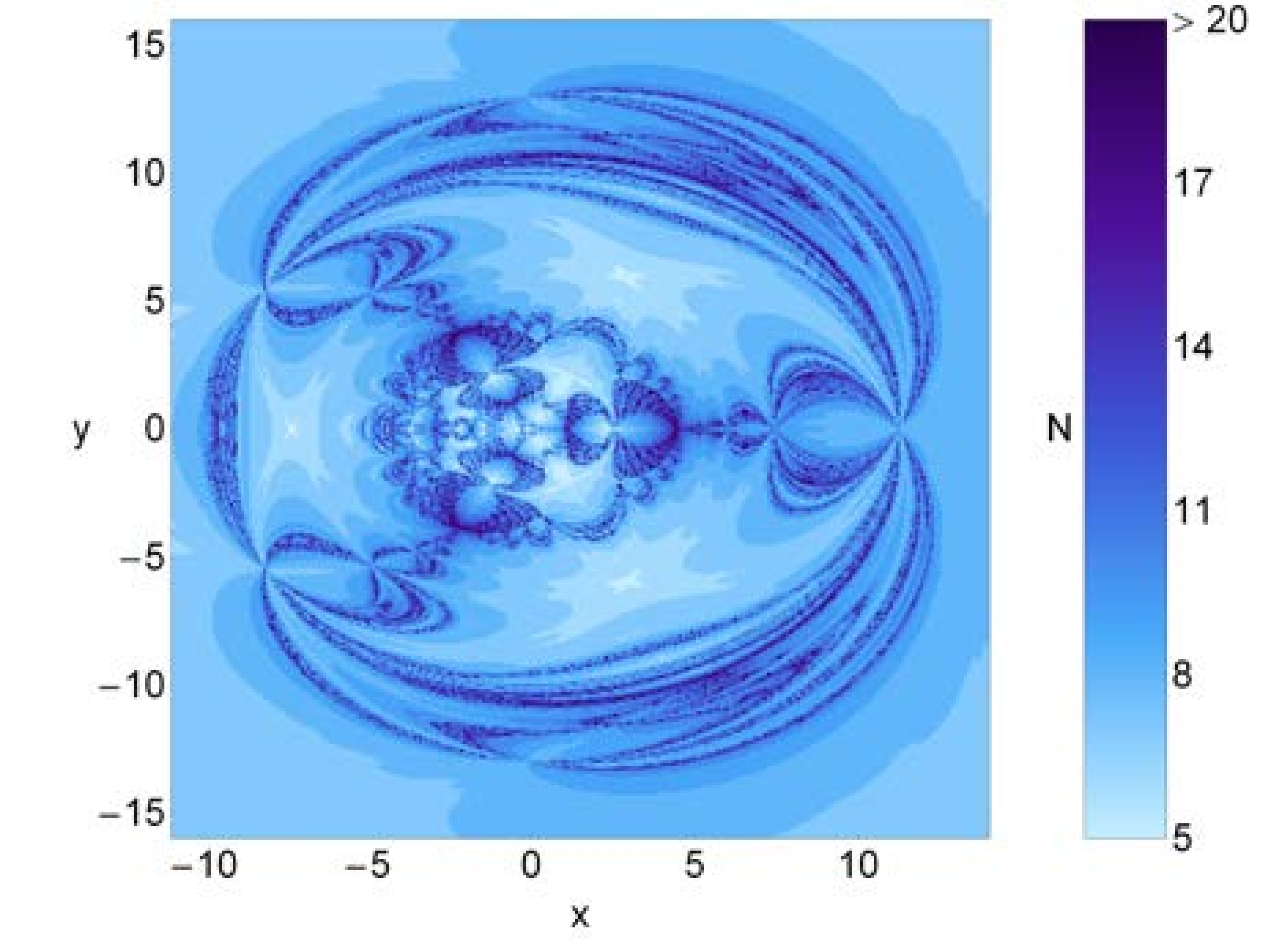}\\
(c)\includegraphics[scale=.35]{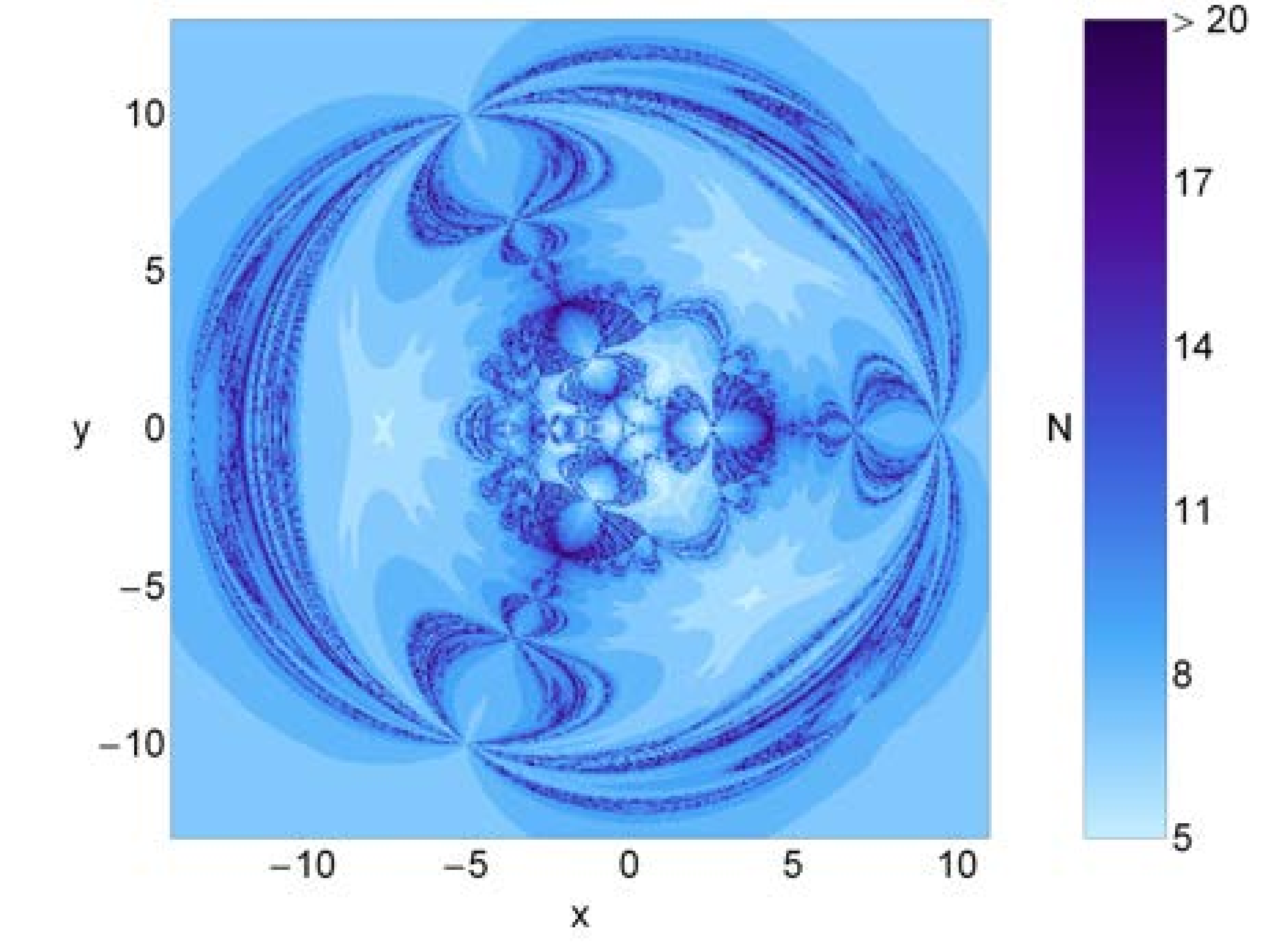}
(d)\includegraphics[scale=.35]{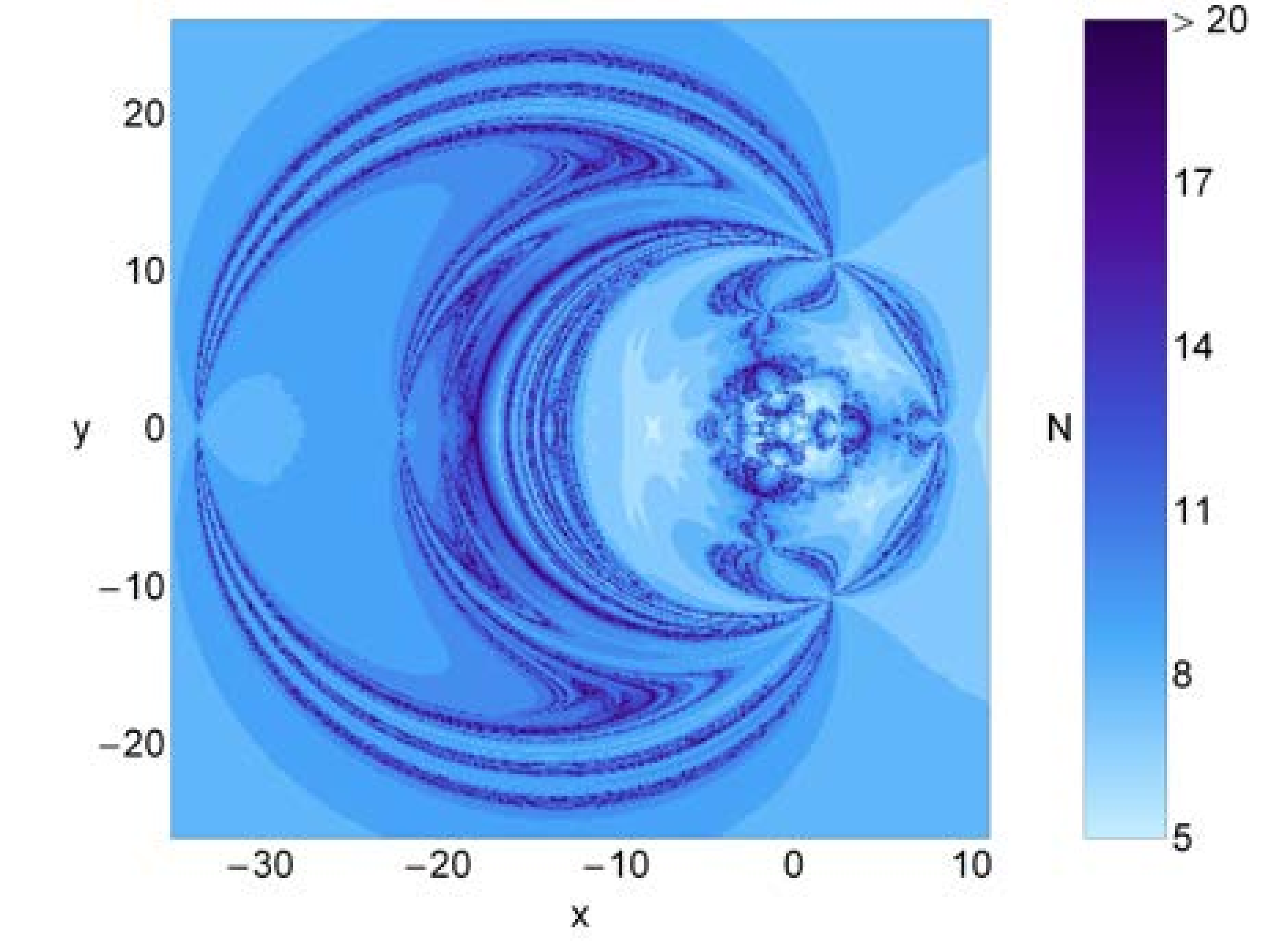}\\
(e)\includegraphics[scale=.35]{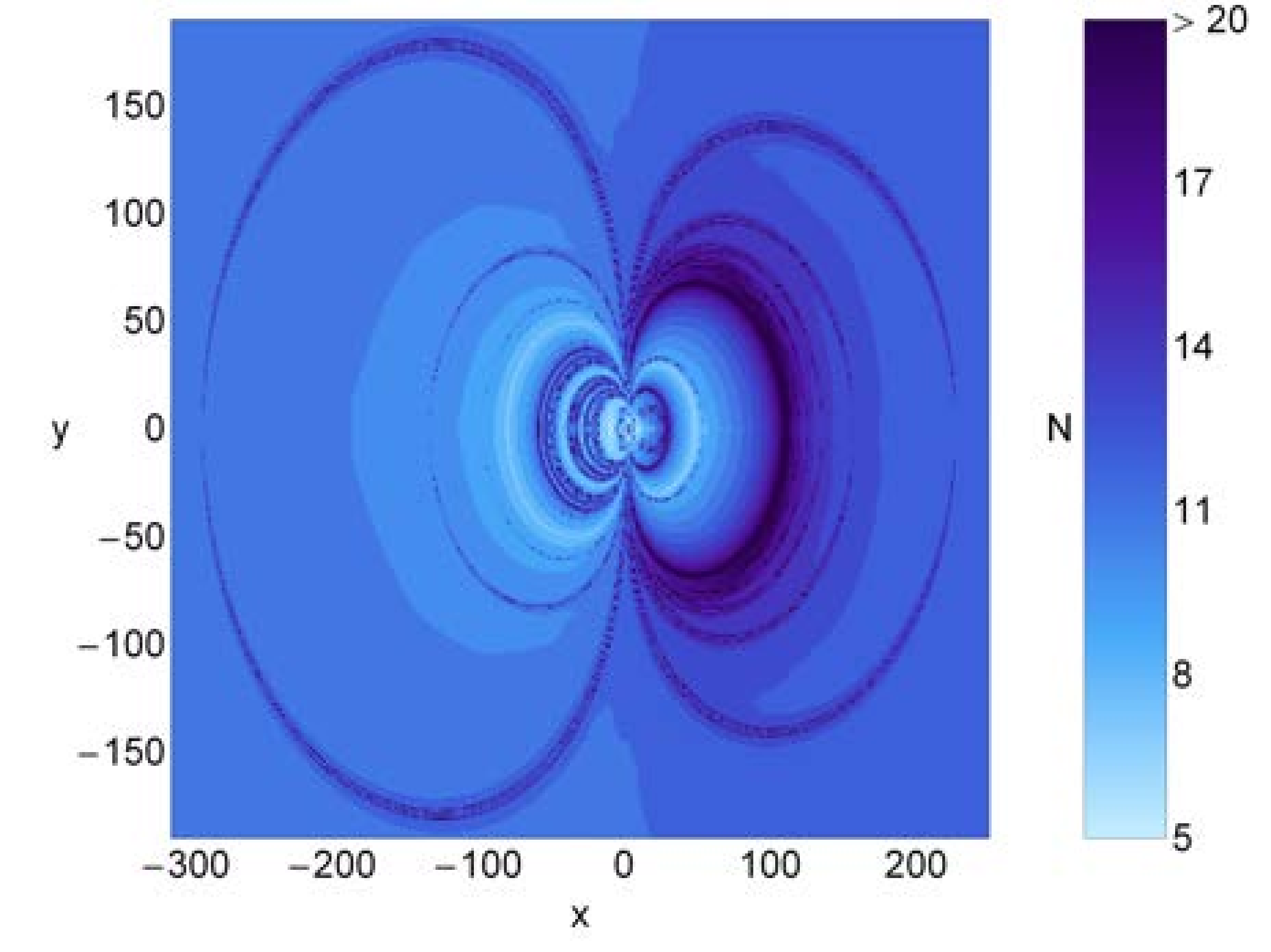}
\caption{The distribution of the corresponding number $N$ of the required iterations for obtaining the Newton-Raphson basins of attraction shown in Fig.\ref{NR_Fig_9}.  (Color figure online).}
\label{NR_Fig_9a}
\end{figure*}

\begin{figure*}[!t]
\centering
(a)\includegraphics[scale=.4]{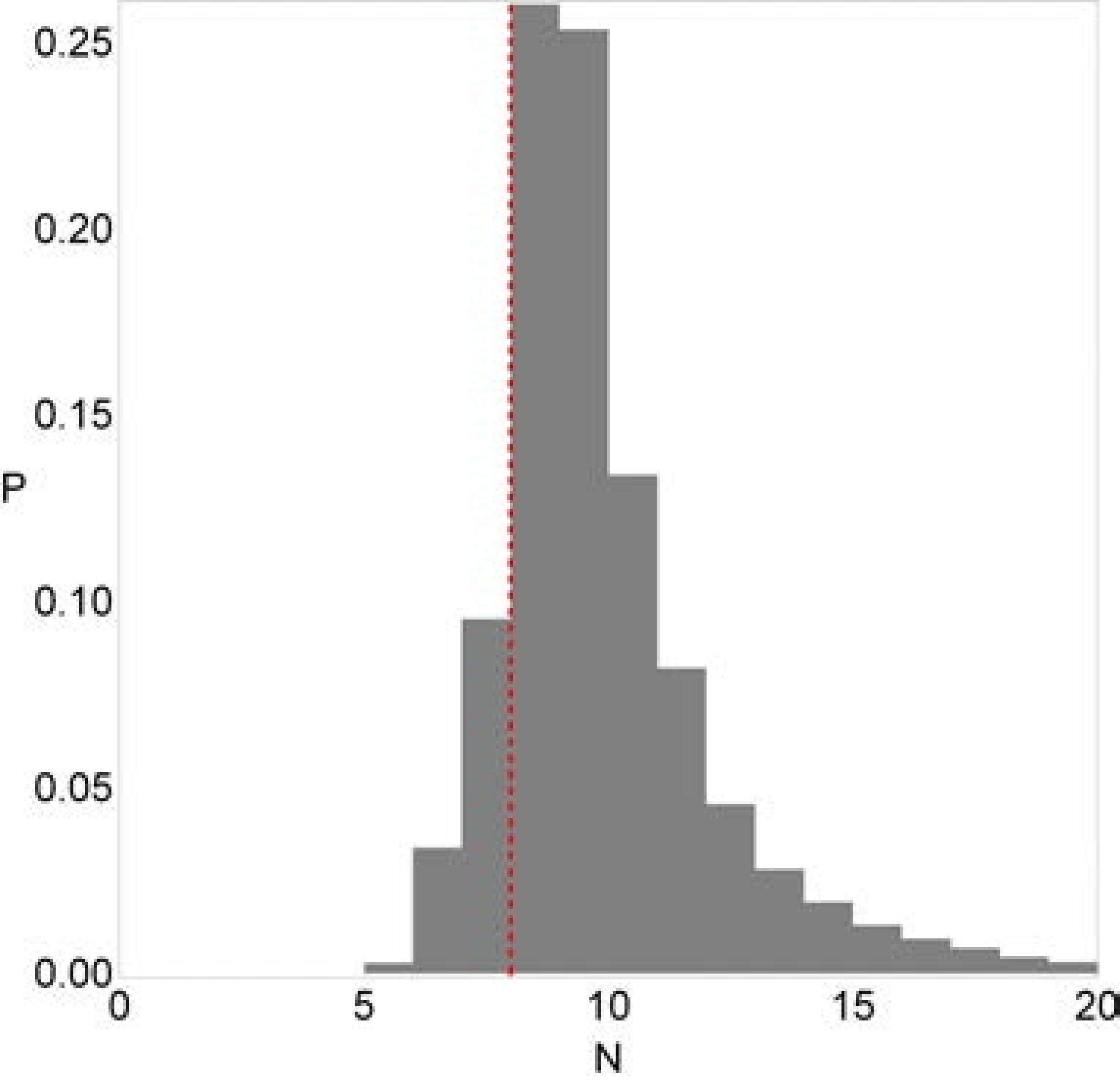}
(b)\includegraphics[scale=.4]{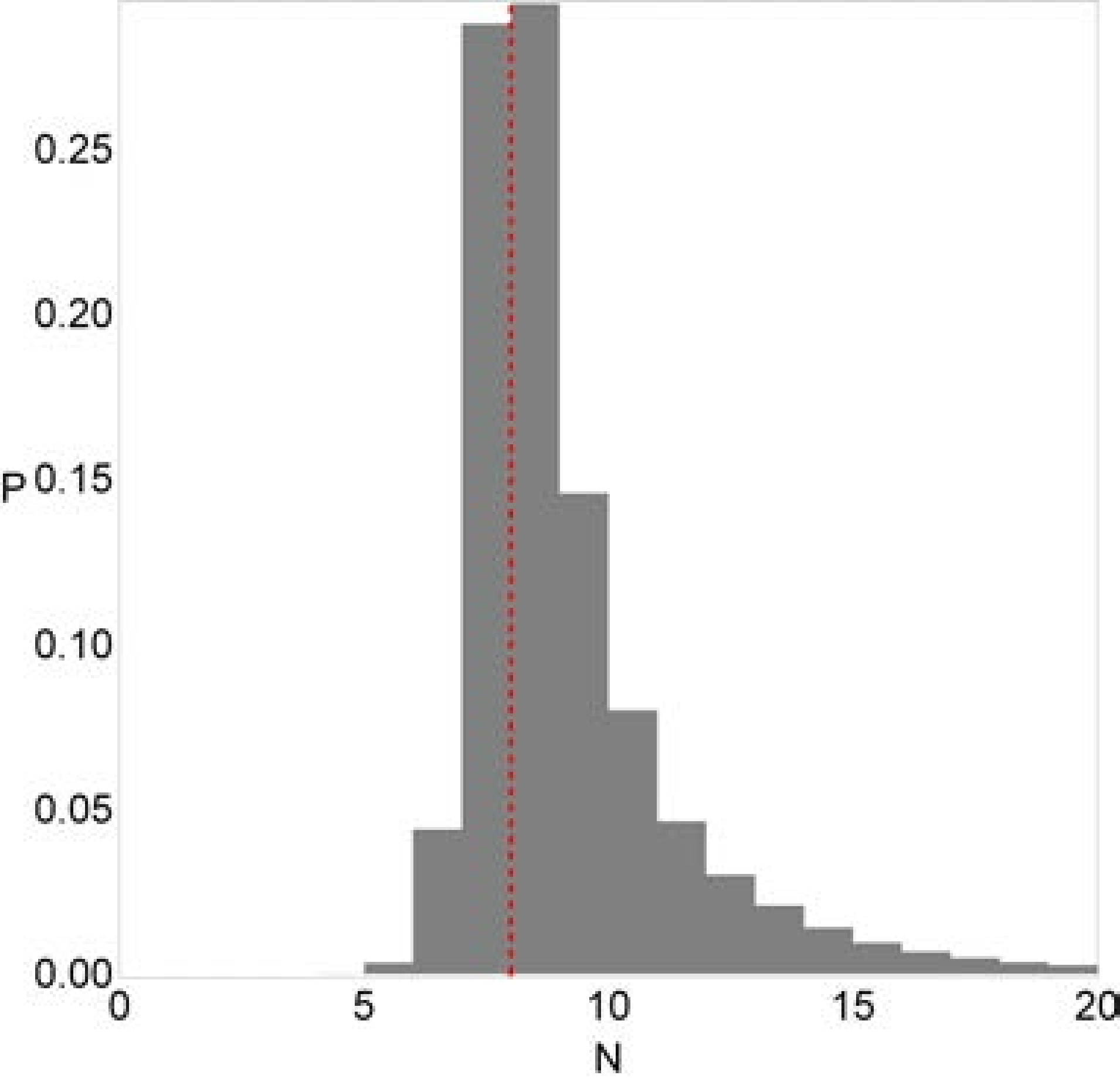}\\
(c)\includegraphics[scale=.4]{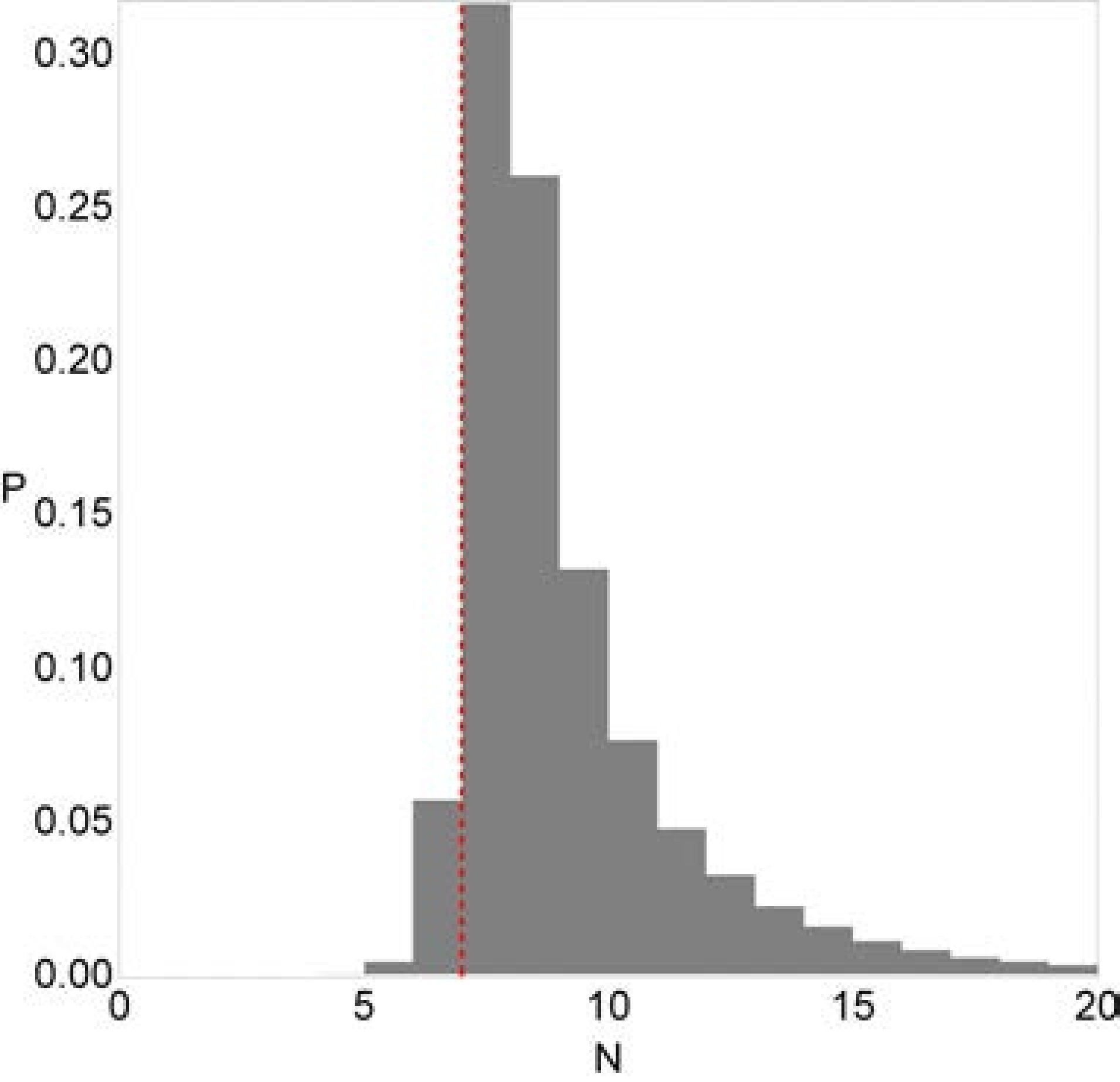}
(d)\includegraphics[scale=.4]{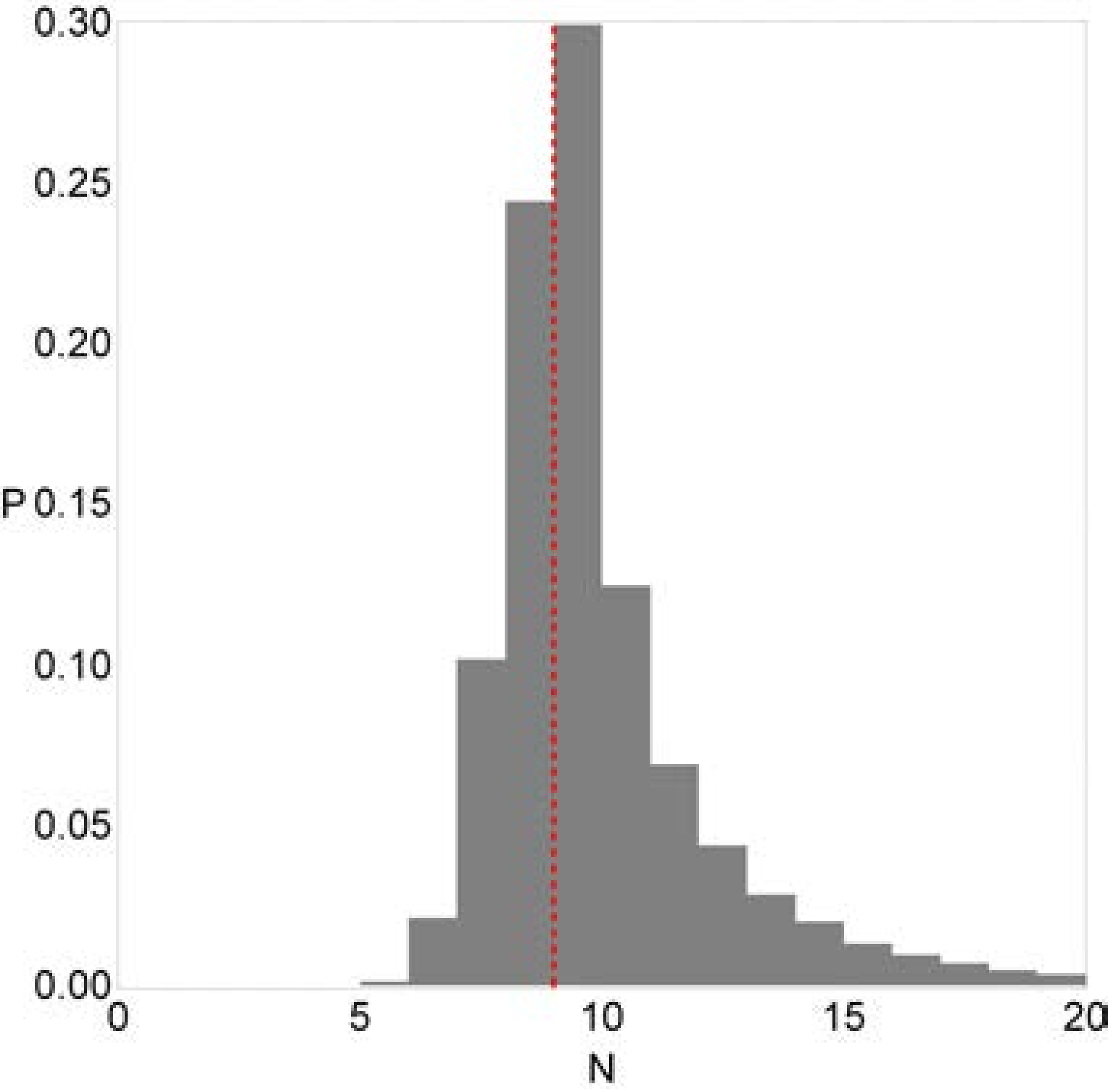}\\
(e)\includegraphics[scale=.4]{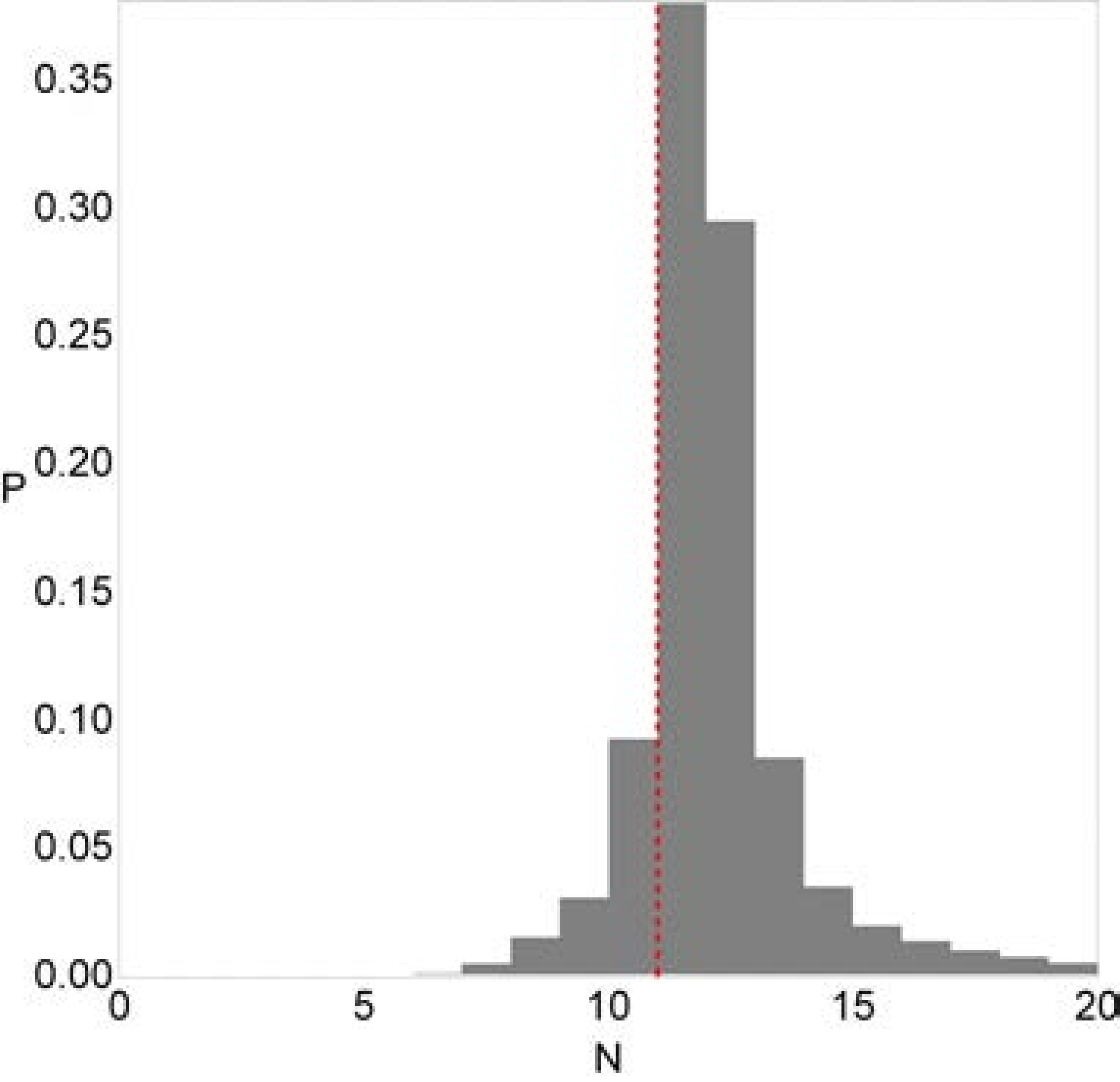}
\caption{The corresponding probability distributions of the required iterations for obtaining the Newton-Raphson basins of attraction shown in Fig. \ref{NR_Fig_9}. (Color figure online).}
\label{NR_Fig_9b}
\end{figure*}

The present subsection deals with the case when $\alpha=59\degree$. In Fig. \ref{NR_Fig_9}, the Newton-Raphson basins of convergence are depicted for five values of the angle parameter $\beta$. It is obvious that the extent of the basins of convergence corresponding to libration point $L_2$ is infinite where as the extent of the basins of convergence corresponding to all the remaining libration points are finite and well shaped. We have observed that the topology of the basins of convergence changes drastically when the angle parameter $\beta$ increases. Moreover, the topology of the finite extent of the basins of convergence is neither increasing nor decreasing uniformly. In Fig. \ref{NR_Fig_9a}a-e, we have discussed the associated number $(N)$ of required iterations to  obtain the predefined  accuracy, whereas  the probability distributions of iterations are illustrated in Fig. \ref{NR_Fig_9b}a--e .

\section{Concluding remarks}
\label{concluding remarks}

In the present study, the Newton-Raphson basins of convergence were numerically unveiled for the convex configuration case of the axisymmetric five-body problem. In particular, we revealed how the angle parameters $\alpha$ and $\beta$ influence the basins of convergence associated with the libration points. The multivariate version of the Newton-Raphson iterative scheme was used for unveiling the basins of convergence associated with the equilibrium points on the configuration $(x, y)$ plane. These convergence regions play a very crucial role, since they unveil how each point of the configuration $(x, y)$ plane is numerically attracted by the libration points which act as attractors.

The most notable outcomes of this numerical analysis can be summarized as follows:
\begin{description}
  \item[-] There exist either eleven, thirteen or fifteen libration points for different combination of the angle parameters.
  \item[-] None of the libration points was found to be stable for the considered values of angle parameters.
  \item[-] In almost all the cases it has been observed that the basins of convergence corresponding to the libration point $L_2$ have infinite extent. Moreover, for some combination of the angle parameters $\alpha$ and $\beta$, the libration points $L_{1, 3}$ have also infinite extent.
  \item[-] It is necessary to note that there is no initial conditions on the configuration $(x, y)$ plane which act as non-converging nodes (i.e., the initial conditions which do not converge) or false-converging nodes to final states different, with respect to the libration points of the system.
  \item[-] It was found that the average value of the required iterations $(N^*)$ for obtaining the predefined accuracy is not fixed, i.e., it decreases, increases or remains constant with increasing value of the angle parameter $\beta$, for different combination with angle parameter $\alpha$.
\end{description}

The graphical illustration of the paper was completed with the help of the latest version 11.3 of Mathematica$^\circledR$ \cite{wol03}. For the classification of every set of the initial conditions on the configuration $(x, y)$ plane, we required about 2 hours of CPU time, using a Quad-Core i5 2.67 GHz PC. It should be emphasized that the iterative procedure is effectively ended when an initial condition converge to one of the libration points and the iterative scheme proceeds to next available initial condition.

\section{Future work}
\label{Future work}

In this study, we investigated the basins of convergence associated with coplanar libration points of the axisymmetric five-body problem in the convex case only whereas, in \cite{gao17}, the libration points were also found in the concave case of the axisymmetric five-body problem. On this basis, a future paper will be based on the fractal basins of convergence associated with libration points in the concave case. Moreover, the existence of the \emph{out-of-plane} libration points in both types of configurations, i.e., concave and convex cases of the axisymmetric five-body problem will be explored in future. In particular, we will try to reveal how the angle parameters $\alpha$ and $\beta$ influence the stability as well as the overall basins of convergence, associated with these out-of-plane equilibrium points.

\section*{Acknowledgments}
\footnotesize
\begin{description}
  \item[*] The authors are thankful to Center for Fundamental Research in Space dynamics and Celestial mechanics (CFRSC), New Delhi, India for providing research facilities.
  \item[*] The authors would like to express their warmest thanks to the anonymous referee for the careful reading of the manuscript and for all the apt suggestions and comments which allowed us to improve both the quality and the clarity of the paper.
\end{description}
\section*{Compliance with Ethical Standards}
\begin{description}
  \item[-] Funding: The authors state that they have not received any research
grants.
  \item[-] Conflict of interest: The authors declare that they have no conflict of
interest.
\end{description}

\end{document}